\documentclass[11pt]{article}

\RequirePackage[OT1]{fontenc}
\usepackage{amsthm,amsmath,natbib}
\RequirePackage[colorlinks,citecolor=blue,urlcolor=blue]{hyperref}
\usepackage{xcolor}
\usepackage{color}

\def\bi{\begin{itemize}}
\def\ei{\end{itemize}}
\def\be{\begin{equation}}
\def\ee{\end{equation}}

\usepackage[english]{babel}
\usepackage{amsfonts}
\usepackage{amssymb}
\usepackage{graphicx}
\usepackage{enumerate}
\usepackage{bm}
\usepackage{umoline}
\usepackage{algorithm}
\usepackage{algorithmic}

\newtheorem{theorem}{Theorem}

\def\modif#1{{{#1}}}   
\def\newmodif#1{{\textcolor{black}{#1}}}

\title{\modif{Semi-parametric} resampling with extremes}

\author{Thomas Opitz$^a$, Denis Allard$^a$, Gr\'egoire Mariethoz$^b$, \\
\small{$^a$Biostatistics and Spatial Processes (BioSP), INRAE, 84914 Avignon, France}\\
\small{$^b$Facult\'e des g\'eosciences et de l'environnement, UNIL, Lausanne, Switzerland}
}
\date{\today}
\begin{document}
	\maketitle
	
\begin{center}
\begin{minipage}{0.8\linewidth}
{\textbf{Abstract}}
Nonparametric resampling methods such as Direct Sampling are powerful tools to simulate new datasets preserving important data features such as spatial patterns from \newmodif{observed datasets} while using only minimal assumptions. However, such methods cannot generate extreme events beyond the observed range of data values. We here propose using tools from extreme value theory for stochastic processes to extrapolate observed data towards yet unobserved high quantiles. Original  data are first enriched with new values in the tail region, and then classical resampling algorithms are applied to enriched data. 
In a first approach to enrichment that we label ``naive resampling", we generate an independent sample of the marginal distribution while keeping the rank order of the observed data. We point out inaccuracies of this approach around the most extreme values, and therefore develop a second approach that works for datasets with many replicates. It is based on the asymptotic representation of  extreme events through two stochastically independent components: a magnitude variable, and a profile field describing spatial variation. To generate enriched data, we fix a target range of return levels of the magnitude variable, and we resample magnitudes constrained to this range. We then use  the second approach to generate heatwave scenarios of yet unobserved magnitude over France, based on daily temperature reanalysis training data for the years 2010 to 2016.


\end{minipage}
\end{center}

\section{Introduction}\label{sec:intro}

Nonparametric resampling procedures for multidimensional data, including analogue methods \citep{Lorenz.1969,Yiou.2014} or geostatistical approaches based on training images \citep{Mariethoz.al.2014}, have become powerful tools for problems such as filling in missing data or generating new data scenarios. The general context of such applications is that training data are available, often stemming from \modif{observations of very high dimension, such as images}. \modif{Although not being a model in the strict sense, the training data can be used as} nonparametric models used for generating new data, which should carry the same features as the original data. Common applications include the generation of inputs for impact models and sensitivity analyses, or the use of the newly generated data for bootstrap-like estimation procedures of \modif{application-relevant} parameters in the original data.

A strong restriction of purely nonparametric procedures, which make minimal assumptions on the data structure, is their inability to generate new data points with values beyond the range of observed values, especially in  climate-related applications where measuring the impact of extreme events and new records is crucial. Morever, when tails are rather heavy in the data distribution, then the observed data points close to the extremes typically have higher inter-point spacing as opposed to the dense scattering in the center of the distribution \modif{\citep{dehaan007extreme}}, which calls for modifications to the simulation algorithm for generating new values lying between those near-extreme points in the lower and upper tails. In addition, the dependence structure at high quantiles can be different from that in the body of the distribution, and functional extreme value theory (\emph{i.e.}, the theory of extremes of stochastic processes)  provides appropriate dependence characterizations and statistical tools \citep{Ferreira.deHaan.2014,Thibaud.Opitz.2016,davison2015statistics}.  
        
In this contribution, we focus on a family of resampling algorithms called multiple-point statistics (or MPS), which aim at generating realizations of a spatial phenomenon (where ``space" can also refer to ``time")  based on the spatial dependencies observed in a given training dataset. The approach is based on nonparametric resampling of spatial patterns, which can be small patches or scattered values. As such, it has the advantage of reproducing very complex dependencies observed in the data, given that those dependencies are sufficiently represented in the training set. While these methods were initially designed to simulate categorical attributes \citep{Guardiano1993,Strebelle2002}, in recent years they have been  extended to the simulation of continuous variables \citep{Mariethoz.al.2010, Zhang2006,Kalantari2016}. In this context, the question of how to include values in the simulations that are outside the range of the training data has become acute in the practice of continuous-variable multiple-point simulations. This naturally extends to handling the occurrence of extreme values, especially when the simulation domain considered is much larger or spanning longer return intervals than the training domain. 

Typically, nonparametric resampling shows good performance for reproducing data characteristics such as correlation and marginal distributions in simulated data when we consider the ``central" part of the data distribution where training data are densely scattered. However, resampling extreme quantiles is more challenging since data become less dense close to the extremes \citep[e.g.][]{Beirlant.al.2006}, and it is simply impossible to sample quantiles beyond the observed range of data when  no assumptions are made on how to extrapolate data towards more extreme values. To address this issue, we propose in this work a \emph{lifting} mechanism that enriches the training data such that they have more extreme values than only those observed in a limited dataset, while accounting for the dependence structure at these high quantiles. Our approach uses three main ingredients: 1) the knowledge or an estimation of the univariate marginal  probability distribution $F$, especially of its tail behavior characterized by values $F(x)$ for high quantiles $x$ as $F(x)$ gets close to $1$ (here, by default, we consider the upper tail); 2) minimal but robust assumptions on the dependence structure between values of the stochastic process at high quantiles; 3) non-parametric resampling techniques, in our case Direct Sampling \citep{Mariethoz.al.2010}. Our approach thus combines nonparametric methods able to account for complex dependence structures with a theoretically founded parametric model to properly account for univariate extremes. \modif{Previous approaches to lifting observed extreme episodes, using extreme-value theory similar to our method but without further resampling steps, have been proposed in \citet{Ferreira.deHaan.2014,Chailan.al.2017,Palacios.al.2019}.}

We assume that  training data have been generated by a stationary stochastic process $\{X_i,\ i\in \mathcal{I}\}$, where the index set $\mathcal{I}$ could refer to  positions in time, space or space-time, such as regularly spaced observation times, the spatial grid of a spatial random field, or a space-time grid. The assumption of stationarity is without loss of generality if appropriate pretransformations are available for nonstationary data, see the example of  heatwave simulations over France in Section~\ref{sec:application}.  To report results and formulas where the specific structure of the support is not important, we can assume (without loss of generality) that \newmodif{$\mathcal{I}=\{1,2,\ldots,n_{\mathcal{I}}\}$; when useful, we will give a more specific definition of $\mathcal{I}$.} 
The theoretical results from extreme value theory underpinning the method that we propose do not require data on a regular grid, but the standard setup of nonparametric resampling techniques uses gridded data. We  will write $X_{ij}$, $i\in \mathcal{I}$, $j=1,\ldots,m$, if observations at point $i$ are repeated $m$ times. Specifically, the case where values for $j_1\not=j_2$ are stochastically independent is relevant. 

\newmodif{
Our resampling procedure based on extreme-value theory assumes that data  show relatively strong dependence at high quantiles when they are observed at closely located points in the support $\mathcal{I}$. This property is known as \emph{asymptotic dependence} \citep[\emph{e.g.},][]{Davison.al.2013,Huser.al.2017}. Given two variables $X_i$ and $X_j$, $i,j\in\mathcal{I}$, it corresponds to a positive limit value of the conditional exceedance  probability
\begin{equation}\label{eq:ad}
\lim_{u\rightarrow\infty} P(X_i >u \mid X_j >u)>0, \quad i,j\in\mathcal{I}.
\end{equation} 
For spatial data, an interpretation of this condition is that the spatial extent of clusters of high values should remain comparable when looking at increasingly high quantile levels. For environmental and meteorological variables, this may not always be the case as shown in the recent literature \citep{Davison.al.2013,Opitz.2016, Huser.al.2017,Bacro.al.2019}, but the asymptotic framework still provides a useful approximation in practice. Moreover, the assumption of asymptotic dependence can be considered as realistic when very extreme events tend to impact large spatial areas simultaneously, for instance heatwaves or thunderstorms \citep[\emph{e.g.},][]{deFondeville.Davison.2020}. 
}

The remainder of this paper is organized as follows. In Section \ref{sec:margins} we present a univariate model of the marginal distribution with minimal assumptions, yet able to model accurately the bulk and the tail of the distribution. In Section \ref{sec:naive} we present a first, straightforward approach for resampling extremes, which will prove to show inaccuracies close to the highest quantiles. For this reason we call this approach the {\em naive approach}. Based on fundamental results exposed in a  short survey of extreme-value theory for stochastic processes in Section~\ref{sec:evt}, we develop a procedure for lifting observed extreme episodes in training data to more extreme quantiles, where we exploit the property of threshold stability arising in extreme value limits. This data enrichment step is the foundation for our second, more sophisticated resampling algorithm presented in Section~\ref{sec:uplifting}. We illustrate this second approach in Section \ref{sec:application} for simulating heatwaves of unprecedented magnitude in France, using  gridded daily reanalysis data provided by the French weather service (M\'et\'eo France) for the $2010$-$2016$ period as training sample. A discussion of our approach with an outlook to follow-up work concludes the paper in Section~\ref{sec:conclusion}. \newmodif{\texttt{R} code implementing procedures proposed in this paper is available at \url{https://github.com/thopitz/resampling-extremes}.}

\section{Modeling and estimating the marginal distribution}\label{sec:margins}

For data that are away from extremes, or whose modeling may be of minor importance (\emph{e.g.}, low values in many contexts), the empirical distribution of training data is usually sufficiently accurate for resampling. Especially in the central region of the distribution, data are typically very dense and provide a good coverage of possible values.  Nevertheless, to improve on the empirical distribution by allowing for generating new values between already observed ones, we propose to generate new values based on a kernel density estimator of the  univariate density  function of data. This also allows decreasing data dimension to  reduce numerical computations and memory requirements, provided that the estimated density has a lower-dimensional numerical representation than the original, potentially very high-dimensional data, which is usually the case. Therefore, we can use the empirical distribution or the kernel density estimation below a high threshold $u$. \newmodif{Kernel density estimators for spatial processes are consistent under very general conditions \citep{hallin2004kernel}. In what follows, the kernel estimate of the probability density function will be computed using a standard setting with a Gaussian kernel and the ``rule of thumb" of \citet{Silverman.1986}, as implemented in the base package of the \texttt{R} statistical software.} Above the threshold $u$, the decreasing availability of observations may lead to problematic artefacts, even more in the presence of correlation-induced clustering of values in space, \modif{as detailed in Section \ref{sec:limitations}. Appropriate extrapolation beyond the observed maximum is thus uncertain when using standard kernels.} 

To generate new values beyond the observed range of data, and to avoid problems with resampling within the range of observations but close to the observed extremes where empirical coverage is only sparse, we need a model for the univariate distribution close to and beyond  the extremes. Univariate extreme value theory provides the asymptotically motivated generalized Pareto distribution (GPD) as the theoretical limit distribution of positive threshold exceedances $Y\stackrel{d}{=} X-u\mid X>u$ above a high threshold $u$ \citep{davison2015statistics}. Its survival function is
\begin{equation}
    P(Y > y) = \left\{ 
    \begin{array}{ll}
      (1+\xi y/\sigma)_+^{-1/\xi},    &  \xi \neq 0, \\
        \exp(-y/\sigma), &  \xi=0,
    \end{array}
    \right.
    \label{eq:Pareto}
\end{equation}
where $(x)_+ = \max\{0,x\}$, $\sigma >0$, and the support for $y$ is such that the right-hand side is not superior to $1$.  We denote the density of the GPD by $f_{\text{GP}}(y\mid \sigma,\xi)$. The scale parameter ($\sigma_u>0$) and the shape parameter ($\xi_u\in\mathbb{R}$) have to be estimated from the data.  The tail of the univariate distribution is thus fully characterized by these two parameters and the probability $p_u=P(X>u)$ of exceedance  over the threshold $u$. In our model, we require continuity of the estimated density $\hat{f}$ of data at the threshold $u$; see \citet{Carreau.Benio.2009} \modif{and \citet{Scarrott.MacDonald.2012} (especially their Section~6.3) for closely related approaches} of such "piecing together", and  we therefore determine the parameter $\sigma_u$ based on the kernel density estimate. This leaves the shape parameter $\xi_u$ to be estimated by using one of a large variety of tail index estimators proposed in extreme value theory \citep{Scarrott.MacDonald.2012}, for instance the  moment estimator of \citet{Dekkers.al.1989}, known to be quite robust and accurate in almost all of practically relevant cases, or the likelihood estimator, or the Hill estimator in the heavy-tailed case with $\xi_u>0$; see \citet{dehaan007extreme} for a comparative discussion of estimators. \modif{The parameters of the GPD and of the kernel density estimate may further vary according to covariates or position in space. For notational and conceptual simplicity, we here describe the procedure only for the stationary setting.}
In summary, we proceed as described in Algorithm~\ref{algo:estim_f} below for estimating the univariate model.

\medskip

\begin{algorithm}[htb]
\caption{Estimation of the univariate probability density function} 
\label{algo:estim_f}
\begin{algorithmic}[1]
\REQUIRE  $p_u$, a small probability of threshold exceedance, \emph{e.g.} $p_u \in \{0.01,0.05,0.1 \}$
\STATE Calculate a kernel density estimation $\overline{f}(x)$ of data.
\STATE Set $u$ such that $\int_{u}^\infty \overline{f}(x)\,\mathrm{d}x=p_u$.
\STATE Set $\hat \sigma_u=p_u/\overline{f}(u)$.
\STATE Estimate $\hat \xi_u$ using an appropriate tail index estimator.
\STATE Define the univariate density model $\hat{f}$ for data:
\begin{equation}
 \hat{f}(x)=
 \begin{cases} 
  \overline{f}(x), & x\leq u, \\
  p_u\, f_{\text{GP}}(x\mid \hat \sigma_u, \hat \xi_u), & x>u.
 \end{cases}
 \label{eq:marginal}
\end{equation}
\RETURN $\hat{f}(x)$.
\end{algorithmic}
\end{algorithm}

\medskip

\begin{figure}
\centering
\begin{tabular}{cccc}
	\includegraphics[width=.25\linewidth]{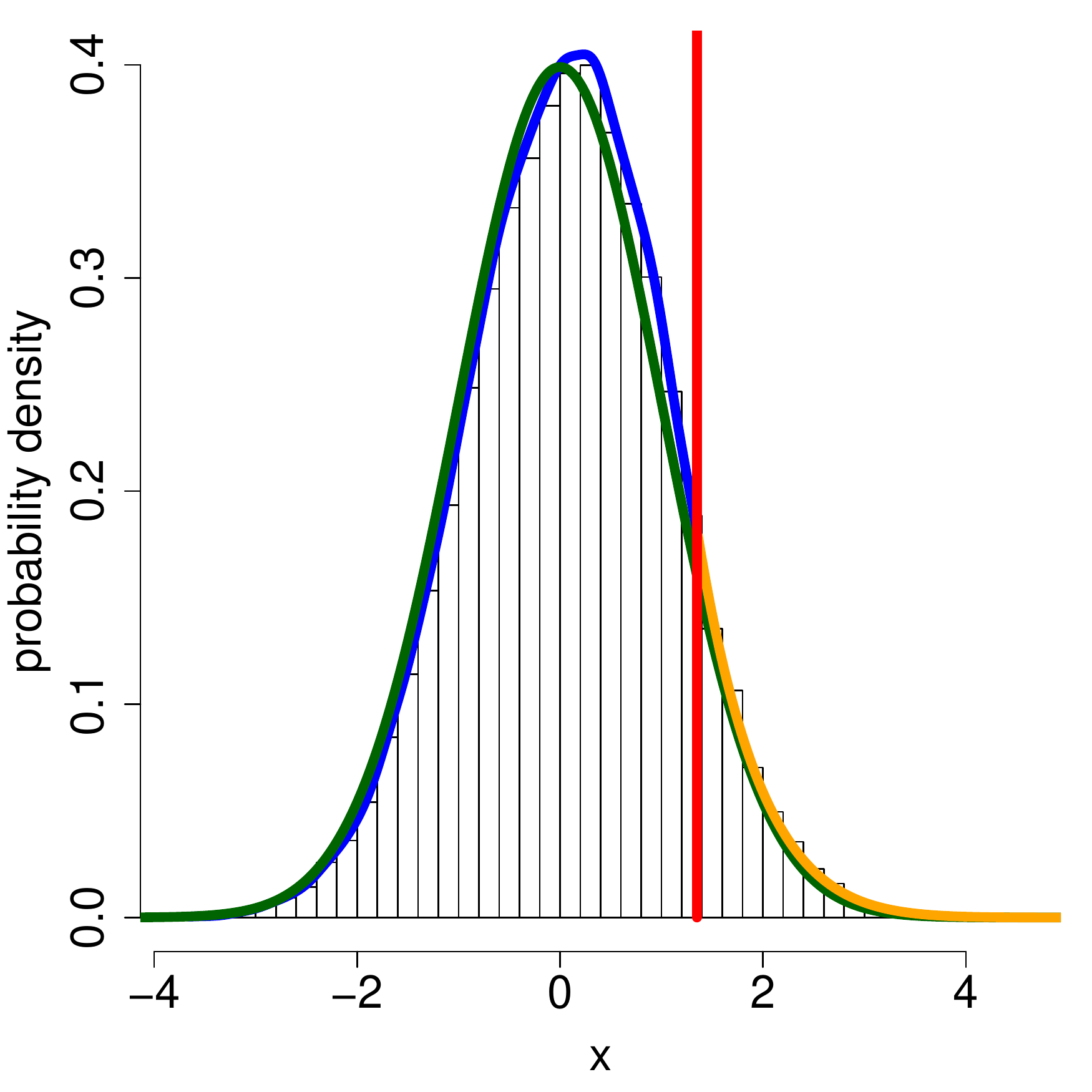} &
	\includegraphics[width=.225\linewidth]{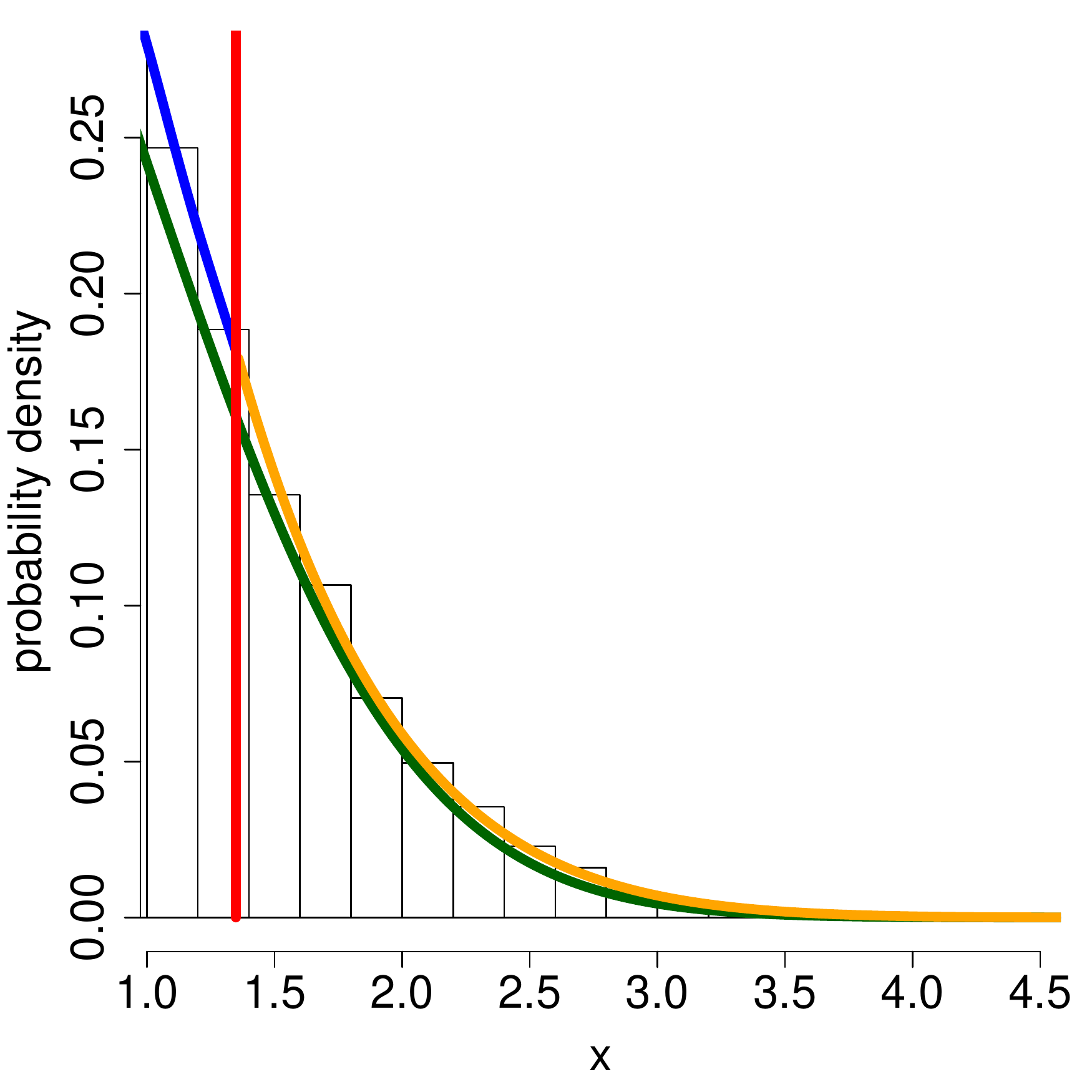} &
	\includegraphics[width=.225\linewidth]{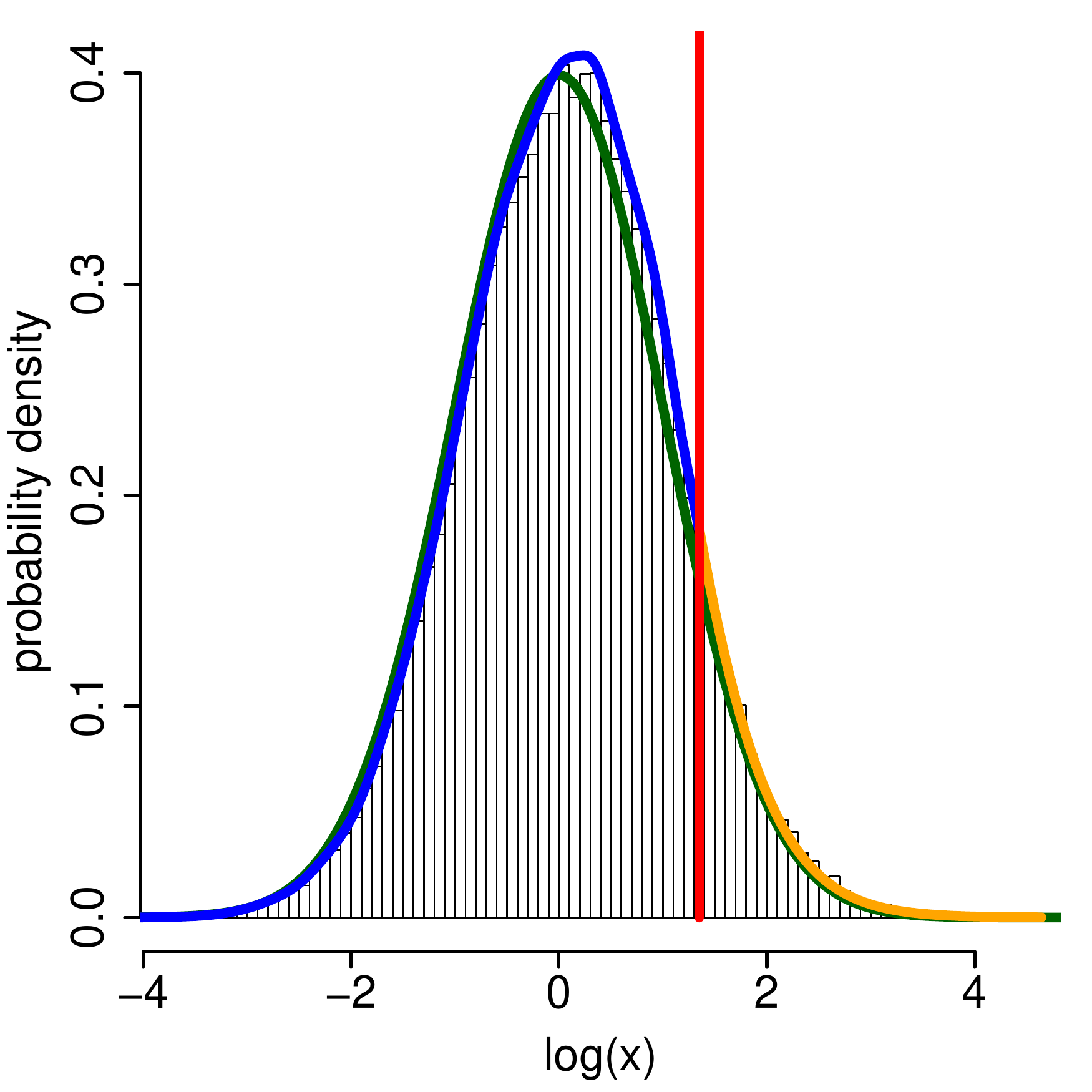} &
	\includegraphics[width=.225\linewidth]{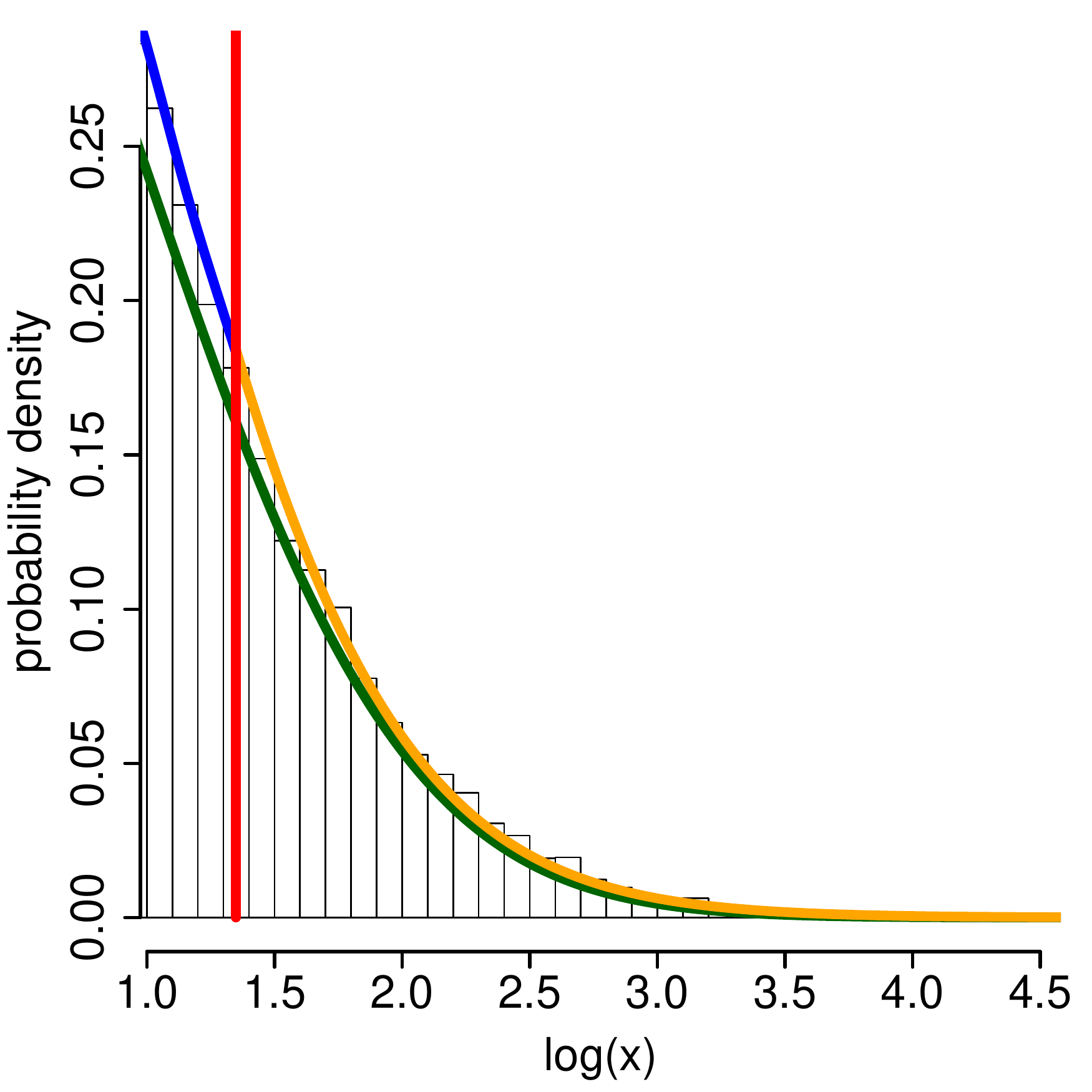} \\
\includegraphics[width=.25\linewidth]{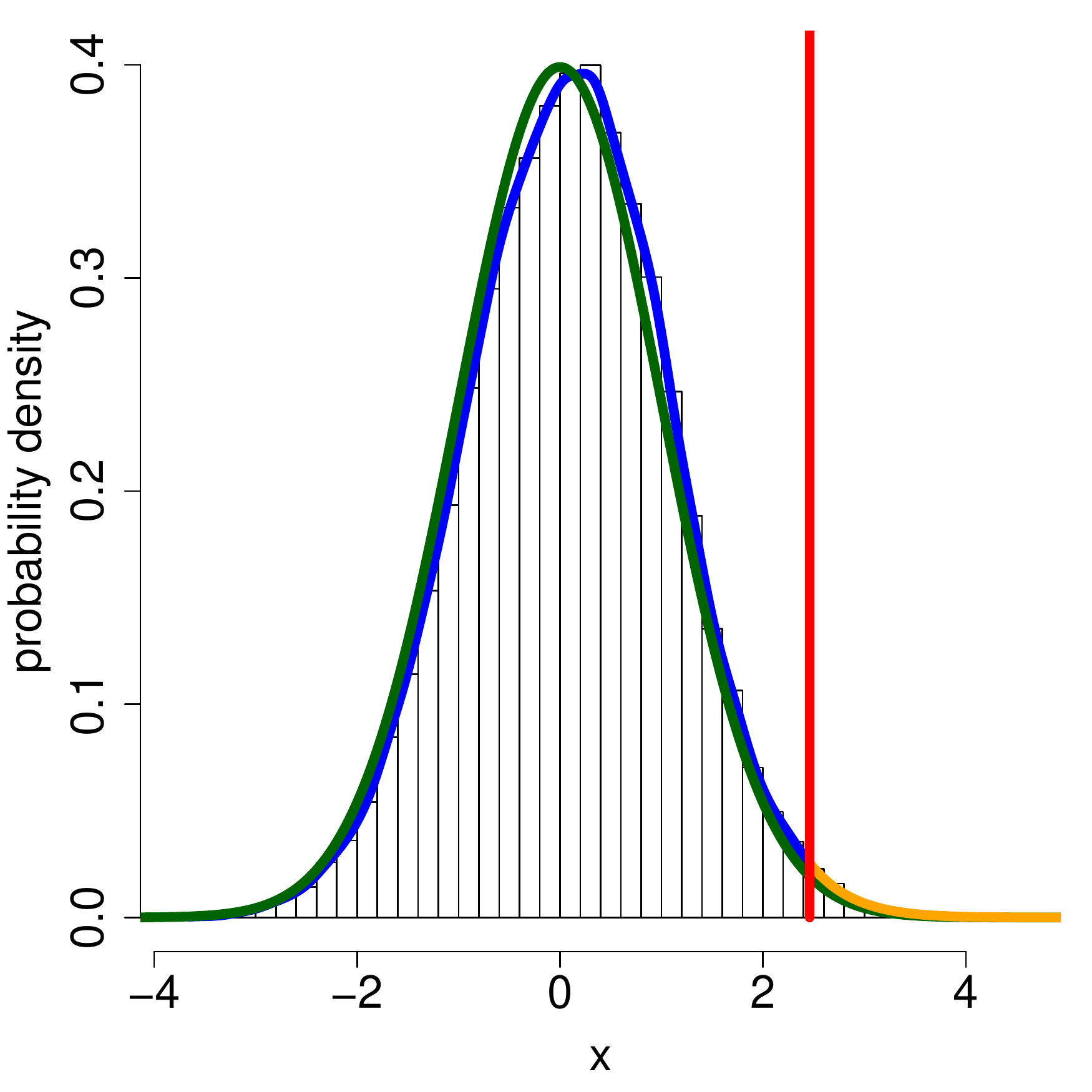} &
    \includegraphics[width=.225\linewidth]{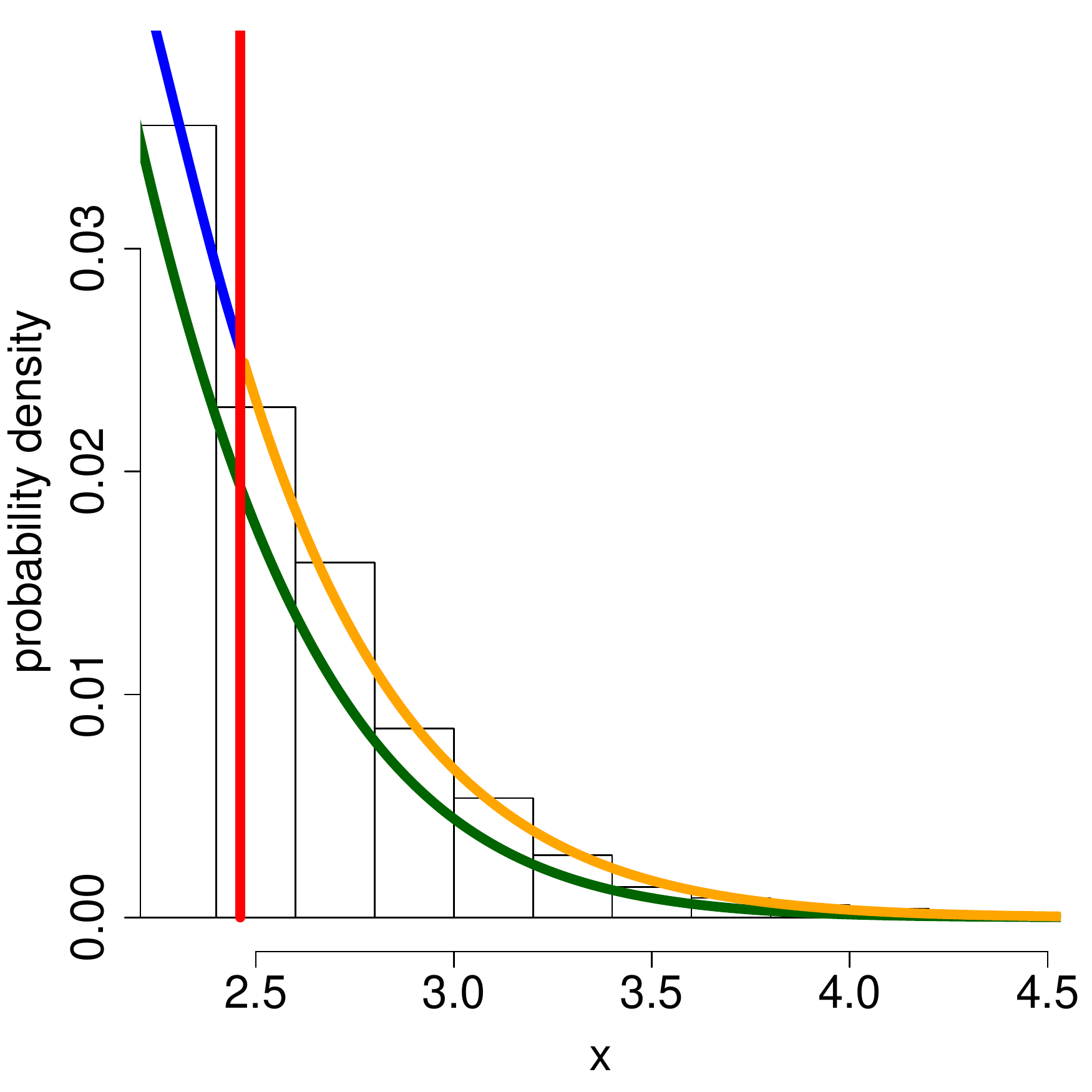} &
    \includegraphics[width=.225\linewidth]{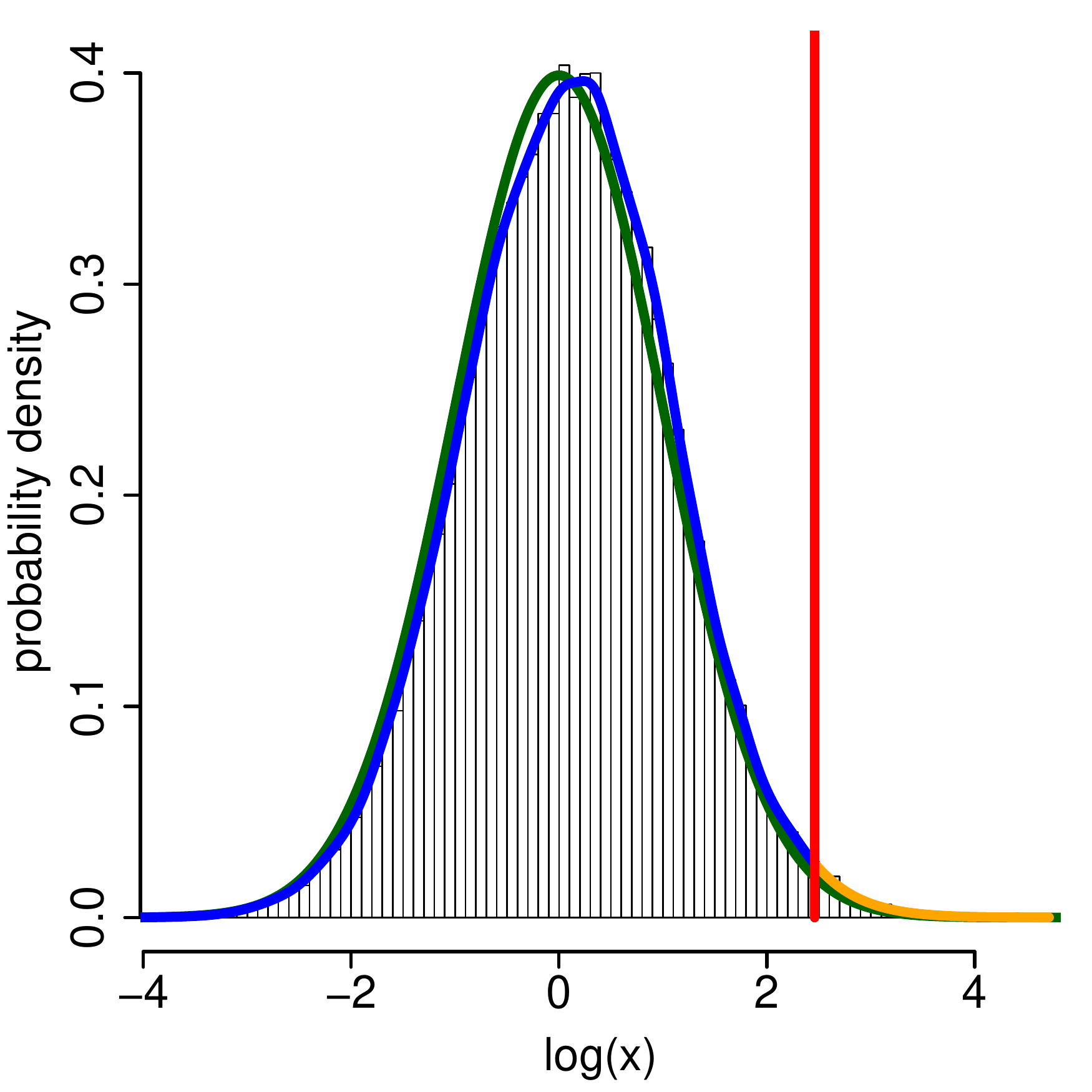} &
    \includegraphics[width=.225\linewidth]{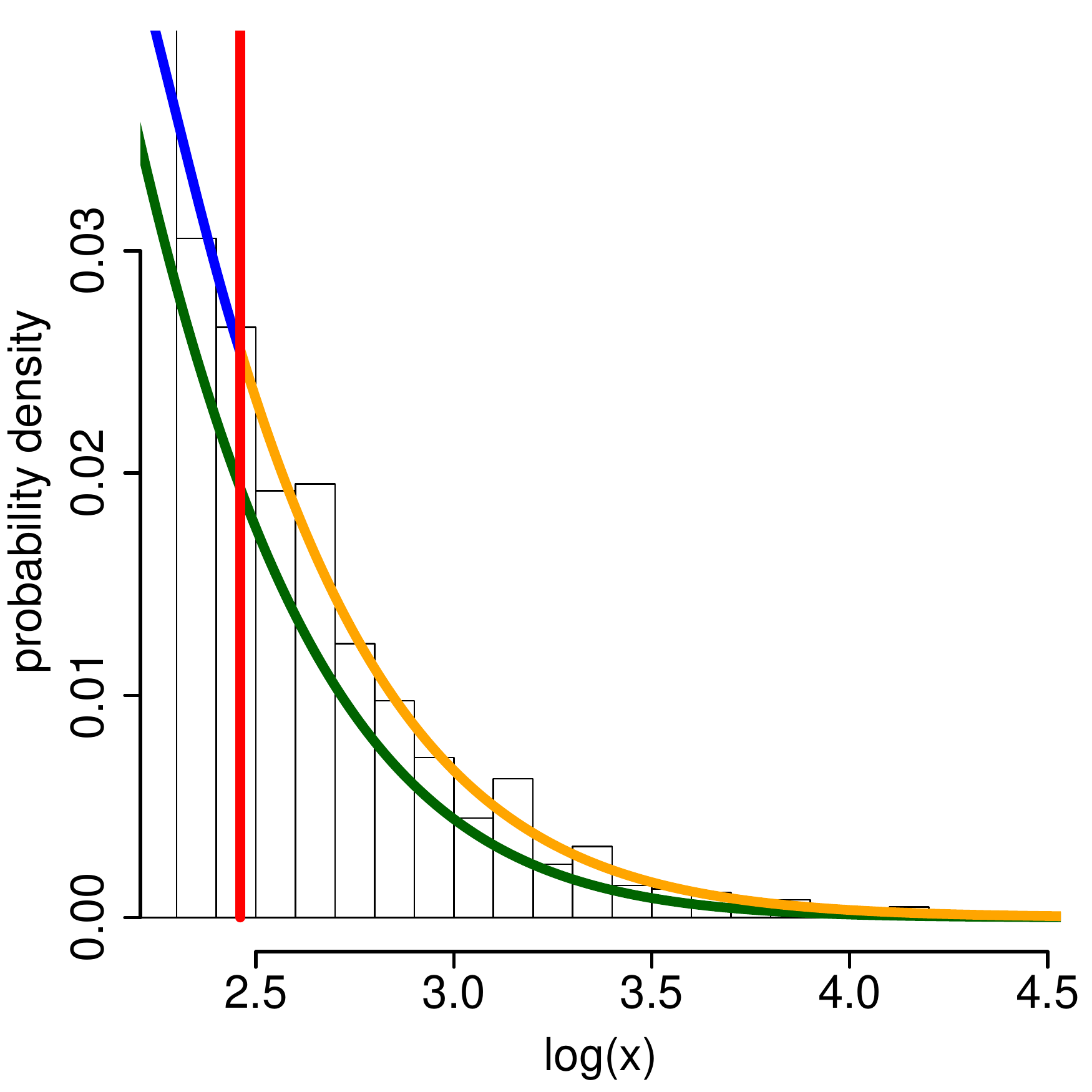} \\
    \includegraphics[width=.225\linewidth]{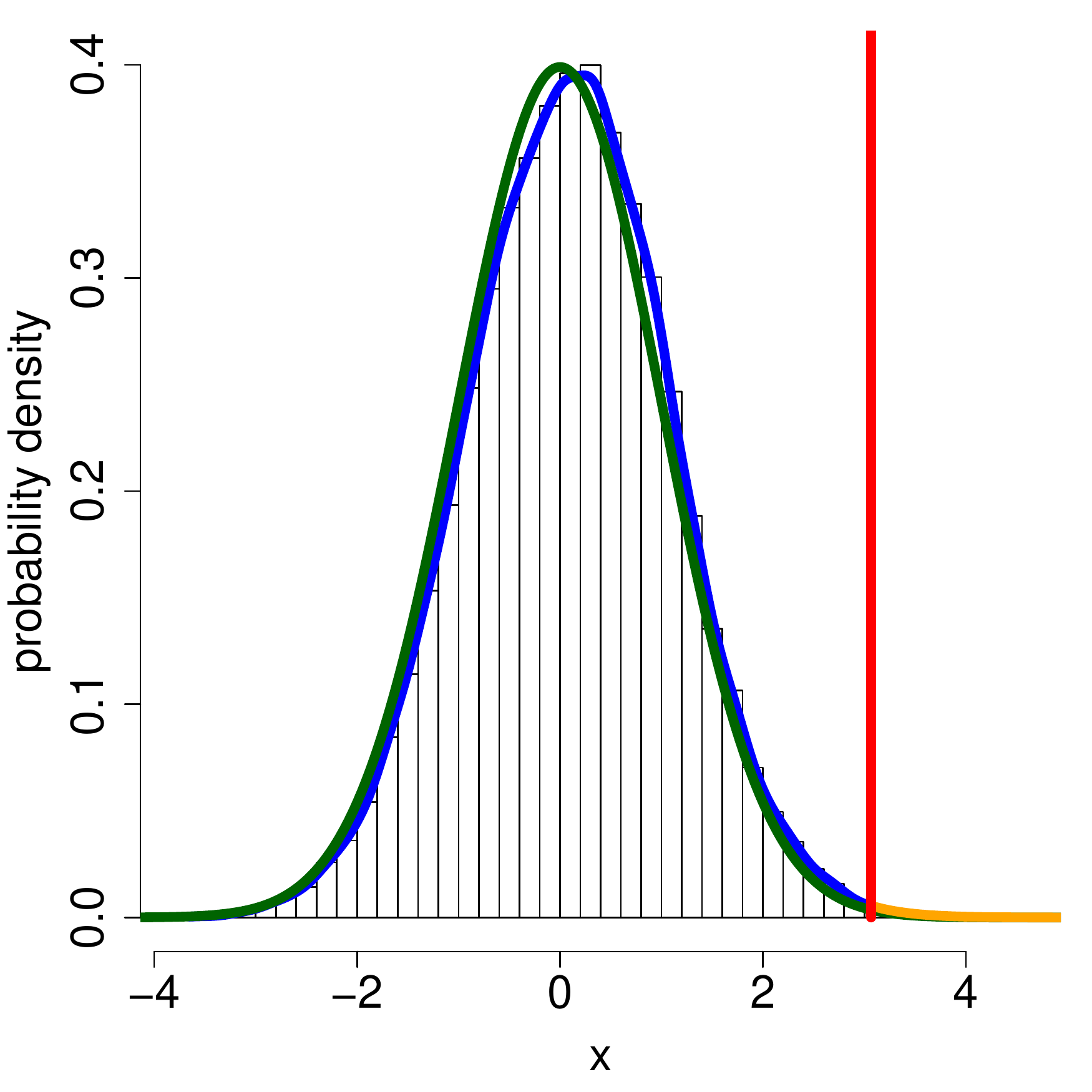} &
    \includegraphics[width=.225\linewidth]{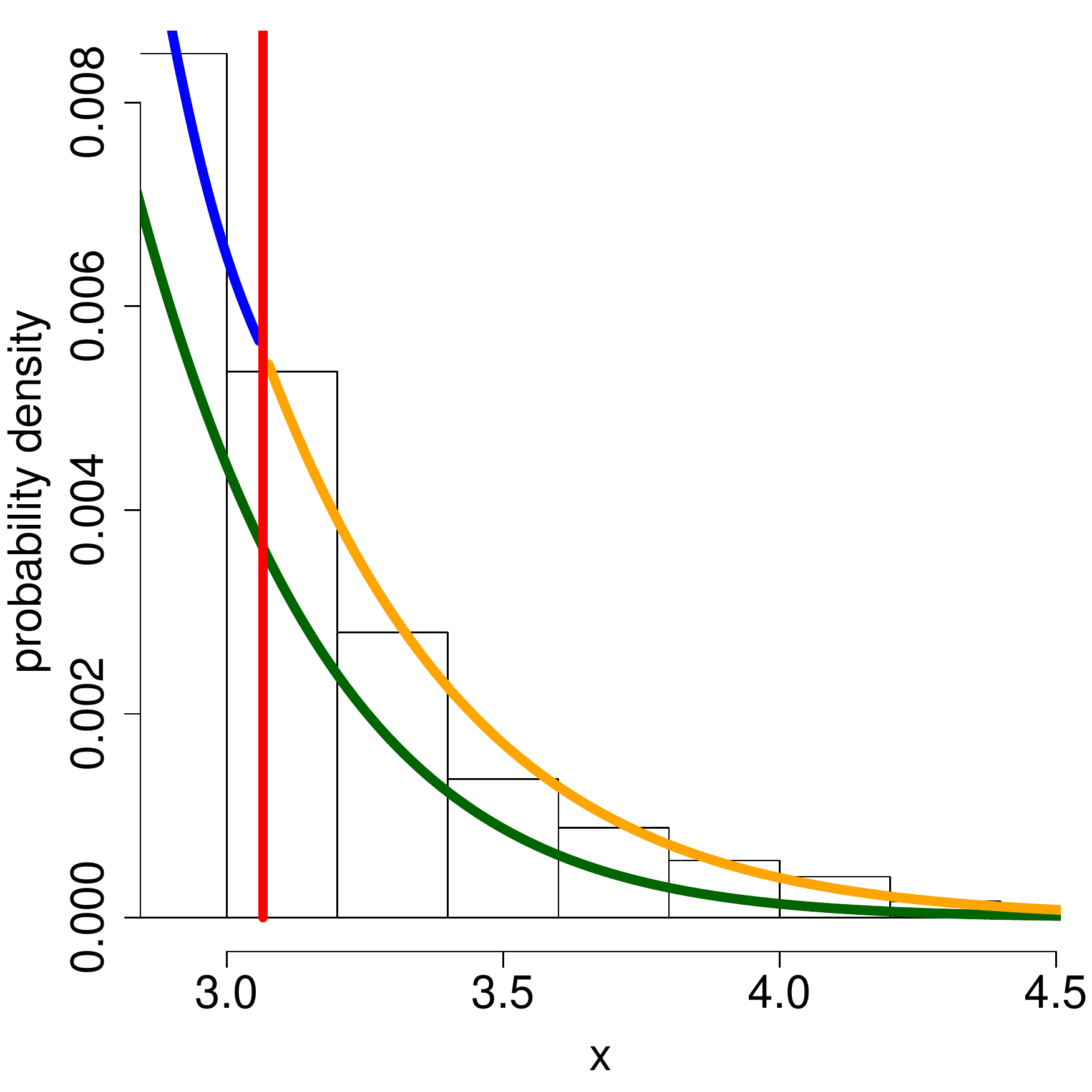} &
    \includegraphics[width=.225\linewidth]{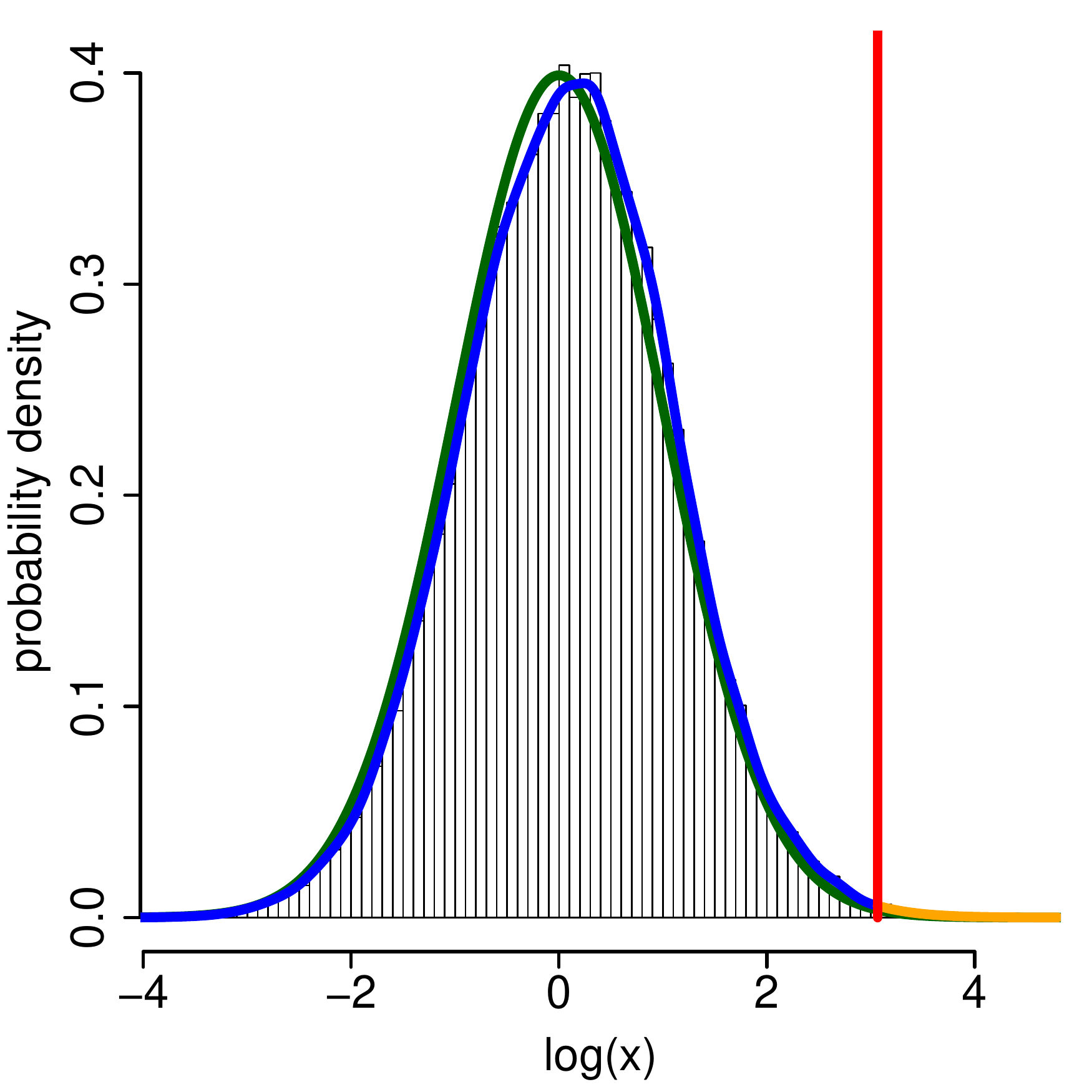} &
    \includegraphics[width=.225\linewidth]{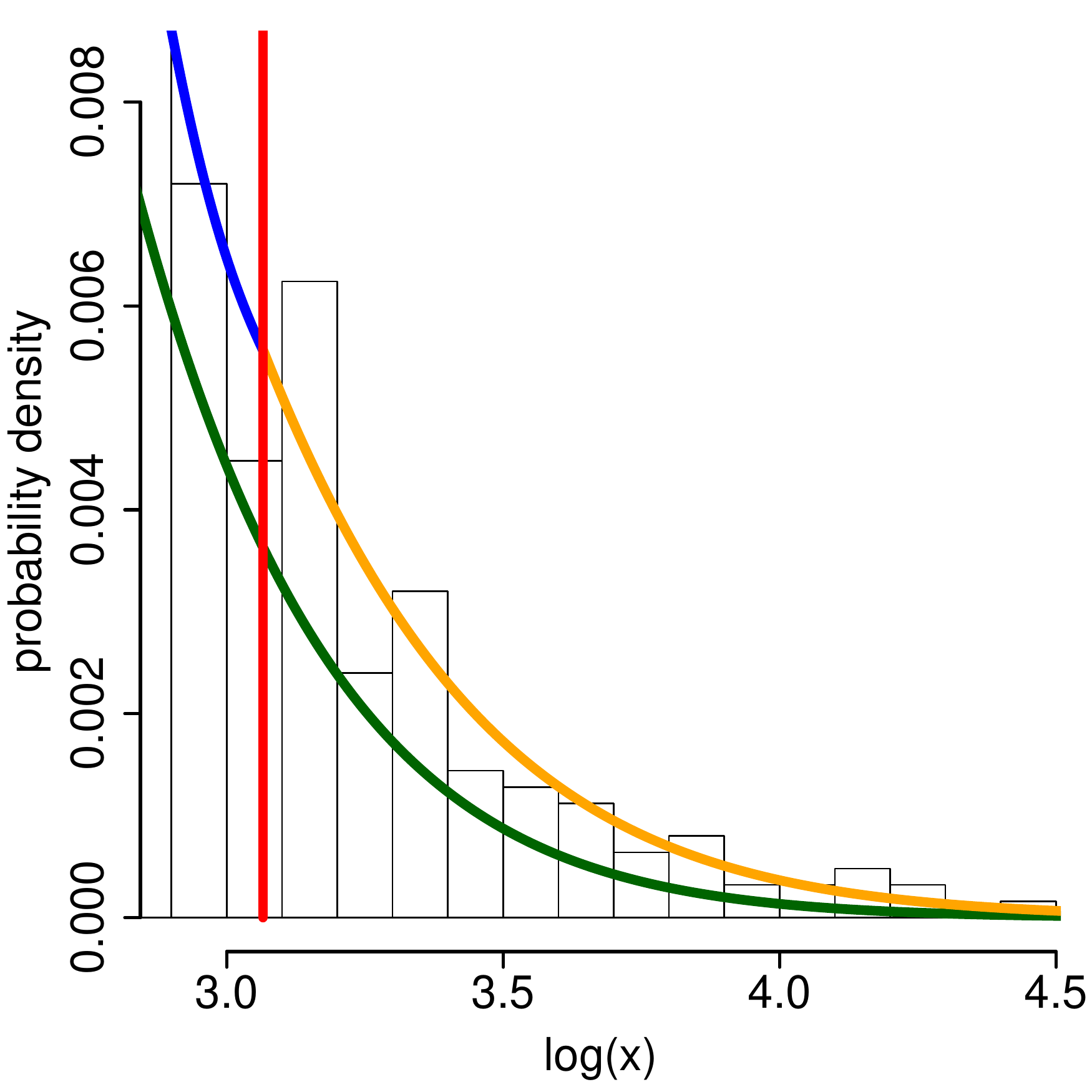} 
\end{tabular}
	\caption{Univariate tail model $\hat{f}$ for Gaussian (columns 1,2) and log-Gaussian (columns 3,4; with log-scale for abscissas) data. Columns 2 and 4 show a zoom into the tail region. Exceedance probabilities are $p_u=0.1$ (top), $0.01$ (middle) and $0.002$ (bottom). Other graphical elements are the threshold position (red), the kernel density (blue), the generalized Pareto density $f_{\text{GP}}$ (orange), and the true density (green).}
	\label{fig:gp}
\end{figure}

Figure~\ref{fig:gp} illustrates the tail model and its estimation for data sampled from the Gaussian distribution or the log-Gaussian distribution. We assume that the true distribution is unknown, as it is usually the case for real data modeling. Data consist of a single replication ($m=1$) of a standard Gaussian random field on the unit square $[0,1]^2$ with exponential covariance characterized by a scale parameter $0.03$, simulated on a regular $250\times 250$ grid.  We consider $p_u\in\{0.1,0.01,0.002\}$. The two left columns in Figure~\ref{fig:gp} show results for the data with Gaussian margins, while the two right columns correspond to log-Gaussian margins, obtained by exponentiating the Gaussian observations. For better readability of the displays in these log-Gaussian case, their abscissas are given on log-scale. In all cases, the fitted tail model performs well, as can be seen from comparison with the histogram of observations. A complicating circumstance is that the Gaussian and the log-Gaussian distributions represent cases where the GPD is only an asymptotic approximation and not exact. Nevertheless, we can see that its fit provides an accurate tail model without making more specific assumptions on the marginal distributions, even at a relatively low threshold such as the one associated to $p_u=0.1$. \modif{We here used the
moment estimator of \citet{Dekkers.al.1989} for the tail index. For $p_u\in\{0.1,0.01,0.002\}$,  estimates of $\xi_u$ were $-0.08$, $-0.02$, $0.04$ respectively in the Gaussian case, and $0.33$, $0.34$, $0.38$ respectively in the log-Gaussian case.}

\section{Naive resampling of extremes}\label{sec:naive}

\subsection{Naive resampling algorithm}
This section develops a first data enrichment technique that is based on two components only: 1) a model to sample unobserved values from the marginal distribution, which requires a univariate model of $F$  estimated on the training data under minimal assumptions; 2) a mechanism, such as a multiple-point simulation algorithm, to generate a process of the same size as the training data while \modif{preserving the spatial, temporal or spatio-temporal dependence between ranks}. In this procedure, the ranks of the training data are considered as the dependence model.  We label this approach \emph{naive lifting} since the independent sampling in step~1 may lead to certain inaccuracies, especially when the spacings between the highest-ranking values in simulated data become too large. We will discuss this issue in  the following Subsection~\ref{sec:limitations}. 
\newmodif{We suppose that data $x_i$, indexed by $i\in\mathcal{J}=\{1,\ldots,n\}$, are given, where $\mathcal{J}=\mathcal{I}$ if $m=1$, and in the case of replication ($m>1$) the index $\mathcal{J}$ runs through all $m\times|\mathcal{I}|$ data points.}
We use the notation $r[i]$ to refer to the rank of $x_i$ among $x_1,\ldots,x_n$, and we write $r^{-1}[k]$ for the index $i$ of the $k$-th ranking value in $x_1,\ldots,x_n$. Algorithm~\ref{algo:naive} summarizes the procedure for naive resampling. If data are replicated, we here assume that the indices in $\mathcal{I}$ run through all observations. 

\begin{algorithm}[htb]
\caption{Naive resampling} 
\label{algo:naive}
\begin{algorithmic}[1]
\STATE Generate a vector of ranks $\tilde{r}[i]$, $i=1,\ldots,n$, of the same size, \modif{$n$}, as the training data and reproducing the dependence patterns (spatial, temporal or spatio-temporal) of the training data ranks $r[i]$, $i=1,\ldots,n$. Nonparametric resampling techniques such as MPS-algorithms can be used.
\STATE Estimate the univariate density function $\hat{f}$ according to Algorithm~\ref{algo:estim_f}.
\STATE Sample $n$ new data values $y_1,\ldots,y_n \sim \hat{f}$.
\STATE Put new data values in non decreasing order: $\tilde{y}_1\leq \tilde{y}_2\leq \ldots \leq \tilde{y}_n$. 
\STATE Define new data $\tilde{x}_1,\ldots,\tilde{x}_n$ such that
$$\tilde{x}_{\tilde{r}^{-1}[k]}=\tilde{y}_k, \quad k=1,\ldots,n.$$
\RETURN $(\tilde{x}_1,\ldots,\tilde{x}_n)$.
\end{algorithmic}
\end{algorithm}

\subsection{Limitations of naive resampling}\label{sec:limitations}

For ease of presentation we here suppose that the stationary marginal distribution of data is standard exponential. This assumption is without loss of generality since the marginal distribution is assumed to be known\modif{, and results regarding quantiles of the exponential distribution can be reformulated for the corresponding quantiles of any other continuous distribution.} Therefore, data can always be transformed to standard exponential scale using the probability integral transform. We use the  index $(i)$ to refer to the $i$-th ranking value in a vector of values. The third step of the naive approach in Algorithm~\ref{algo:naive} consists in sampling $n$ independent and identically distributed (i.i.d.) standard exponential random variables, and sorting them from lowest to highest value $Y_{(1)},\ldots,Y_{(n)}$ in Step~4.
According to R\'enyi's theorem \citep[p. 37]{dehaan007extreme} we have the following equality in law: 
\begin{equation}
Y_{(i)}\stackrel{d}{=}\sum_{l=1}^i\frac{Z_k}{n-l+1}, \quad Z_l\stackrel{\text{i.i.d.}}{\sim} \mathrm{Exp}(1),\quad k=1,\ldots,n,
\label{eq:Reny}
\end{equation}
\modif{see also \citet[Section~4.4]{Beirlant.al.2006}.}
Several conclusions can be drawn for the naive resampling procedure. First, if the data process is mixing (\emph{i.e.}, it has long-range independence in the observation window), simulated quantiles will correspond well to original quantiles in the central part of the distribution where data are dense, owing to the law of large numbers. \newmodif{Similar behavior is obtained for replicated data with sufficiently large $m$.} 
However, simulated quantiles can differ substantially from data quantiles close to the extremes. Spacings between $Y_{(i)}$ and $Y_{(i+1)}$ are relatively large when $i$ is close to $n$. For instance, as per Equation~\eqref{eq:Reny} the final spacing $Y_{(n)}-Y_{(n-1)}$ follows a standard exponential distribution, which means that it has the same distribution as the original data $Y_i$. For other choices of distributions, such spacings may be smaller, but a large difference still arises in the probability levels of the corresponding quantiles. \newmodif{As a consequence, too large maxima will arise in the naive resampling, and it would be difficult to target specific ranges of return periods for summary statistics correlated with the maximum.}
 Another consequence is that Algorithm~\ref{algo:naive} produces very strong spatial variability in the pixels having a value close to the maximum value $Y_{(n)}$. Indeed, even if a training image shows relatively smooth behavior around the pixel containing the maximum value, naive resampling will tend to produce a very rugged surface around the maximum in simulated images. \newmodif{Finally, we mention that the naive resampling procedure depends strongly on the grid resolution. While measures of effective sample size for dependent data may be comparable for different resolutions, they can strongly vary in the case of an i.i.d. sample. We exemplify some of these shortcomings  through simulations in the Appendix Section~\ref{sec:examples-naive}}.

\newmodif{In summary, naive resampling is an acceptable approximation when the data show relatively weak asymptotic independence, and if the spatial resolution is coarse enough such that the mesh size is close to the spatial range of dependence in the most extreme values.}

A workaround for taking into account the spatial clustering of high order statistics $Y_{(i)}$ when $i$ is close to $n$ could be to model the cluster structure explicitly. Various techniques for such cluster modeling  could be developed, such as adding dependence between exponential variables $Z_i$ in Equation~\eqref{eq:Reny}, but we think that there is no simple parametric choice with good theoretical motivation that avoids rather arbitrary modeling assumptions. Instead, we take advantage of extreme value theory for stochastic processes, which suggests techniques for extrapolating values of dependent variables beyond the observed range while avoiding any parametric assumptions on the dependence structure of extremes.

\section{Lifting based on functional extreme value theory}
\label{sec:evt}

\subsection{Scale-profile decomposition of Pareto processes}
\modif{From now on, we adopt the following convention: random variables and random functions are denoted with upper case letters. We use lower case letters for scalars, dummy variables as well as for datasets and realizations of random values and random functions.}
Extreme value theory for stochastic processes provides an asymptotic decomposition $Y_{\mathcal{I}}\stackrel{d}{=}S\eta_{\mathcal{I}}$ of extreme events $Y_{\mathcal{I}}$ into a scale variable $S$ and a normalized profile process $\eta_{\mathcal{I}}$ for certain choices of marginal distribution in data,  with scale and profile being stochastically independent \citep{Ferreira.deHaan.2014,Thibaud.Opitz.2016,Dombry.Ribatet.2015,Opitz.al.2015,deFondeville.Davison.2016,Engelke.al.2018b,Palacios.al.2019}.  The  processes presenting such factorization of scale and profile are known as \emph{Pareto processes}. While it is easy to generate new realizations of the scaling variable $S$, whose distribution is parameter-free, it is more intricate to provide new realizations of the profile process $\eta_{\mathcal{I}}$ owing to the inherent dependence structure.

Recall that the training data are modeled as a stationary stochastic process $X_i$, $i\in \mathcal{I}$, with $X_i\sim F$. While the upper section of the univariate distribution $F$ can be conveniently modeled by the GPD as explained in Section~\ref{sec:margins}, our focus is now only on dependence of different values $X_i$. It is useful to abstract away from the specific shape of $F$. Theory is most easily presented by assuming a normalized distribution function $F^\star$ with non-negative support and standard Pareto tails $1-F^\star(x) = 1/x$ for large $x$, which is at the origin of the notion of Pareto processes. We define normalized data with standard Pareto tails through the transformation
\begin{equation}\label{eq:transform}
X_i^P=1/(1-F(X_i)),
\end{equation}
which establishes the standard Pareto distribution $F^\star$ for $X_i^P\sim F^\star$ if $F$ is continuous. It is easy to verify that $X_i^P$ possesses a standard Pareto tail, since for $x \in [1,+\infty)$,
$$P(X_i^P > x) = P((1-F(X_i))^{-1} > x) = P(1-F(X_i) < 1/x) = P({\cal U} < 1/x )=1/x;$$
where $\cal U$ denotes a uniform random variable on the interval $(0,1)$.

The fundamental assumption in extreme value theory of stochastic processes is the functional maximum domain of attraction condition, which we outline here, while  technical details can be consulted in the literature \citep[\emph{e.g.},][and the above-listed references]{dehaan007extreme}. \newmodif{We suppose that $m$ independent replicates $X_{ij}$, $i\in\cal{I}$, $j=1,\ldots,m$, of the stochastic process $X_{\cal I}$ are given.  For the sake of generality of the following theoretical results, the set $\mathcal{I}$ can now be any nonempty compact subset of $\mathbb{R}^d$.} We assume that deterministic normalizing sequences $\alpha_{im}$ (for scale) and $\beta_{im}$ (for location) exist such that the componentwise maximum converges to a nondegenerate limit process, that is, 
\begin{equation}\label{eq:mda}
\max_{j=1,\dots,m} \frac{ X_{ij}-\beta_{im}}{\alpha_{im}} \to Z_i, \quad i\in \mathcal{I}, \quad m\rightarrow\infty, 
\end{equation}
where convergence takes place in an appropriately defined function space, such as the space of continuous functions over  $\mathcal{I}$.  

In equivalence to the convergence of the dependence structure of extremes under the domain of attraction condition \eqref{eq:mda}, extreme value theory states a certain convergence of the normalized data $X^P_{\mathcal I}$ when  a \emph{summary functional} $r$ (also termed \emph{aggregation functional}, \emph{risk functional}, or \emph{loss functional}) exceeds a high threshold $u^P$, which tends towards infinity. Therefore, we fix a homogeneous function 
\begin{equation}\label{eq:riskfunc}
r:[0,\infty)^{\mathcal{I}}\rightarrow [0,\infty), \quad r(a x_{\mathcal{I}})=a r(x_{\mathcal{I}}), \quad a>0,
\end{equation}
that is continuous at $0$, where $0$  represents a dataset with index running through $\mathcal{I}$ that has value $0$ everywhere. For instance, $r$ could be one of $\max$, $\min$, $\mathrm{sum}$, $\mathrm{mean}$, the value at a fixed site $s_0$, the median,  or the order statistics $x_{(i)}$ for fixed $i$, but we exclude the trivial case where $r\equiv 0$. The choice of $r$  is usually driven by the type of extreme event that one wants to consider. We are now ready to define \emph{$r$-Pareto processes}.

\begin{theorem}[\cite{Dombry.Ribatet.2015}]\label{theor:gp}
Let the distribution of $X_{\cal I}$ be in the functional maximum domain of attraction as described in Equation~\eqref{eq:mda}, and denote by $X_{\cal I}^P$ the dataset after normalization to standard Pareto tails using transformation \eqref{eq:transform}. Let $r$ be a summary functional as defined in Equation~\eqref{eq:riskfunc} and denote $R_{\cal I} = r(X_{\cal I}^P)$. Then,
\begin{equation}\label{eq:ppconv}
\hbox{Conditional on\ } R_{\cal I} \geq q,\quad \frac{X_{\mathcal{I}}^P}{q} \rightarrow Y_{\mathcal{I}}^P, \quad \hbox{as\ } q\rightarrow\infty,
\end{equation}
where the limit $Y_{\mathcal{I}}^P=\{Y_i,\ i\in \mathcal{I}\}$ is an \emph{$r$-Pareto process}.

Any $r$-Pareto process factorizes into a standard-Pareto-distributed \emph{scale variable} $S^P=r(Y_{\mathcal{I}}^P)$, independent of the \emph{profile process} $\eta^P_{\mathcal{I}}=Y_{\mathcal{I}}^P/S^P$ satisfying $r(\eta_{\mathcal{I}})=1$: 
\begin{equation}\label{eq:ppfactor}
Y^P_{\mathcal{I}}\stackrel{d}{=} S^P\,\eta^P_{\mathcal{I}}, \qquad S^P \perp \eta^P_{\mathcal{I}},
\end{equation}
where $\perp$ means "independent of".
\end{theorem}
The result holds for any summary function $r$ as defined above. A converse result states dependence structure convergence of maxima in \eqref{eq:mda} if convergence \eqref{eq:ppconv} holds with $r$ chosen as the maximum  \citep{Dombry.Ribatet.2015}
The existence of the Pareto process limit in \eqref{eq:ppconv} requires a certain regularity of the limit process, such as having realizations that are continuous in space in the case where $\mathcal{I}$ is a subset of $\mathbb{R}^2$. This implies in particular that the data must be asymptotically dependent at small distances in the sense of Equation~\eqref{eq:ad}. 

The summary variable $R_{\mathcal{I}}$ for original data is known to satisfy \newmodif{$P(R_{\mathcal{I}}>q)\sim \theta_r/q$, $q\rightarrow\infty$, such that}
\begin{equation}\label{eq:extcoef}
 q\, P(R_{\mathcal{I}}>q)\sim \theta_r, \quad q\rightarrow \infty,
\end{equation}
where $\theta_r\geq 0$ is the \emph{$r$-extremal coefficient} \citep{Engelke.al.2018b}. Its theoretical value is known beforehand in some cases, such as for the mean aggregation where $\theta_r=1$, but in other cases it depends on the (unknown) extremal dependence structure. The value  $\theta_r=0$, indicating a situation where very extreme aggregated events $R_{\mathcal{I}}$ do not occur, can arise only for very specific dependence structures in $X_{\cal I}$ and specific choices of $r$ but is usually not of practical importance. Often, $\theta_r$ is not known beforehand, but we want to fix a threshold $q$ corresponding to a specific exceedance probability $p=P(R_{\mathcal{I}}>q)$ by assuming equality in the approximation \eqref{eq:extcoef}. Then, we have to estimate $\theta_r$ from the data. This step is described later in Section~\ref{sec:extcoef}. 

\subsection{Lifting mechanism}

According to Equation~\eqref{eq:ppconv}, the distribution of the process $qY^P_{\mathcal{I}}$ provides a good approximation to the distribution of extreme events in $X_{\mathcal{I}}^P$ if $q$ is high and $R_{\mathcal{I}}>q$.  Given $q$ and $\eta^P_{\mathcal{I}}$ \modif{and given a realization $x_{\mathcal{I}}$}, we can generate a new dataset satisfying $r(\tilde{x}^P_{\mathcal{I}})>\tilde{q}$ with a  threshold $\tilde{q}>0$, possibly different from $q$,  by generating a new scale $s^P$ according to the standard Pareto distribution. \modif{The new scale could either be drawn randomly according to the Pareto distribution, or it could be the Pareto quantile associated to a given probability. Because $s^P$ leads to a new realization, we opt for lower case notations here. We then set}   $\tilde{x}^P_{\mathcal{I}}=\tilde{q} s^P \eta^P_{\mathcal{I}}.$ In this case, if $r(x^P_{\mathcal{I}})=q$ and $\eta^P_{\mathcal{I}}=x^P_{\mathcal{I}}/r(x^P_{\mathcal{I}})$, we have transformed an event with return level $q$ of the summary $r$ into an event with return level higher than $\tilde{q}$. From Equation~\eqref{eq:extcoef}, we further see that a return level of $q$ corresponds approximately to a return period of $q/\theta_r$. For a concrete example of the lifting mechanism, consider a rank summary $r(x^P_{\mathcal{I}})=x^P_{(i)}$ with fixed rank $i\in\mathcal{I}=\{1,2,\ldots,n\}$, which includes the minimum, the median and the maximum as special cases. Then, the new sample can be written $\tilde{q} s^P x^P_{\mathcal{I}}/x^P_{(i)}$.

\modif{The normalization to a Pareto scale is natural for establishing limit results in extreme-value theory \citep{Klueppelberg.Resnick.2008}, but is much less used in ``classical" statistical analyses.}  For describing further our resampling approach, we reformulate the above theoretical results by switching to a uniform scale  $X^U=F(X)-1$ on the interval $[-1,0]$, which may lend itself to  easier visual analysis and interpretation, especially when readers are familiar with the copula literature \citep{Joe.2014}.  Since $X^U=F(X)-1=-1/X^P$, the threshold on this scale is $u=-1/q<0$ where $q$ is the threshold on the Pareto scale, $S=-1/S^P$ has uniform distribution on $[-1,0]$, and $\eta_{\mathcal{I}}=1/\eta^P_{\mathcal{I}}$. The convergence \eqref{eq:ppconv} now becomes:
\begin{equation}\label{eq:convunif}
\hbox{Conditional on\ } - R_{\cal I}^{-1} \geq u, \quad -\frac{X^U_{\cal I}}{u} \rightarrow Y_{\cal I}, \quad \hbox{as\ }  u\uparrow 0.
\end{equation}
We now skip superscripts $^U$ for notational simplicity since we will by default assume the uniform scale $^U$ in the following. As before, we have 
\begin{equation}\label{eq:ppfactor2}
Y_{\mathcal{I}}\stackrel{d}{=} S\eta_{\mathcal{I}}, \qquad S \perp \eta_{\mathcal{I}}. 
\end{equation}
\newmodif{While $S$ follows a uniform distribution, it is noteworthy that the transformed Pareto process $Y_{\mathcal{I}}$ does not have uniform marginal distributions as a result of its support being determined by the choice of the risk functional $r$.}
We will make repeated use of the variable $- R_{\cal I}^{-1}$  and  define the notation
$$
V=- R_{\cal I}^{-1} = - \frac{1}{r(X^P_{\cal I})} = - \frac{1}{r(-1/X^U_{\cal I})},
$$
and we use lower-case notation $v$ when referring to its observed value. \newmodif{Due to the Pareto tail in $\mathcal{R}_{\mathcal{I}}$, see \eqref{eq:extcoef}, we obtain that $V$ follows uniform distribution $P(V>u)=-\theta_r u$ when $u<0$ is close to $0$.}
Algorithm~\ref{algo:r-Pareto} below summarizes the steps to generate new $r$-Pareto processes by adopting the uniform scale given by Equations~\eqref{eq:convunif} and \eqref{eq:ppfactor2}, based on threshold exceedances of the summary functional $r$ in data,  where the summary functional has return level constrained to the interval $[u_1,u_2]$ with $u_1\leq u_2\leq 0$ and $u_1<0$ if $u_2=0$.  Specifically, $u_1=u_2$ corresponds to a fixed return level, and $0<u_1<u_2=0$ to threshold exceedances of the return level $u_1$. \newmodif{Given the observed value of $v$, the new return level $\tilde{V}\stackrel{d}{=}uS$ determines the factor $v/\tilde{V}$ by which the return period of the observed risk functional is multiplied during the lifting step. Both \emph{uplifting} ($v/\tilde{V}>1$) and \emph{downlifting} ($v/\tilde{V}<1$) are possible. For a restriction to uplifting, one may set $[v_1,v_2]=[v,0]$. 
}

\begin{algorithm}[htb]
\caption{Lifting with uniform \newmodif{data} margins} 
\label{algo:r-Pareto}
\begin{algorithmic}[1]
\REQUIRE Data $x_{{\cal I},j}^U$, $j=1,\ldots,m$, on uniform marginal scale in $[-1,0]$
\REQUIRE Summary threshold $u<0$ for extracting extreme events
\REQUIRE \newmodif{Thresholds $v_1,v_2$ for lifting where $v_1\leq v_2\leq 0$, and $v_1<0$ if $v_2=0$}
\REQUIRE Event $x^U_{\cal I}=x^U_{{\cal I},j_0}$ exceeding the threshold such that  $v=-1 / r\left(-1/x^U_{{\cal I},j_0}\right) > u$
\STATE Generate a new scale \newmodif{$\tilde V\sim {\cal U}(u_1,u_2)$}.
\STATE Compute the lifted simulation \newmodif{$\tilde{x}^U_{\cal I}= \tilde V x^U_{\cal I}/v$}.
\RETURN $\tilde{x}^U_{\cal I}$.
\end{algorithmic}
\end{algorithm}

\modif{
\subsection{Estimation of exceedance probabilities of the summary functional}
\label{sec:extcoef}
With replications $x_{\mathcal{I},j}$, $j=1,\ldots,m$, we can estimate the $r$-extremal coefficient $\theta_r$ in $P(V>u) \sim -u\theta_r$ with $u<0$
A simple empirical estimator is given by 
$$\hat{\theta}_r=-(um)^{-1}\, \sum_{j=1}^m 1(V_j>u),\quad  V_j=-1/r(-1/x_{\mathcal{I},j}).$$
If events $j=1,\ldots,m$ are independent and identically distributed, this estimator follows a binomial distribution with success probability $P(V>u)$ and rescaled by the factor $-(um)^{-1}$. When simulating new data with $V>u$, we then assume that the probability of observing such events is $-u\hat{\theta}_r$. 
}

 \subsection{Post-processing non extreme observations}
 \label{sec:postprocess}
We denote by $x^U_{{\cal I},j_0}$  one of the events exceeding the summary threshold, such that 
$$v=-1 / r\left(-1/x^U_{{\cal I},j_0}\right) > u.$$
\newmodif{An important step after lifting observed profile processes using Algorithm~\ref{algo:r-Pareto} is the post-processing of small or moderately large values, for which the lifting step may create artefacts. Specifically, we may obtain a lifted value $\tilde{x}_{i}<-1$ outside of the support of the uniform data distribution over $[-1,0]$, corresponding to a situation that can occur if $\tilde{v}<v=-1/r(-1/x^U_{{\cal I},j_0})$, \emph{i.e.}, if the scaling factor $s$ in step~3 of Algorithm~\ref{algo:resample} satisfies $s>1$ (\emph{downlifting}). In this situation, values smaller than $-1$ cannot be backtransformed to the original marginal scale. In other cases, corresponding to a situation where $\tilde{v}>v$ and $s<1$ (\emph{uplifting}), the lifted dataset may satisfy $\min_{i\in\mathcal{I}}{\tilde{x}_{ij_0}}>\varepsilon>-1$ with a relatively large interval $[-1,\varepsilon]$ not covered by the lifted data; this may also be unrealistic. For this reason, we do not directly apply the lifting step to data values below the marginal threshold $u^{\text{marg}}=\hat{F}(u^F)-1$, where  $u^F$ is the marginal threshold defined on the original scale of data and $\hat{F}$ is the estimated marginal distribution; see Algorithm~\ref{algo:estim_f}. To set the value $u^{\text{marg}}$, below which post-processing is applied, it would also be possible to use another, not too low  threshold, such as the median $u^{\text{marg}}=-0.5$ or the $75\%$-quantile $u^{\text{marg}}=-0.25$.
}

\newmodif{
For deriving our postprocessing procedure, we consider the  dataset lifted according to Algorithm~\ref{algo:r-Pareto} as a mixture of two populations. The first population consists of the part of the standardized dataset $x^U_{\cal I}$ where  values are below $u^{\text{marg}}$, and we exclude these values from the lifting transformation in step~2 in Algorithm~\ref{algo:r-Pareto}. This first population stems from a uniform distribution on $[-1,u^{\text{marg}}]$ with probability mass $1+u^{\text{marg}}$ (when looking at the full uniform sample comprising extreme and non extreme episodes). The second population corresponds to the part uplifted according to the new scale variable $\tilde{V}$ generated in step~1 of Algorithm~\ref{algo:r-Pareto}.  The second population thus stems from a uniform distribution on $[u^{\text{marg}}\tilde{V}/v,0]$ and probability mass $-u^{\text{marg}}$. At this stage, the probability density of the newly generated dataset on the interval $(\min(u^{\text{marg}},u^{\text{marg}}\tilde{V}/v),\max(u^{\text{marg}},u^{\text{marg}}\tilde{V}/v))$ is either equal to $0$ in the case of uplifting, or equal to $1+v/\tilde{V}$ in the case of downlifting. Without corrections, simulated images resulting from this modification of Algorithm~\ref{algo:r-Pareto} would show artificial discontinuities when we move from areas belonging to the bulk of the distribution to areas corresponding to lifted values. We therefore propose to post-process the first population of non-extreme values, \emph{i.e.}, the values $\tilde{x}^U_{ij_0}$ with $x_{ij_0}<u_F$, $i\in\cal I$.
}

\begin{figure}
    \centering
    \includegraphics[height=4cm]{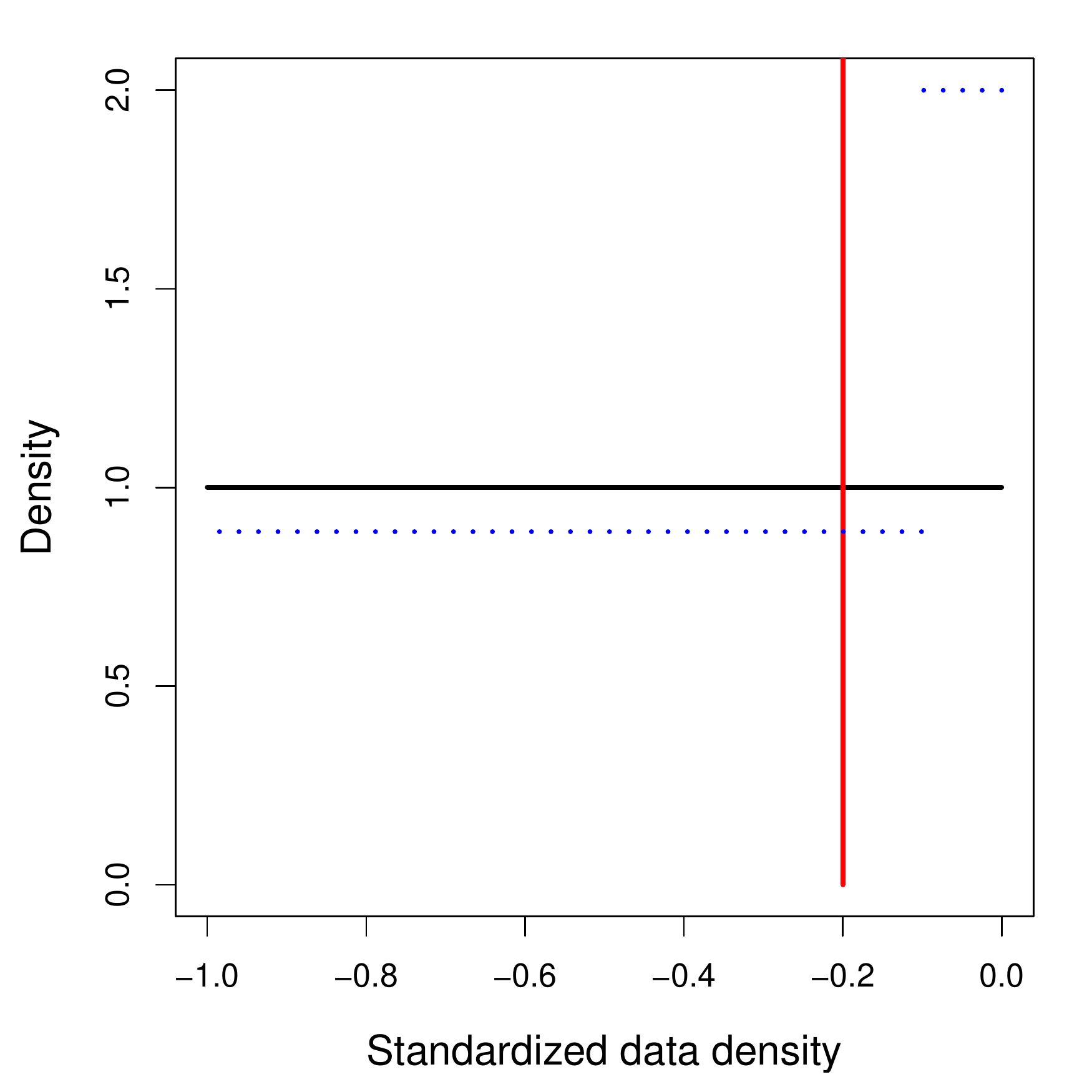} \qquad 
     \includegraphics[height=4cm]{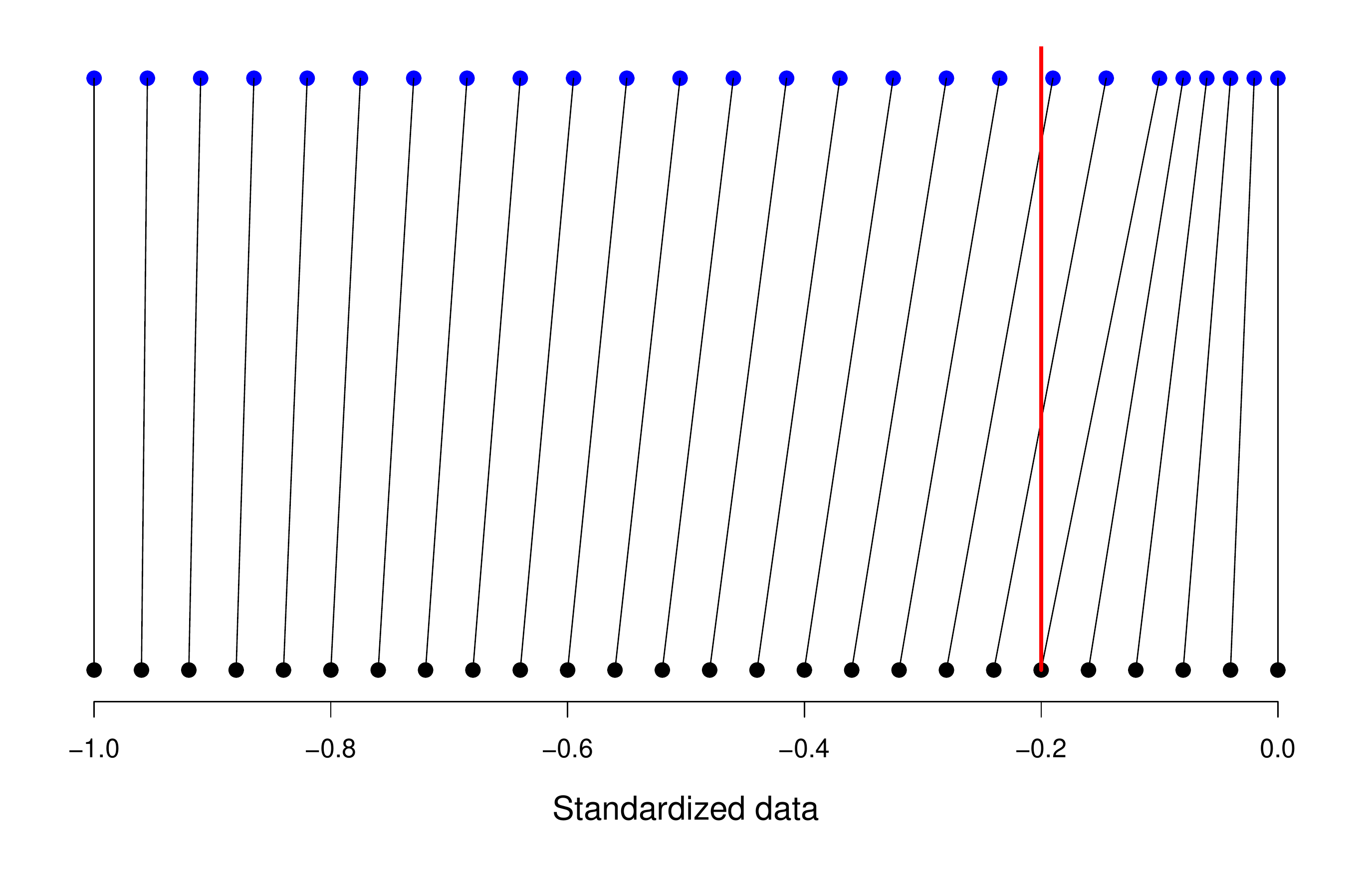}  \\ 
        \includegraphics[height=4cm]{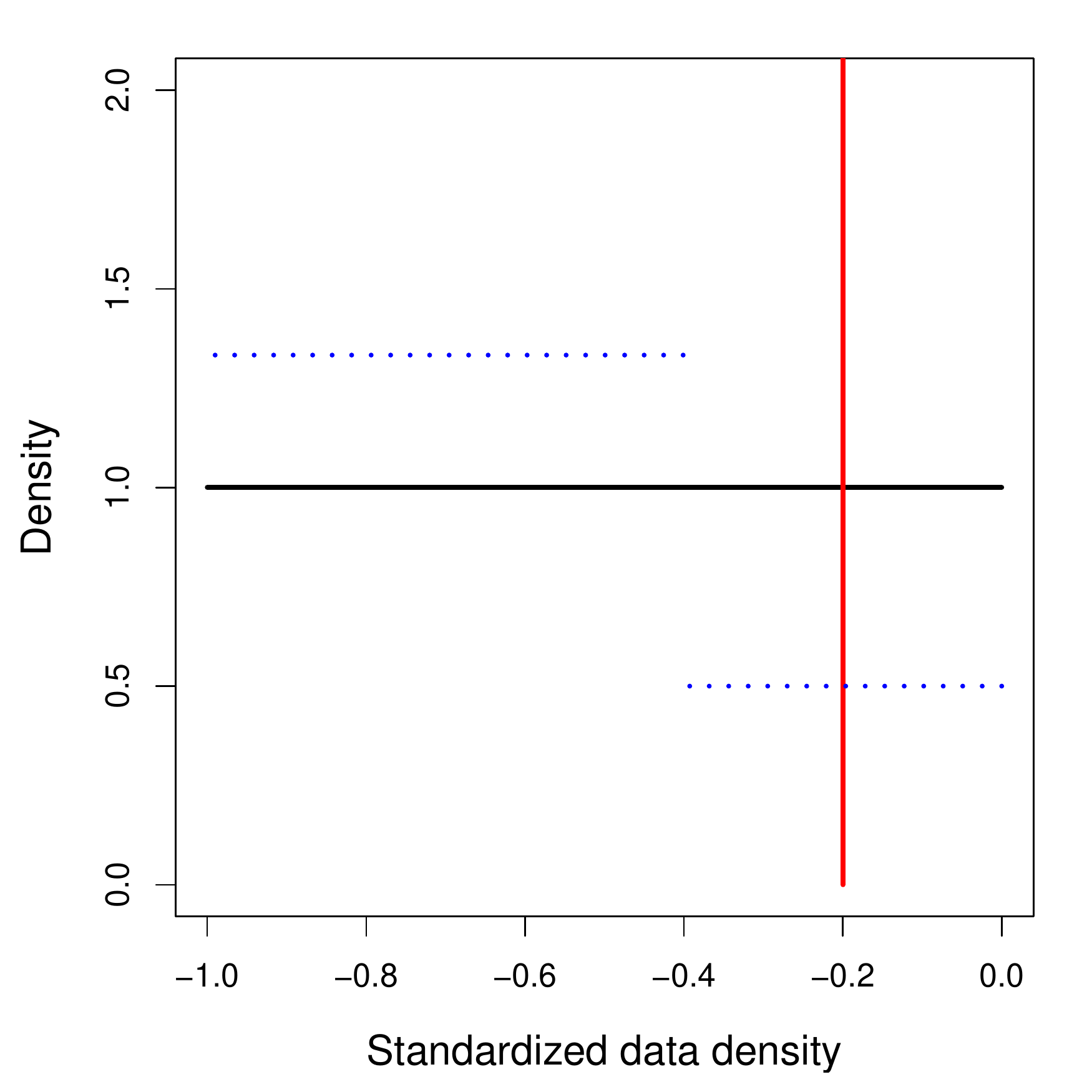} \qquad 
     \includegraphics[height=4cm]{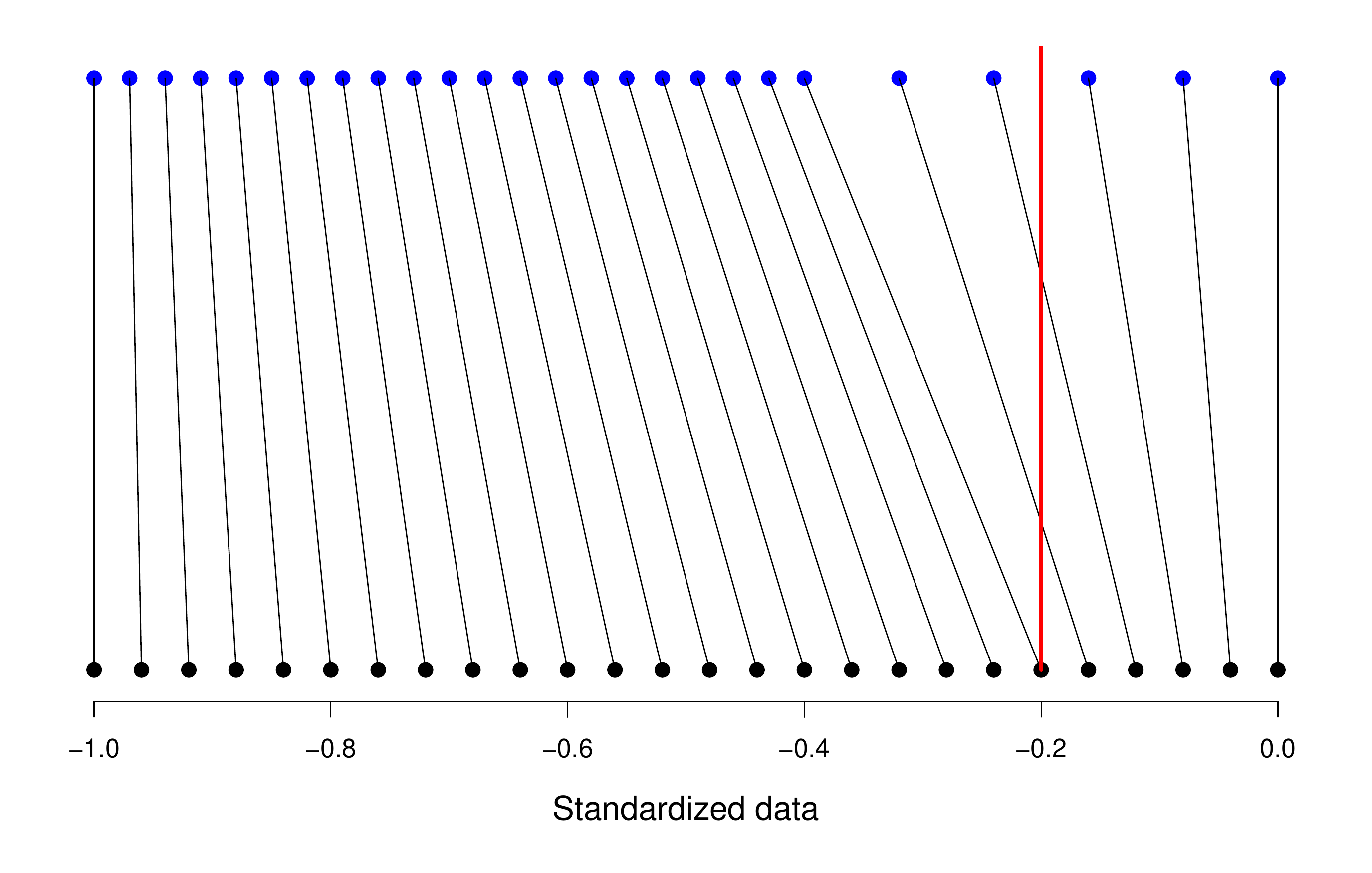}  
    \caption{\newmodif{Illustration of post-processing procedure for lifting factors $s=1/2$ (uplifting, first row) and $s=2$ (downlifting, second row), with the marginal threshold $u^{\text{marg}}=-0.2$ shown in red. Left: A uniform data density before and after transformation. Right: Transformation of original points (black) to post-processed lifted points (blue).}}
    \label{fig:postprocess}
\end{figure}

\newmodif{
Figure~\ref{fig:postprocess} illustrates the general lifting procedure that we explain in the following, for $s=1/2$ (uplifting -- the return period of the observed risk functional is doubled) and $s=2$ (downlifting -- the return period of the observed risk functional is halved). We keep the lifted population fixed now, and we follow a principle of minimum Kullback-Leibler (KL) divergence to post-process non-extreme values. Therefore, we minimize the divergence between a uniform distribution on $[-1,0]$ (corresponding to the initial distribution of standardized values) and the mixture distribution of the two populations. The resulting modified distribution of the first population is uniform on $[-1,u^{\text{marg}}\tilde{V}/v]$. In other words, we  post-process the first population according to the affine transformation that maps the uniform distribution on $[-1,u^{\text{marg}}]$ to the uniform distribution on $[-1,u^{\text{marg}}\tilde{V}/v]$. Algorithm~\ref{algo:r-Pareto2} below summarizes our modified lifting procedure.}  

\newmodif{We here opt for this relatively simple post-processing procedure based on the  minimum KL argument. Though, it would also be possible to provide a smoother transition between the tail and bulk regions, for instance by constructing a continuous and piecewise linear density that avoids  the jumps visible on the left-hand diagrams of Figure~\ref{fig:postprocess}.}

\begin{algorithm}[htb]
\caption{Lifting with uniform \newmodif{data} margins and postprocessing} 
\label{algo:r-Pareto2}
Requirements are identical to \modif{those listed in} Algorithm~\ref{algo:r-Pareto}.  

\begin{algorithmic}[1]
\STATE Set $u^{\text{marg}}=\hat{F}(u^F)-1$ (or some other not too low quantile).
\STATE Generate a new scale $\tilde V\sim {\cal U}(v_1,v_2)$.
\STATE Set the scaling factor $s=\tilde{V}/v$.
\STATE Compute the lifted simulation $\tilde{x}^U_{\cal I}$:
\FOR{$(i \in {\cal I})$}
\IF{$x^U_i>u^{\text{marg}}$}
\STATE Set $\tilde{x}^U_i= sx^U_i$.
\ELSE 
\STATE Set $\tilde{x}^U_{i} = -1 + \frac{1+su^{\text{marg}}}{1+u^{\text{marg}}} \left(1+ x_{i}^U \right)$.
\ENDIF
\ENDFOR
\RETURN $\tilde{x}^U_{\cal I}$.
\end{algorithmic}
\end{algorithm}

\section{Resampling dependent extremes}
\label{sec:uplifting}

\subsection{Replicated data setting}

Statistical analyses in classical extreme value theory are typically applied to datasets  $x_{ij}$, $i\in \mathcal{I}$, $j=1,\ldots,m$ with many replications $m$, such as spatial snapshots taken at regular time steps. In such studies, replications are assumed to be independent, or to correspond to the realization of a stochastic process that is mixing, such that the dependence between distant subsamples of the process becomes very weak, and negligible in practice. Specifically, if $\mathcal{I}$ is given as a domain in $2$D space and $j=1,2\ldots$ are time steps, we assume that  dependence between the processes $X_{\mathcal{I},j_1}$ and $X_{\mathcal{I},j_2}$ ultimately vanishes for increasing time lags $|j_2-j_1|$. Then, the extreme events correspond to the replicates satisfying a threshold exceedance criterion of the summary functional $r$ with a high enough threshold, such that the asymptotic properties stated in  Theorem~\ref{theor:gp} can be expected to provide a good description of the data distribution. 

In other situations, for example relating to geology and geosciences \citep{Chiles2012}, there is no time replicate. If the spatial domain $\mathcal{I}$ is very large enough in comparison to the correlation range, one could divide $\mathcal{I}$ into a large number $m$  of blocks of same size and proceed as above. In cases where we do not observe any useful replication structure in data, it would be awkward to apply the Algorithm~\ref{algo:r-Pareto}, which is based on the large-sample assumptions underpinning Theorem~\ref{theor:gp}. We exclude such cases in the following, since they would require strong assumptions that are difficult to validate in practice. For the remainder of this paper, we assume that the training dataset comprises a large number of replicates $m$, at least several hundreds. 

\subsection{Resampling algorithm for dependent extremes}

In our approach to enrichment of extreme data, we sample from the distribution of the scale variable $S$ in Equation~\eqref{eq:ppfactor2} and generate new, potentially more extreme events by combining resampled univariate scales with observed profile processes. In cases where the univariate distribution $F$ is known beforehand, this mechanism is fully nonparametric, except for the $r$-extremal coefficient $\theta_r$ in Equation~\eqref{eq:extcoef}, which we have to estimate in some cases. 

\newmodif{
In principle, there are two possible approaches for the step at which we apply nonparametric resampling. One solution would be ``resampling first -- lifting second": 1) resample the profile processes $\eta_{\mathcal{I}}$  in \eqref{eq:ppfactor2}, 2) lift them, and finally (3) post-process the lifted images. This could be statistically sound and produce realistic simulations if the constraints on the profile process are preserved. However, we here prefer the solution ``lifting first -- resampling second": (1) lift profile processes, (2) post-process the lifted images, (3) resample using the lifted images; this is for the following reasons.
First, explicitly conserving the constraints on the profile process during resampling would require sophisticated extensions of the standard nonparametric resampling techniques. Second, we wish to properly separate the data enrichment step (lifting followed by postprocessing) and the resampling step. The asymptotic assumptions made by our approach are minimal in a certain sense but still impose validity of asymptotic theory for the finite-sample data, such that the lifting procedure may not be free of introducing some bias, especially for relatively low values. Moreover, one may aim to impose certain local constraints in the resampling simulations, which would be more difficult if resampling is performed before the lifting step.  In summary, to avoid that  biases propagate too strongly to the final resampling simulations, and to facilitate validation of the two steps (enrichment, resampling) in practice, we think that it is more cautious to perform resampling after lifting.}

We now summarize the main steps of the full procedure for resampling with extremes in the Algorithm~\ref{algo:resample}. We first  extract $m'$ extreme events, then lift them to generate $m''$ events \newmodif{(and postprocess them)}, and finally generate $m'''$ simulations with nonparametric resampling. Simulations can be generated on a support $\mathcal{I}'$ that is different from the original support $\mathcal{I}$. We distinguish two cases: In the first case,  we aim to perform simulation for a fixed return level $\tilde{v}$ of $V$ such that  the return level $\tilde{v}$ of the lifted dataset should be  (approximately) equal to $\tilde{v}$. In this case, we set $m''=m'$ since each extreme event is lifted exactly once to the target return level. In the second case, we want to produce simulations for which the summary functional $\tilde{v}$ has values within an interval \newmodif{$[v_1,v_2]$}; see Algorithm~\ref{algo:r-Pareto}. A special case of the interval constraint arises with exceedances of a summary threshold $v_1$ such that $v_2=0$ and $\tilde{v}\geq v_1$. In the second case we can generate any number $m''$ of lifted events using Algorithm~\ref{algo:r-Pareto}. 

It is not easy to provide precise but general rules on how to fix the numbers $m'$, $m''$ and $m'''$. The $m'$ extreme events used as a training sample for data enrichment should correspond to relatively high quantiles of the summary functional $r(X^U_{\mathcal{I}})$ where the asymptotics leading to the scale-profile decomposition in Equation~\eqref{eq:ppfactor2} kick in, and $m'$ should be large enough to provide a training sample that is representative of the extremal dependence patterns in the data-generating process. If we lift to a specific target return value (\emph{i.e.}, $v_1=v_2<0$), then $m''=m'$, and each extreme training event is lifted exactly once. The strategy of setting $m''=m'''$ usually makes sense, such that the extreme event magnitudes in a dataset enriched with $m''$ images of lifted extremes will be representative for a simulated dataset with $m'''$ extreme episodes. We adopt this approach in our data application in Section~\ref{sec:application}. 

To increase flexibility of our procedure, we further allow for nonstationary margins $X_{ij}\sim F_i$ by estimating the density $\hat{f}_i$ of $F_i$ in Algorithm~\ref{algo:estim_f} from a sample of data collected at index $i$ or close to index $i$. By applying a nonparametric resampling method of our choice to the resulting standardized data $x_{\cal I}^U$, we bypass the intricate handling of nonstationary margins during this resampling step. 

Application-specific checks using hold-out data to support validation of the model, such as those used in the data application, are recommended and can help to confirm sound behavior of the extrapolation mechanism. We refer to \citet{Palacios.al.2019} for a detailed discussion. 

\begin{algorithm}[htb]  
\caption{Resampling dependent extremes} 
\label{algo:resample}
\begin{algorithmic}[1]
\REQUIRE Training data $x_{{\cal I},j}$, $j=1,\ldots,m$, on original scale 
\REQUIRE Summary functional $r$
\REQUIRE Marginal thresholds $u^F_{i}$, $i\in \cal I$ 
\REQUIRE Summary threshold $v<0$  for extracting extreme events from uniform data on $[-1,0]$ using  $r$
\REQUIRE Thresholds $v_1,v_2$ for lifting, where $v_1\leq v_2\leq 0$, and $v_1<0$ if $v_2=0$
\STATE Estimate the marginal distributions  $\hat{F}_i$ using Algorithm~\ref{algo:estim_f} with marginal thresholds $u^F_{i}$, $i\in\mathcal{I}$. 
\STATE Standardize data to uniform scale $x^U_{ij}=\hat{F}_i(x_{ij})-1$, $i\in\mathcal{I},\ j=1,\ldots,m$.
\STATE Extract extreme events $x^U_{\mathcal{I},j_\ell}$, $\ell=1,\ldots,m'$, satisfying $v_{j_\ell}=-1/r\left(-1/x^U_{\mathcal{I},j_\ell}\right)>v$.
\STATE Generate a sample $\tilde{x}^U_{\mathcal{I},j}$, $j=1,\ldots,m''$, of lifted extreme events  using Algorithm~\ref{algo:r-Pareto2}. 
\STATE Perform nonparametric resampling based on the enriched data  to produce $m'''$ simulations 
$\overline{x}_{ij}^U$, $i\in \mathcal{I}',j=1,\ldots,m'''$.
\STATE Backtransform resampled data to the original marginal scale: $$\tilde{x}_{ij}= \hat{F}_i^{-1}\left(1+\overline{x}_{ij}^U\right),\quad i\in\mathcal{I}',\quad j=1,\ldots,m'''.$$
\end{algorithmic}
\end{algorithm}

\section{Application to heatwaves in France}
\label{sec:application}

\begin{figure}
    \centering
    \includegraphics[width=7.5cm]{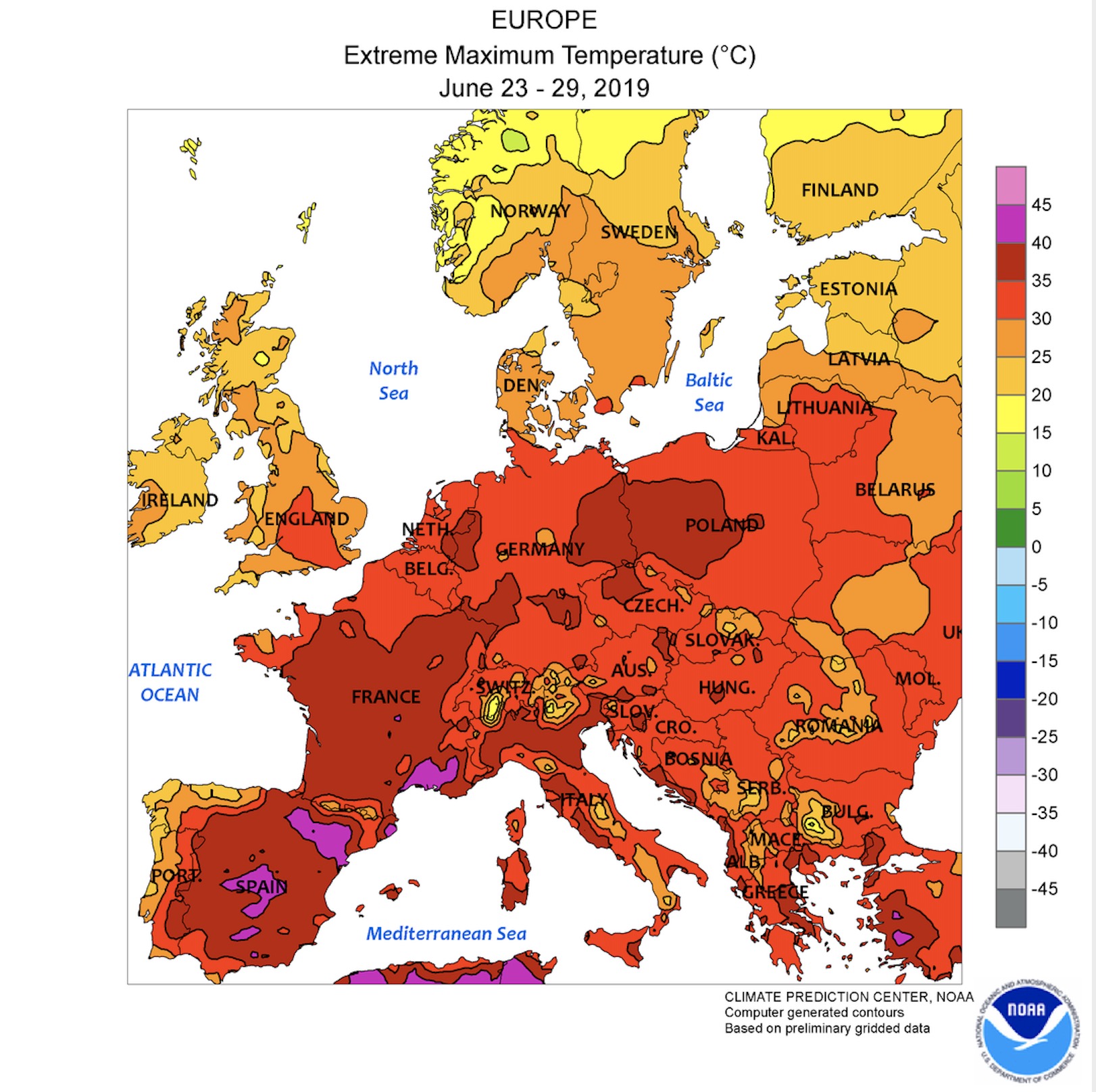}  \qquad \includegraphics[width=7.5cm]{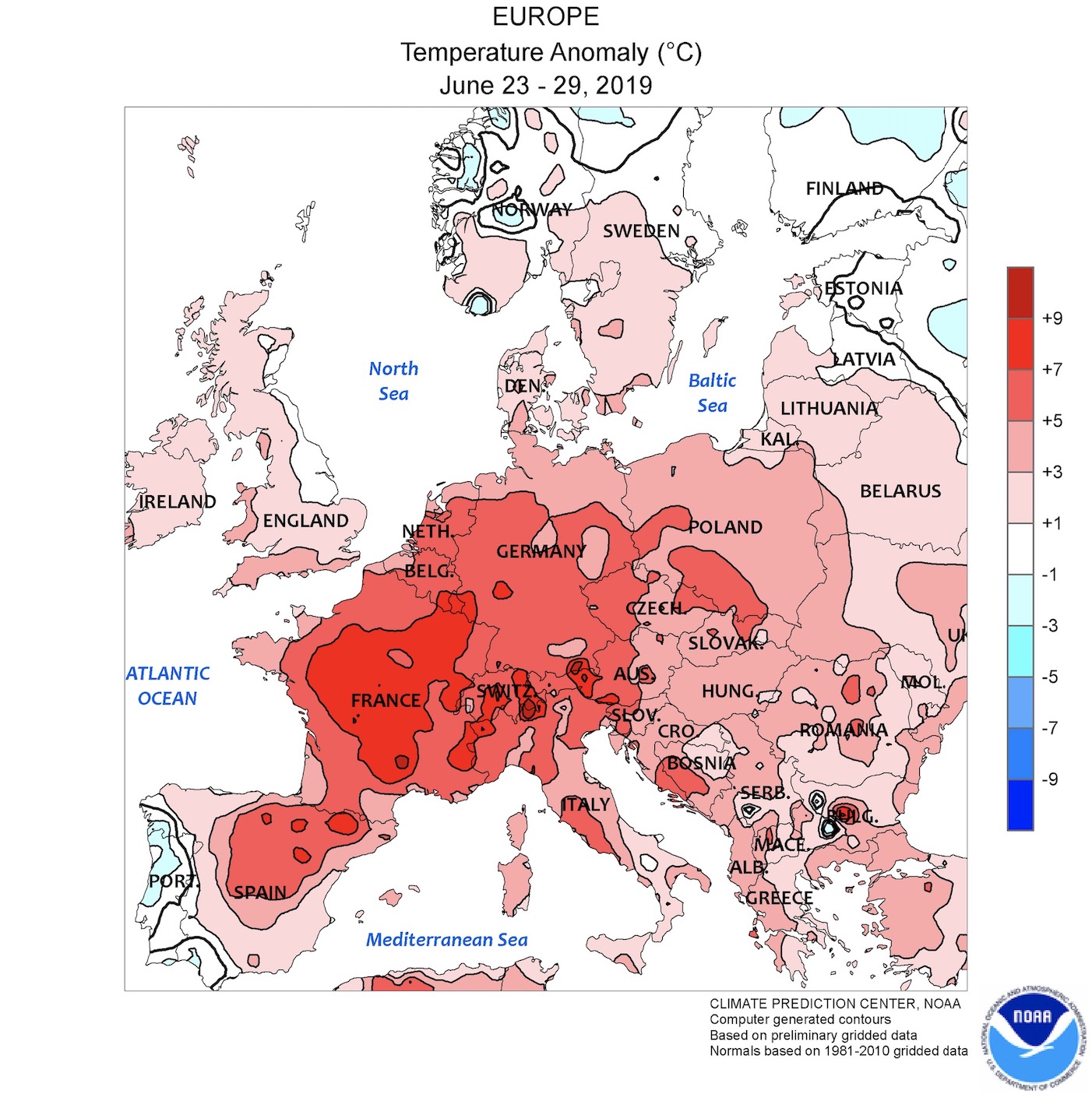}
    \caption{Extreme maximum temperature (in $^{\circ}$C; left display) and temperature anomaly (in $^{\circ}$C; right display) over Europe, June 23rd to 29th of 2019. Illustration provided by the US National Oceanic and Atmospheric Administration (NOAA).}
    \label{fig:2019heatwave}
\end{figure}

The heatwave that hit \modif{France and other parts of Europe} in 2019 from June 23rd to June 30th was one of the most extreme heatwaves ever recorded in Europe. June 27th is the hottest day recorded for June over the period $1900$-$2019$ in France with a daily average temperature of 27.9 $^{\circ}$C,  which is 8.6 $^{\circ}$C "above normal", see Figure \ref{fig:2019heatwave}. \modif{We remind the interested reader that the "normal" is, by  definition, the 30-year average, recomputed every 10 years. Specifically, it is here  the 30 daily average temperatures measured on 27th of June, from 1981 to 2010.}  The temperature maxima on 26th to 29th of June reached an exceptional level over a large part of Europe, equating or exceeding the value of 35$^{\circ}$C . In France, these afternoons were the hottest since the heatwave of August 2003, and a new absolute record was established for 23\% of the weather stations maintained by M\'et\'eo France. \modif{It resulted in an excess of 1435 deaths according to Sant\'e Publique France. Up to 10,000 hectares of vineyards were destroyed in the sole department of H\'erault where the heatwave peaked at 45.9$^{\circ}$C.}

This exceptional event illustrates the need for simulation techniques able to simulate new extreme fields of unobserved magnitude, which can then be fed to process-based impact models to obtain reliable projections for such extreme events of unprecedented magnitude.  

Based on the analysis of $7$ years of daily maximum temperatures, our aim is to simulate new heatwaves corresponding to a given return period. In order to focus on the simulation approach, we make the simplifying assumption of temporal stationarity over years, \emph{i.e.}, we deliberately ignore the presence of trends due to climate change and long-term periodic climate cycles.   

\subsection{Data preprocessing and standardization}

We consider SAFRAN reanalysis data over France and part of Switzerland. SAFRAN  combines ECMWF  global  reanalysis  archives  and  all  available surface  observations from the  climatological database of M\'et\'eo-France to produce high temporal  and  spatial  resolution  (6 hours,  8 km) for several meteorological variables, including  temperature,   humidity,   wind   speed  and cloudiness \citep{vidal201050}. Here, we selected  daily Maximum  Temperature for the months of June to September from 2010 to 2016. 

\begin{figure}
    \centering
    \includegraphics[width=7.5cm]{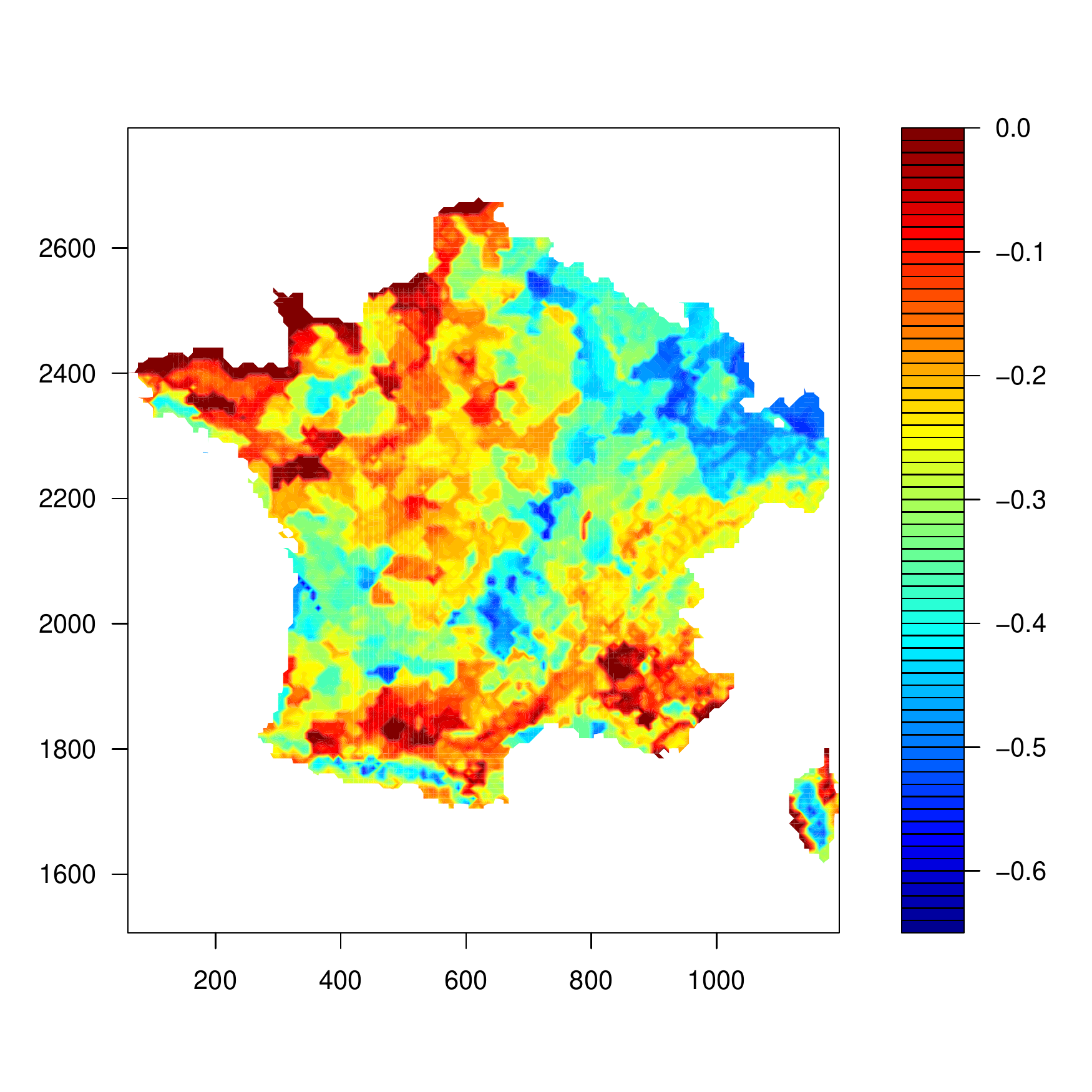}  \qquad \includegraphics[width=7.5cm]{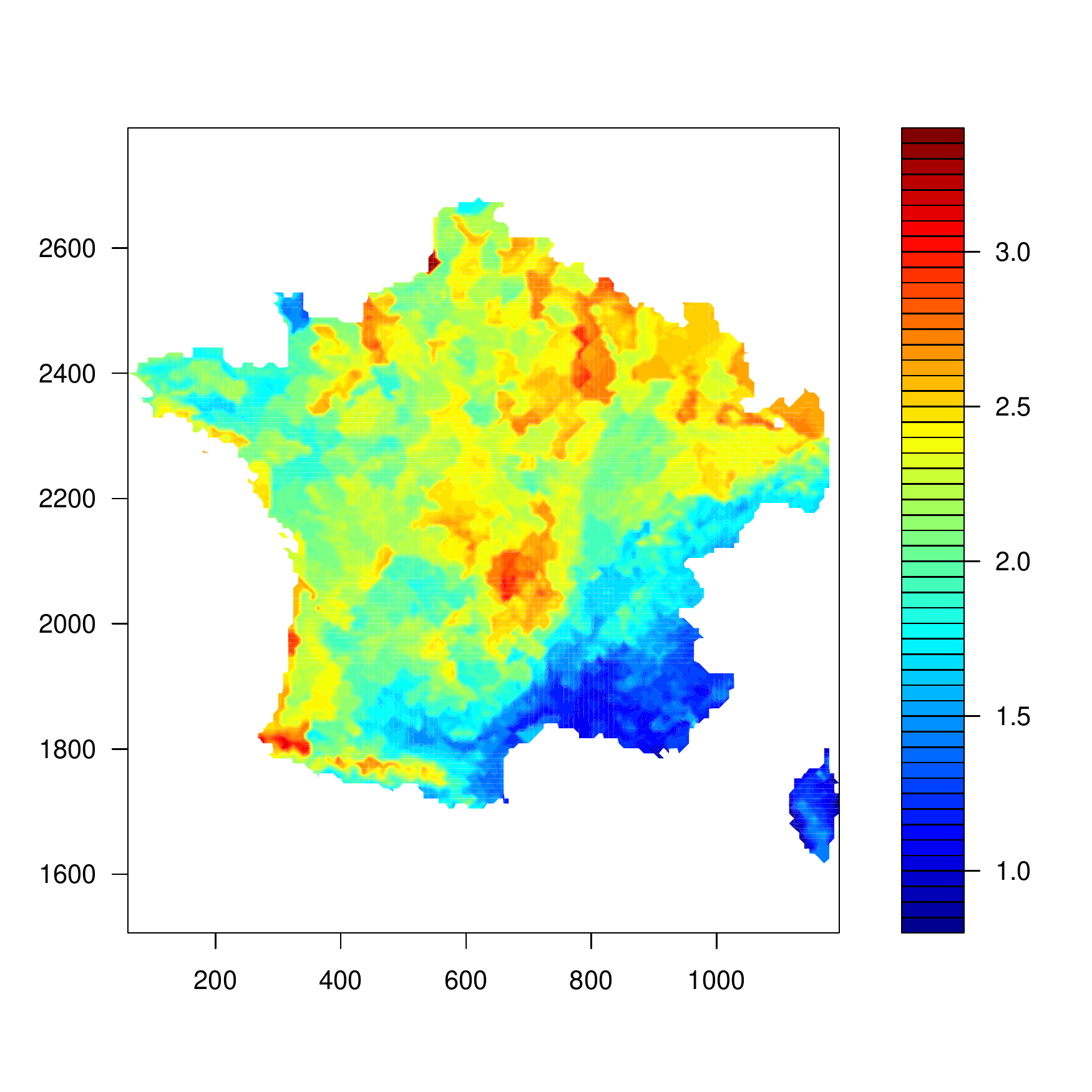}
    \caption{Estimated shape ($\hat{\xi}_u$, left) and scale ($\hat{\sigma}_u$, right) parameters of the Generalized Pareto tail corresponding to the exceedance probability $p_u=0.05$.}
    \label{fig:marginals}
\end{figure}

Data are preprocessed in the following way. For each $8 \times 8$ km$^2$ SAFRAN grid cell $i$, we first fit a marginal probability density function $\hat{f}_i$ with kernel density estimate in the bulk and GPD tail as described in Section \ref{sec:margins} and in Algorithm~\ref{algo:estim_f}, using a threshold corresponding to the local $95\%$ quantile. \modif{The tail indices $\xi(i)$ have been estimated through the maximum likelihood approach with a restriction to \newmodif{non-positive} values since power-law tails are not realistic for temperatures in the European territory of France.} Then, using the fitted marginal distribution, the daily maximum temperatures are transformed into $[-1,0]$ uniforms by applying $X^U_{i,t} = F_{i}(X_t)-1$, where $i$ is the cell index ranging from $1$ to $9892$, and $t$ is the day index ranging from $1$ to $854$. Figure \ref{fig:marginals} reports the estimated shape parameter $\xi_u(i)$ and the estimated scale parameter $\sigma_u(i)$, where $u(i)$ is the local threshold, and maximum likelihood estimation was applied. As summary statistics $r$ for a given day $t$ we use the median value over all cells, \emph{i.e.},
$$r(t) = \hbox{med}\{ X^U(\cdot,t)\}.$$

\subsection{Data enrichment and resampling}

\begin{figure}
    \centering
    \includegraphics[width=5cm]{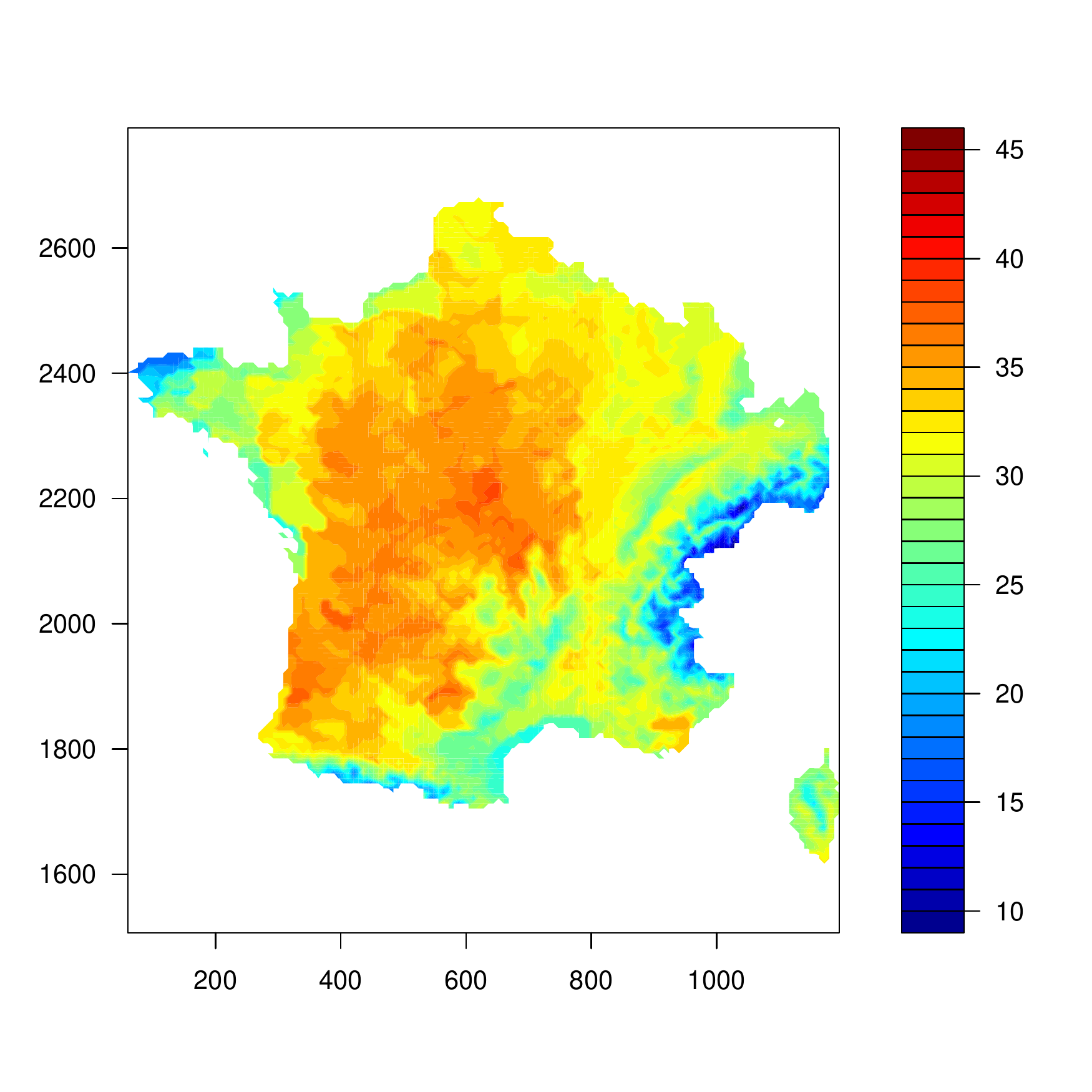}   \quad \includegraphics[width=5cm]{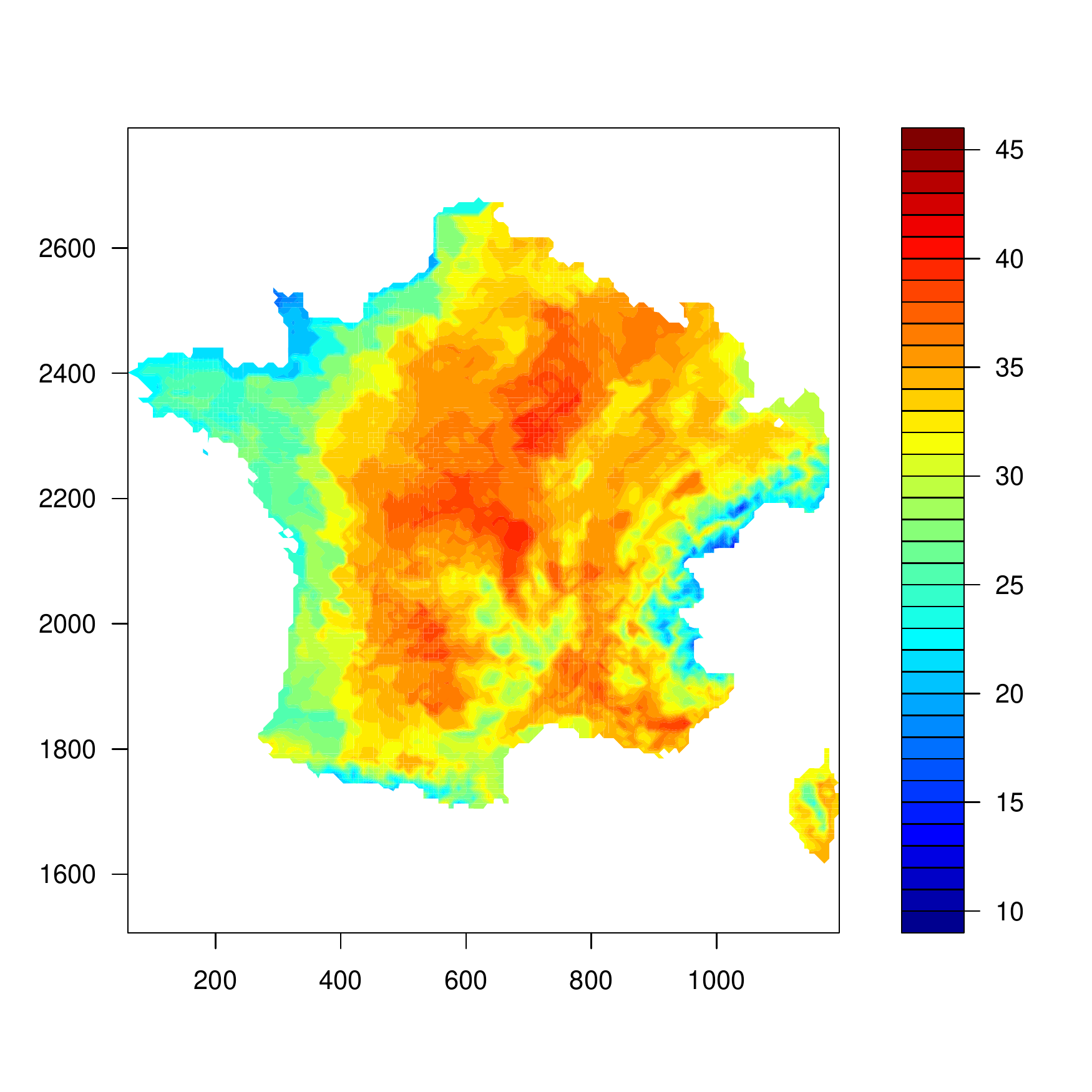}  \quad
        \includegraphics[width=5cm]{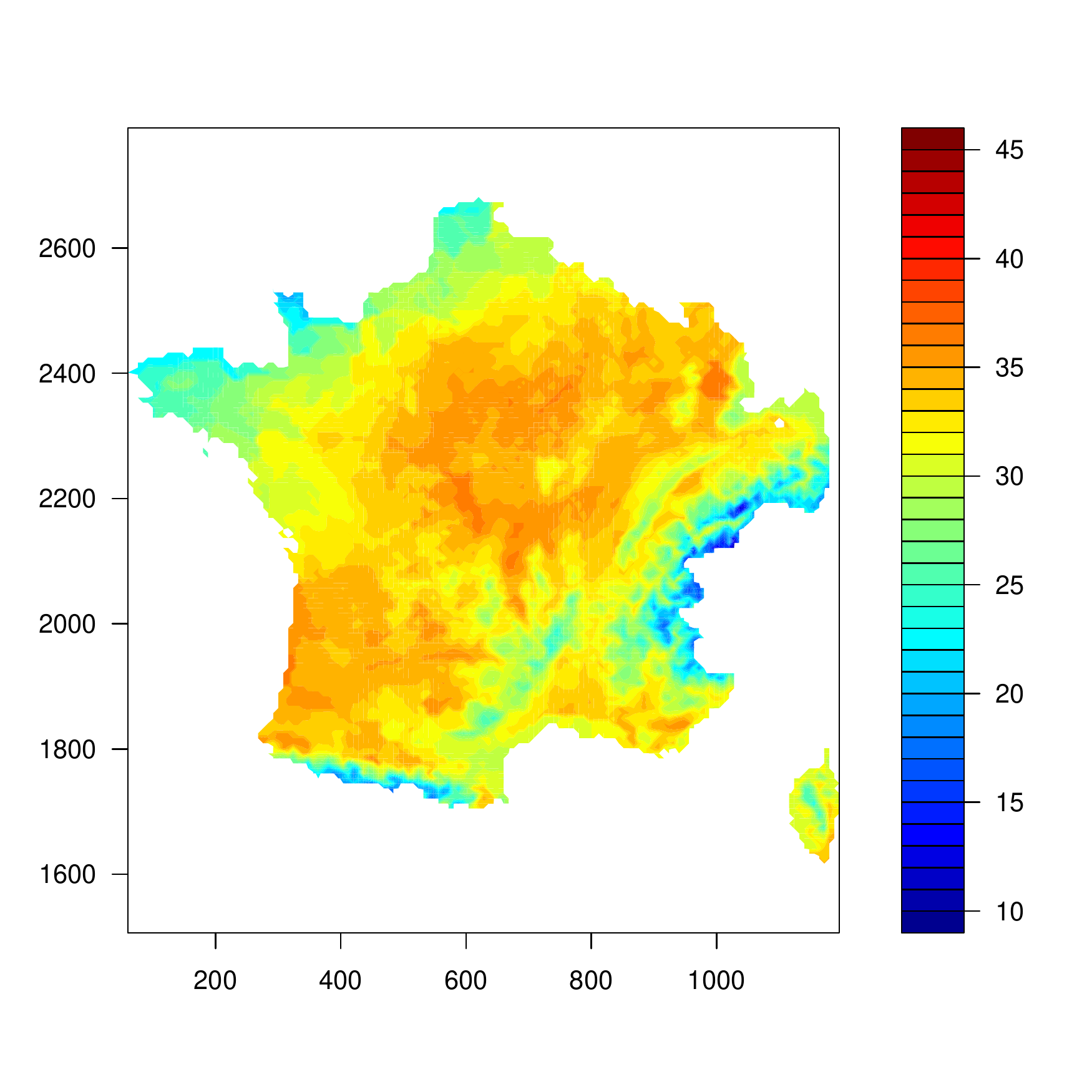} \\   \includegraphics[width=5cm]{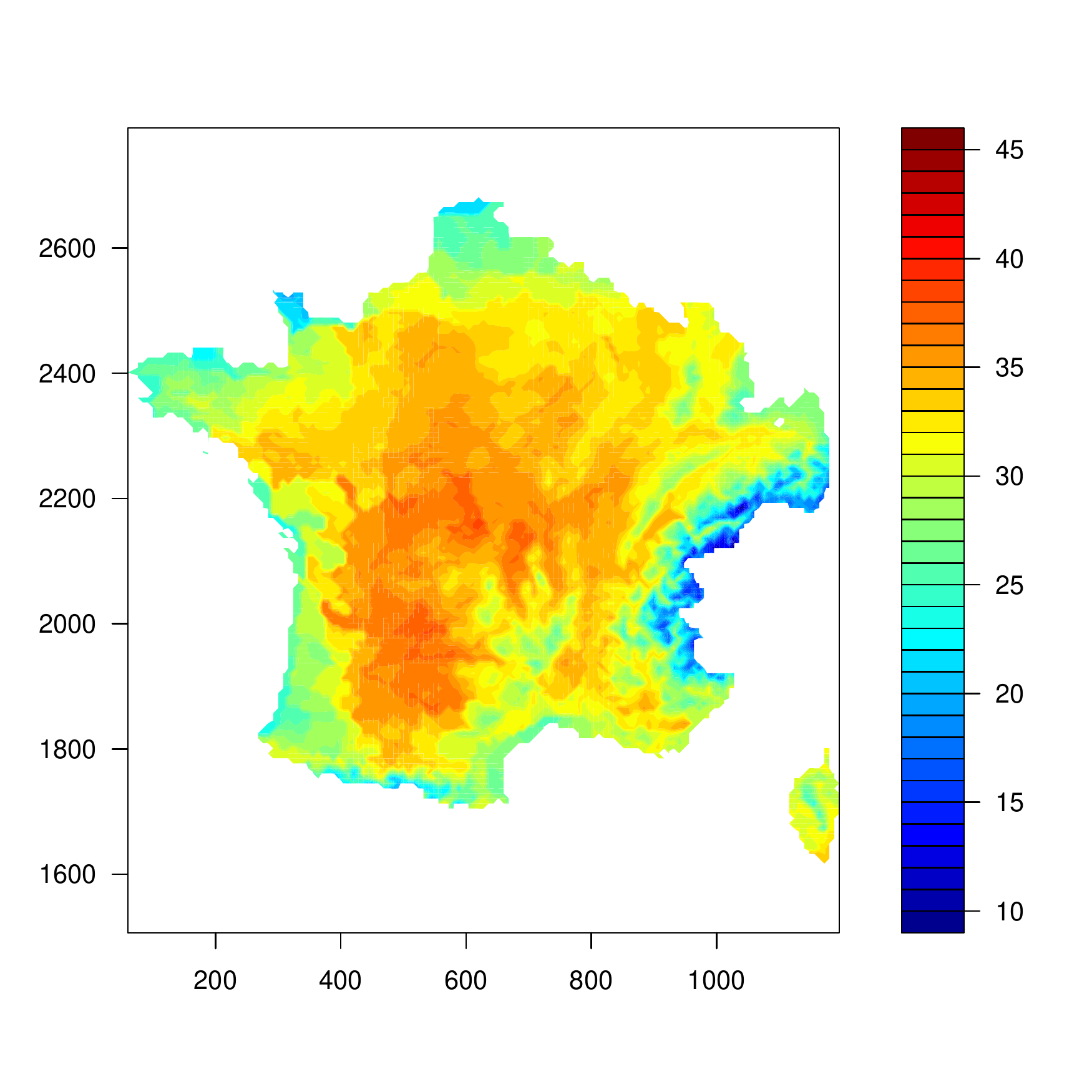}  \quad
            \includegraphics[width=5cm]{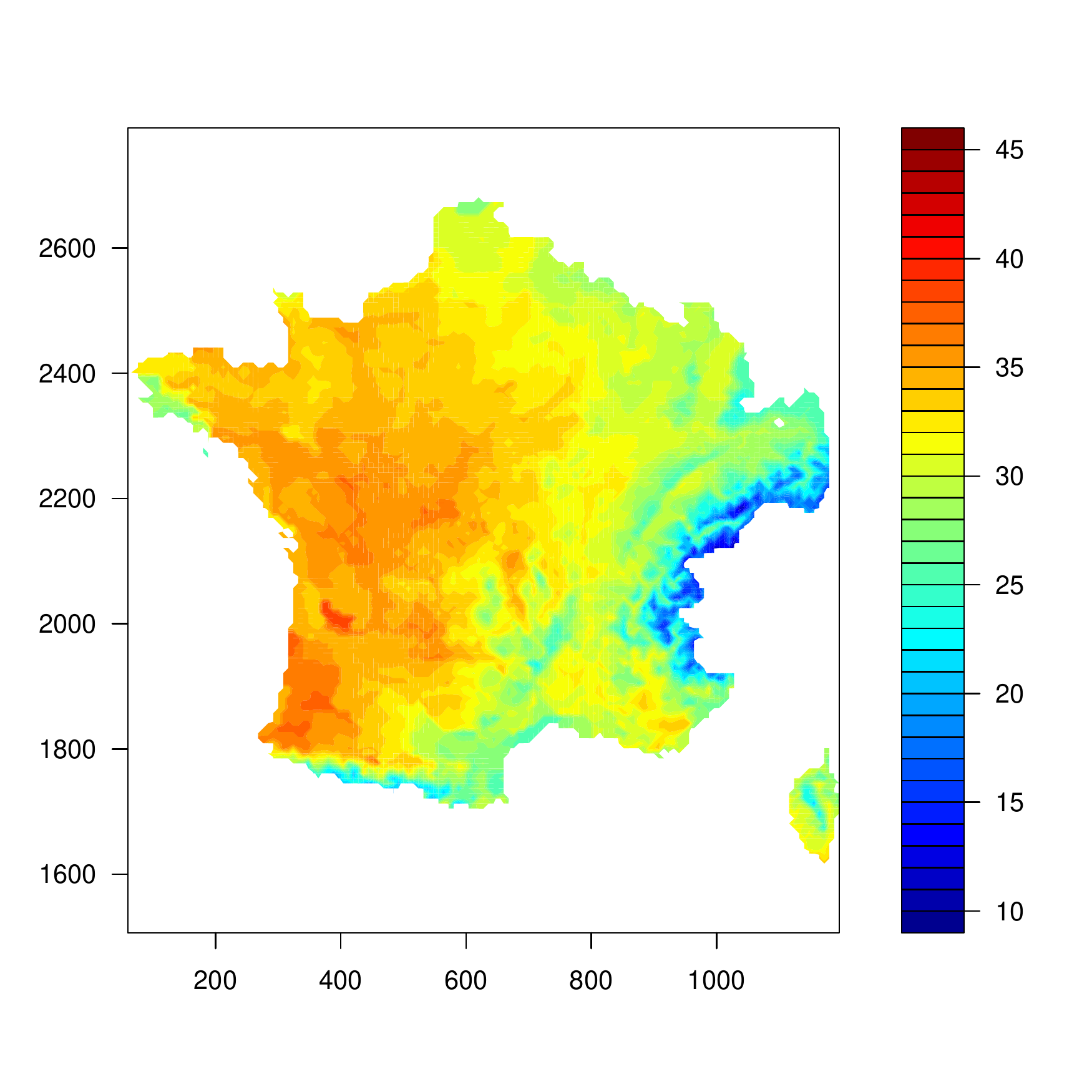}   \qquad \includegraphics[width=5cm]{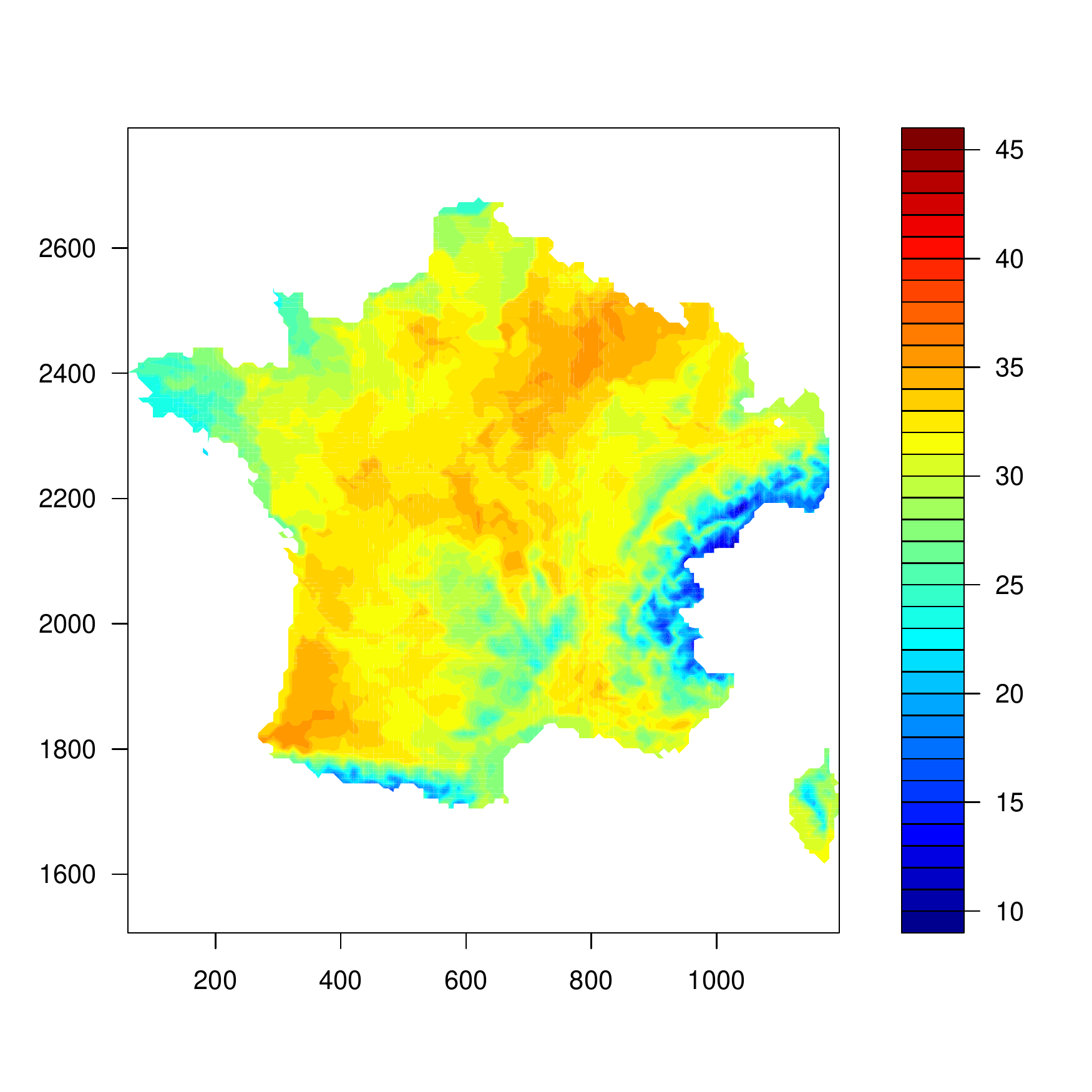}  
    \caption{The six most extreme daily events pertaining to different heat episodes. From left to right: 27/06/2011, 19/08/2012, 03/07/2015 (top); 16/07/2015, 19/07/2016, 27/08/2016 (bottom).}
    \label{fig:6real_episodes}
\end{figure}

We now illustrate how we can create new extreme events with values beyond the range of observed ones based on the analysis of the dataset presented above. We select the $6$ most extreme events, corresponding to the $6$ highest values of $r(t)$. In addition, we impose that these events should be separated by $2$ days at least in order to prevent selecting highly correlated events. The selected events occurred on 27/06/2011, 19/08/2012, 03/07/2015, 16/07/2015, 19/07/2016, 27/08/2016. They are depicted in Figure \ref{fig:6real_episodes}. The amplitude and the spatial extent of zones with high values are quite variable for these events. For example,  the second and most extreme event is characterized by a very large area with very high temperatures. 

\begin{figure}
    \centering
    \includegraphics[width=7cm]{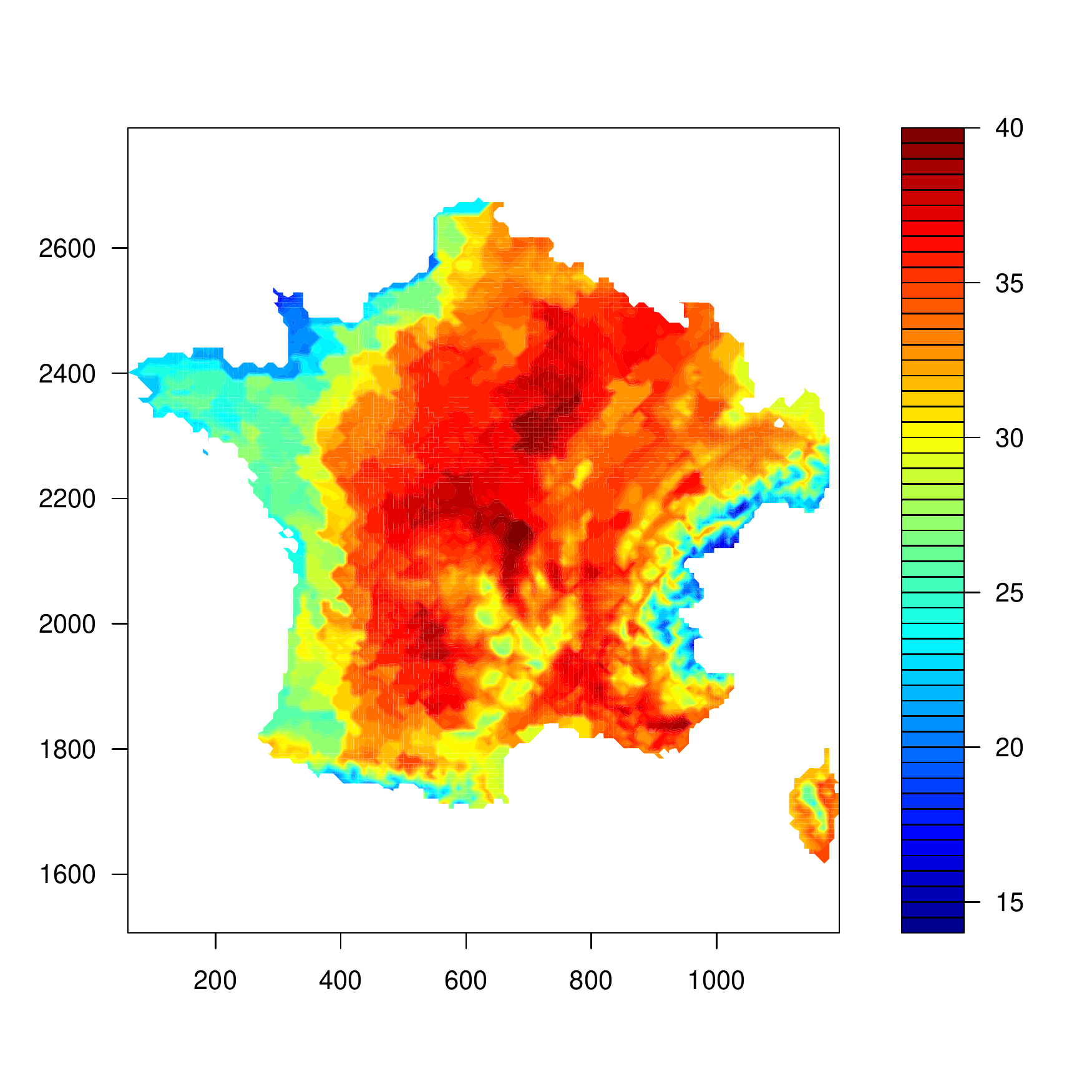}   \quad \includegraphics[width=7cm]{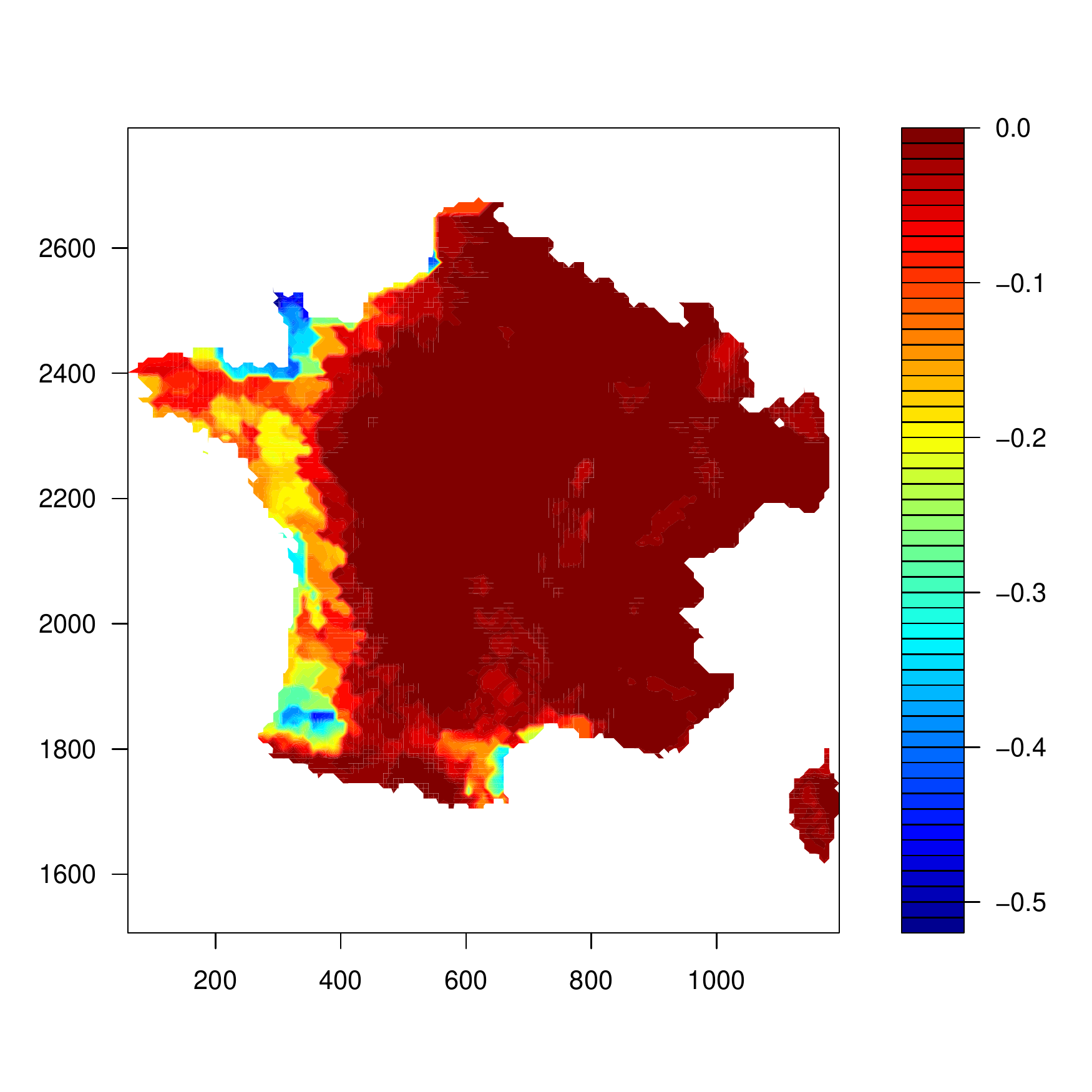} 

    \caption{The most extreme event from the study period (19/08/2012) on the original scale (left) and the standardized uniform scale (right).}
    \label{fig:most_extreme}
\end{figure}

    Figure \ref{fig:most_extreme} represents the most extreme event in the original scale (left display) and on the transformed, uniform scale (right display). On the original scale, the heat wave seems to be concentrated in the center of the country, whilst the ocean shore and also the \newmodif{mountains of the Alps (east) and the C\'evennes (centre-south) show considerably lower temperatures}. The transformed uniform scale tells a  different story. Thanks to the use of local transformations, we  see that the heat wave extends over the whole country with the exception of ocean shores. Most notably, the Alps experience very high temperatures with respect to local distributions.
    
\begin{figure}
    \centering
    \includegraphics[width=5cm]{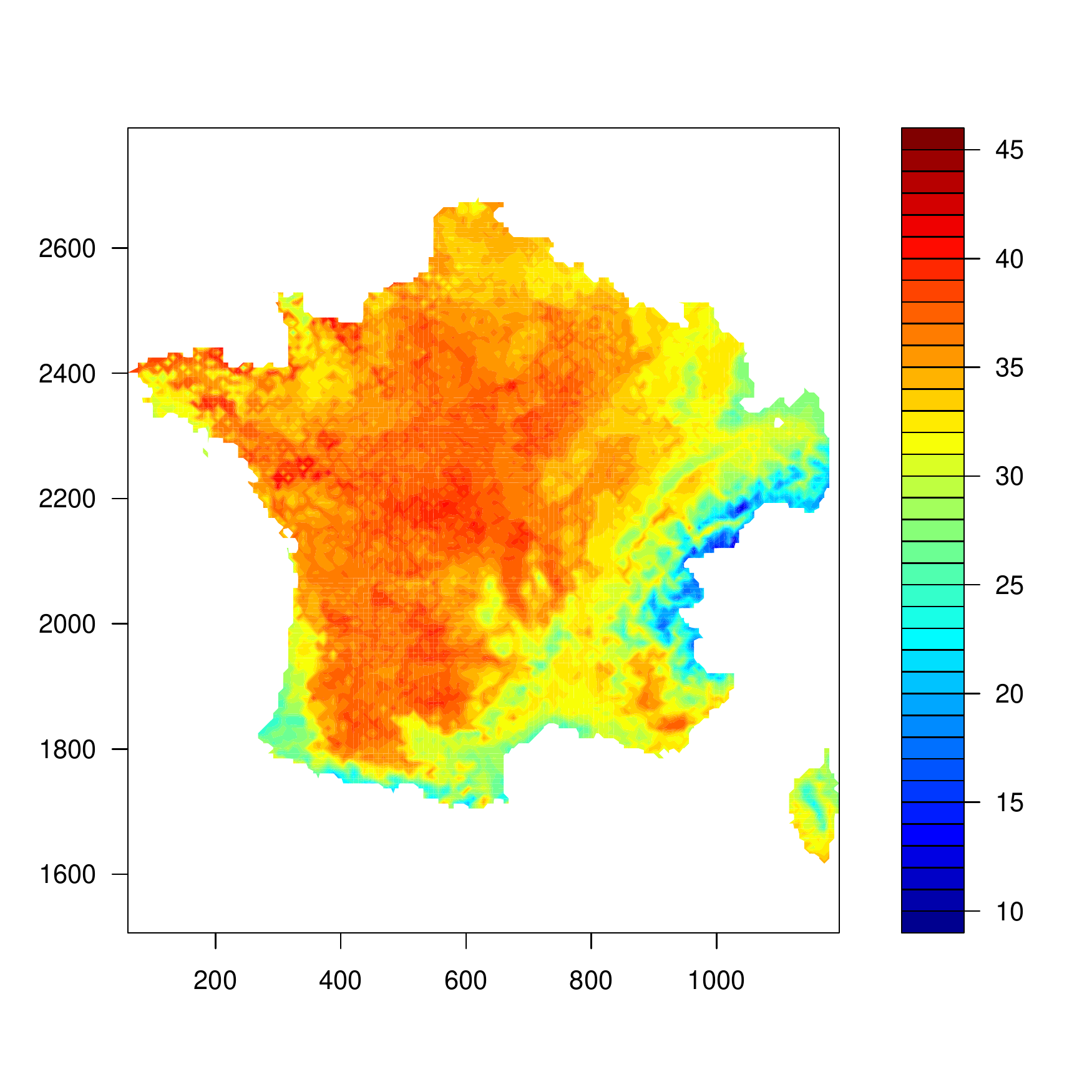}  \quad \includegraphics[width=5cm]{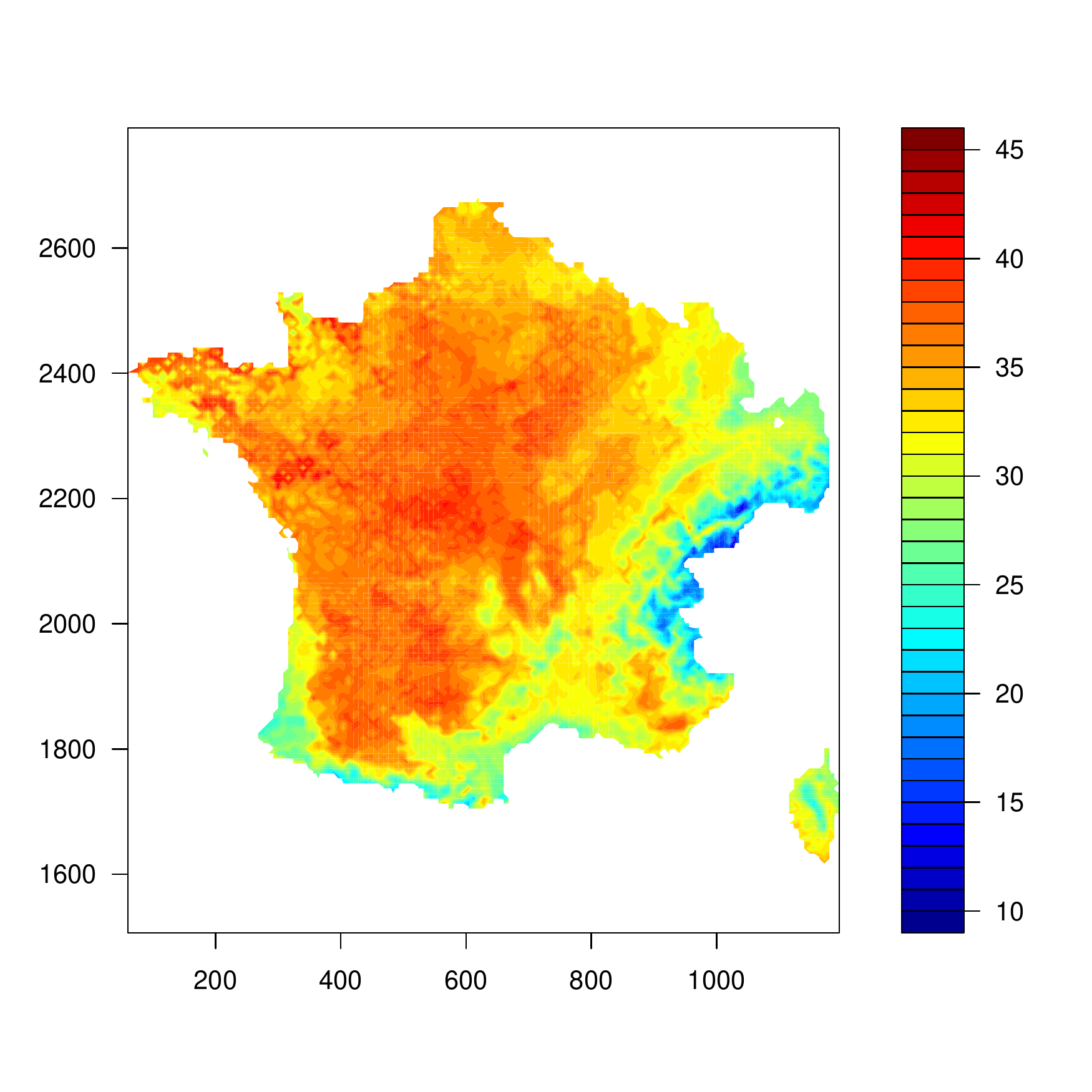} \quad
        \includegraphics[width=5cm]{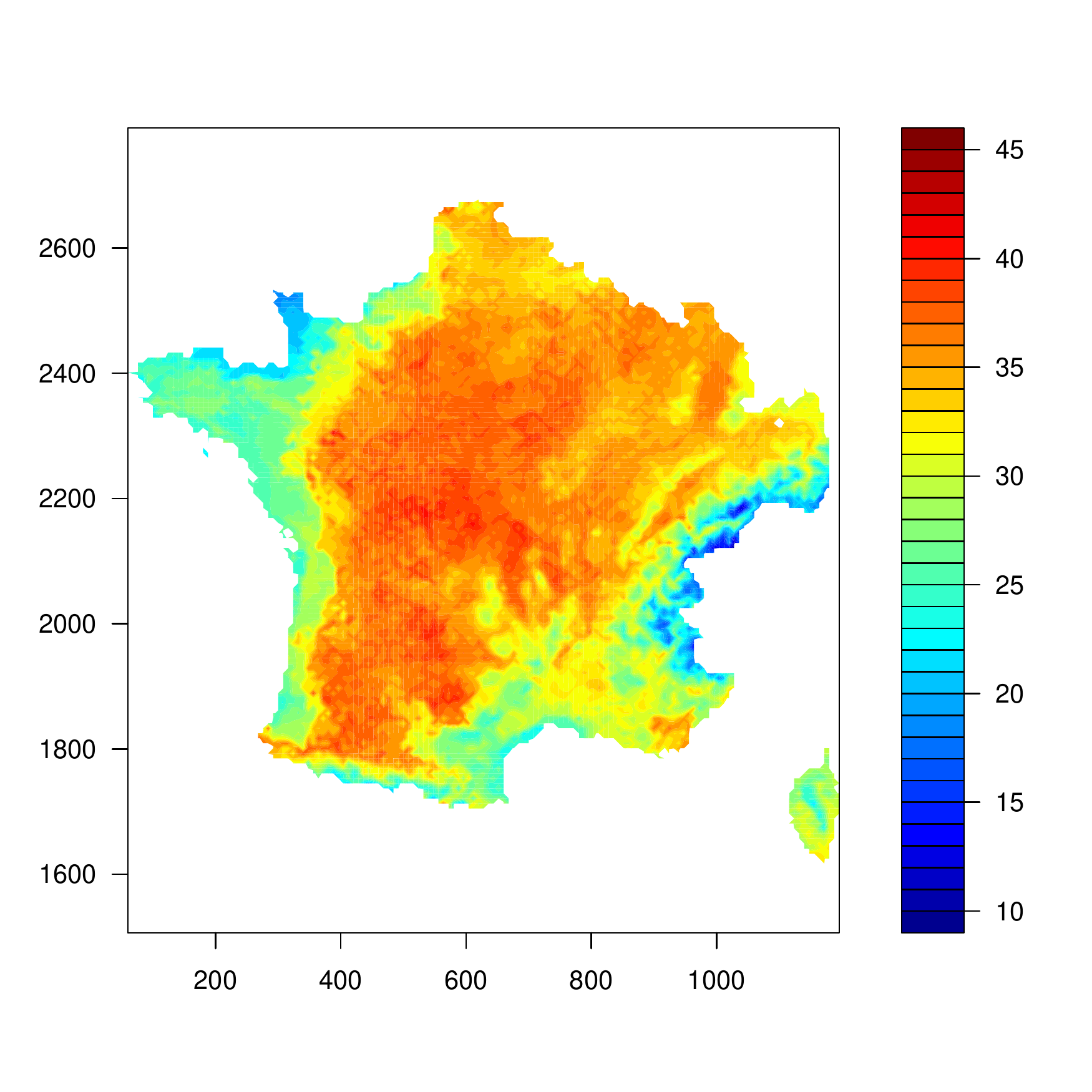}\\   \includegraphics[width=5cm]{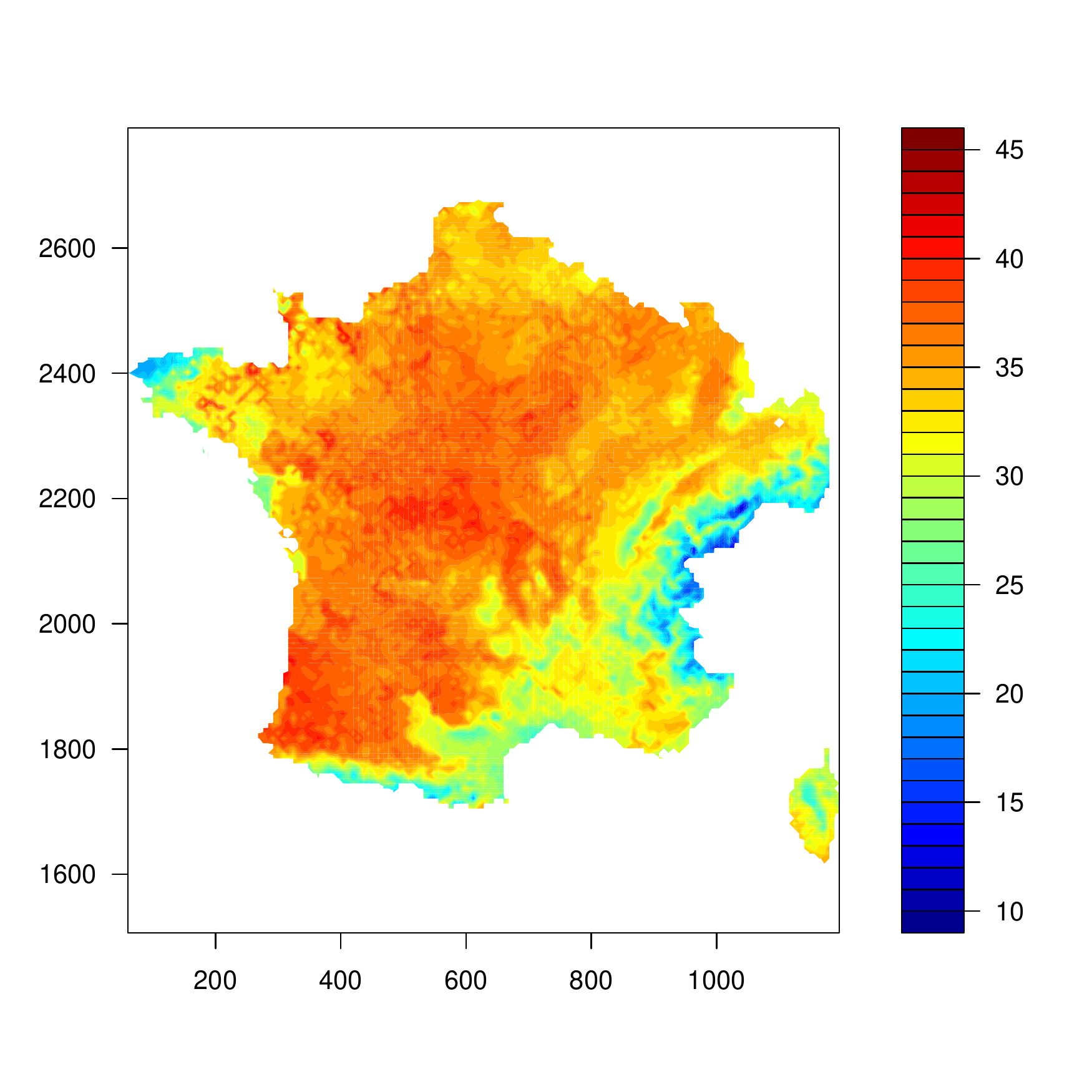} \quad
            \includegraphics[width=5cm]{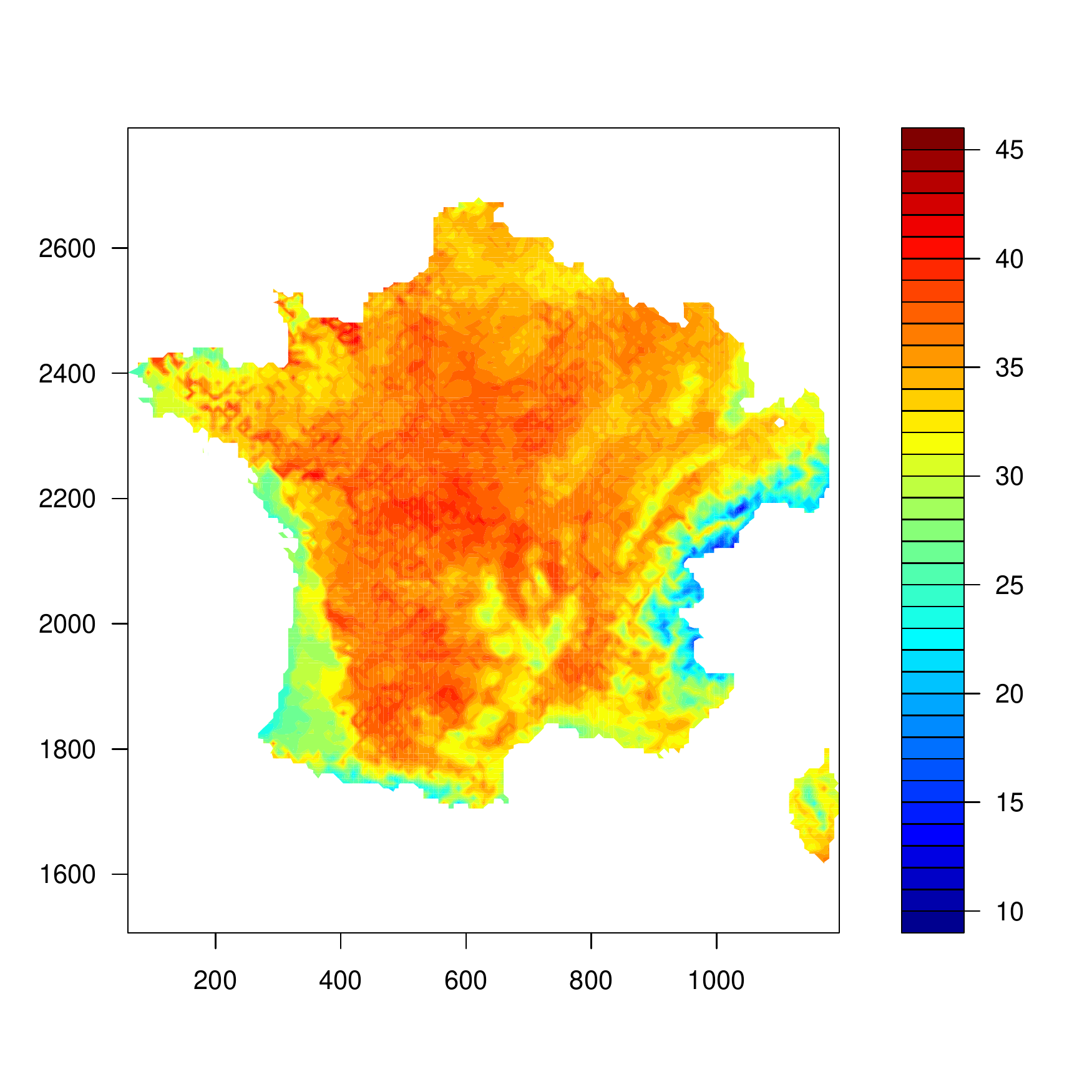}  \qquad \includegraphics[width=5cm]{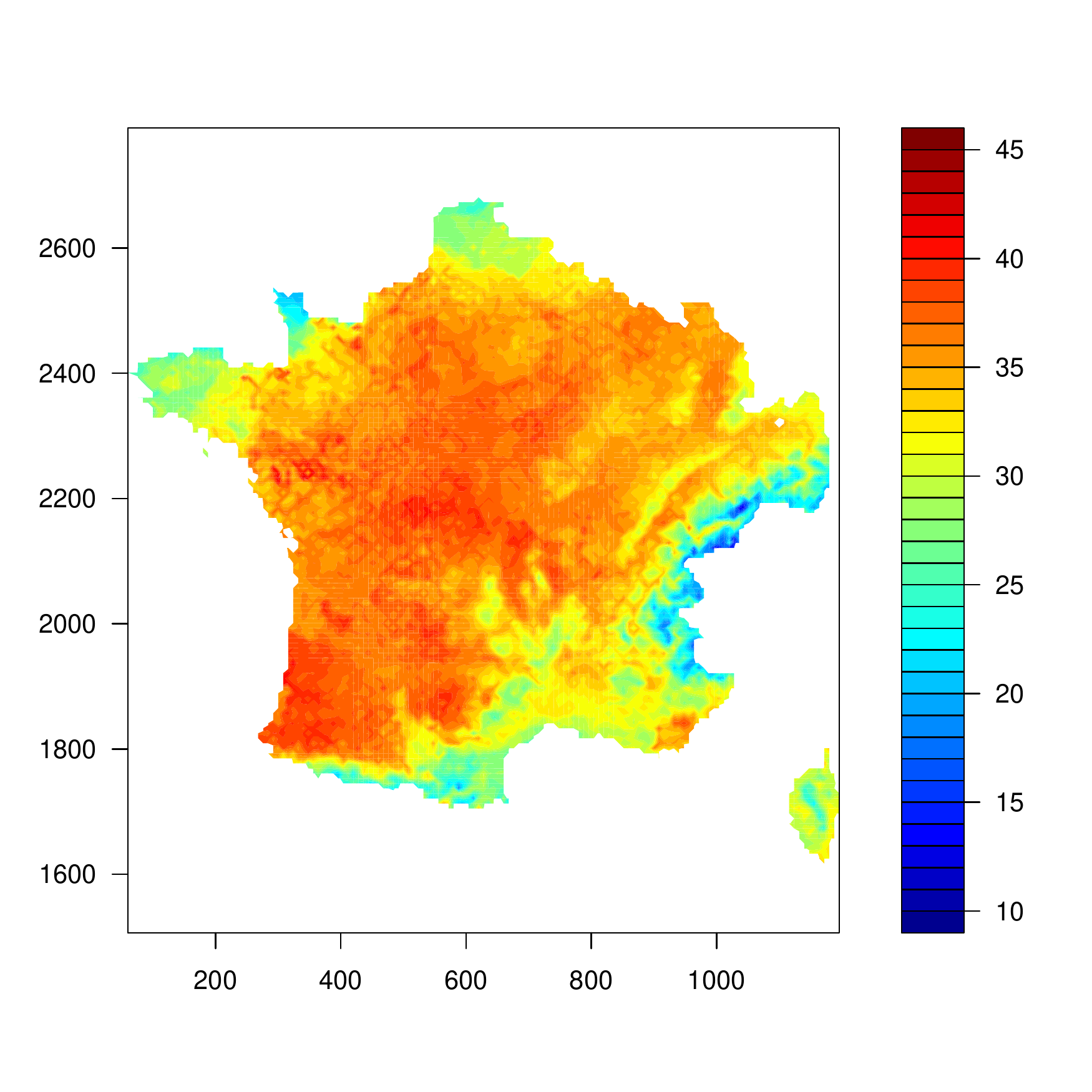} 
    \caption{Six new realizations of extreme heatwaves with $10$-year return period.}
    \label{fig:6new_episodes}
\end{figure}

\begin{figure}
    \centering
    \includegraphics[width=7.5cm]{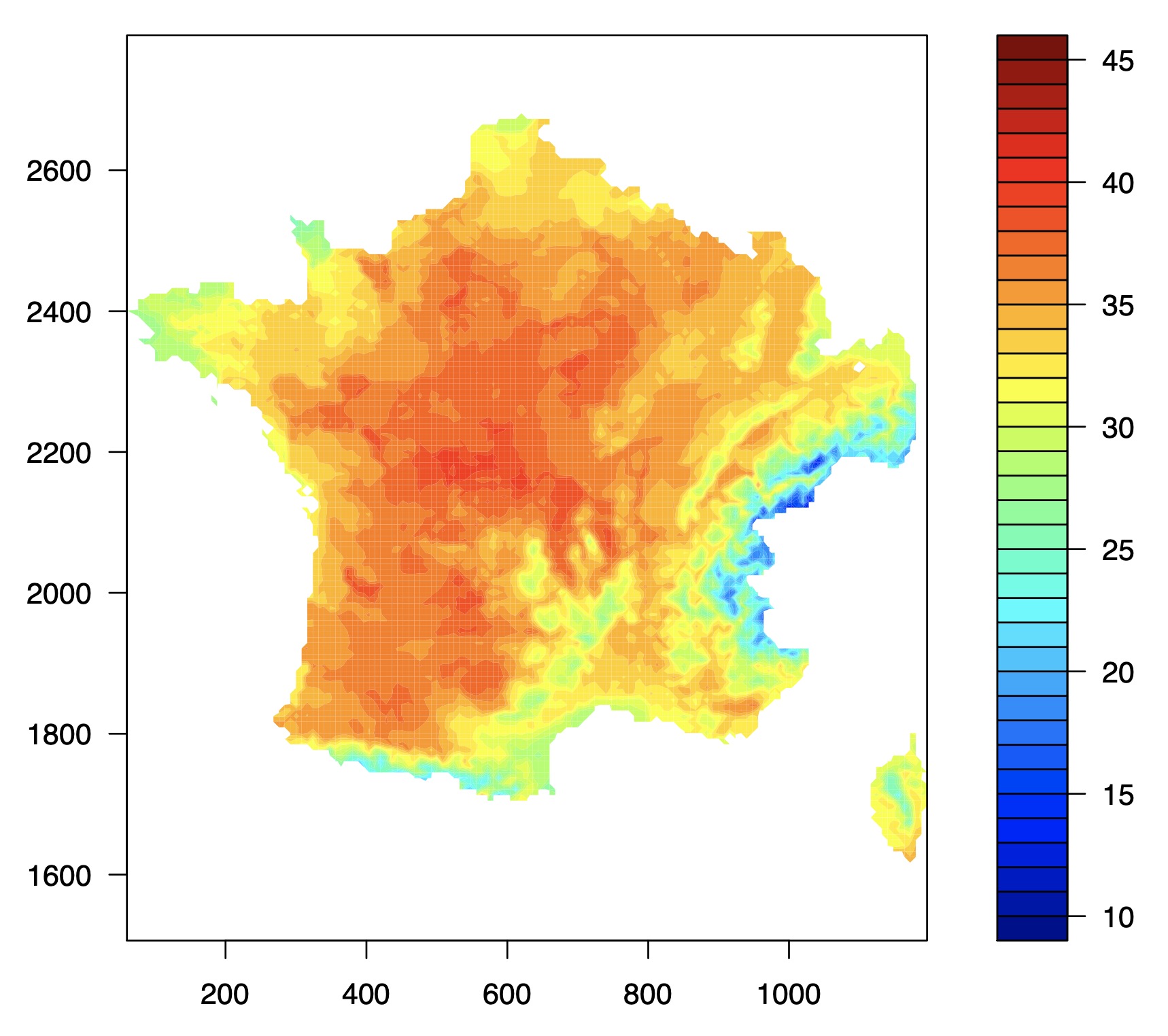} \qquad \includegraphics[width=7.5cm]{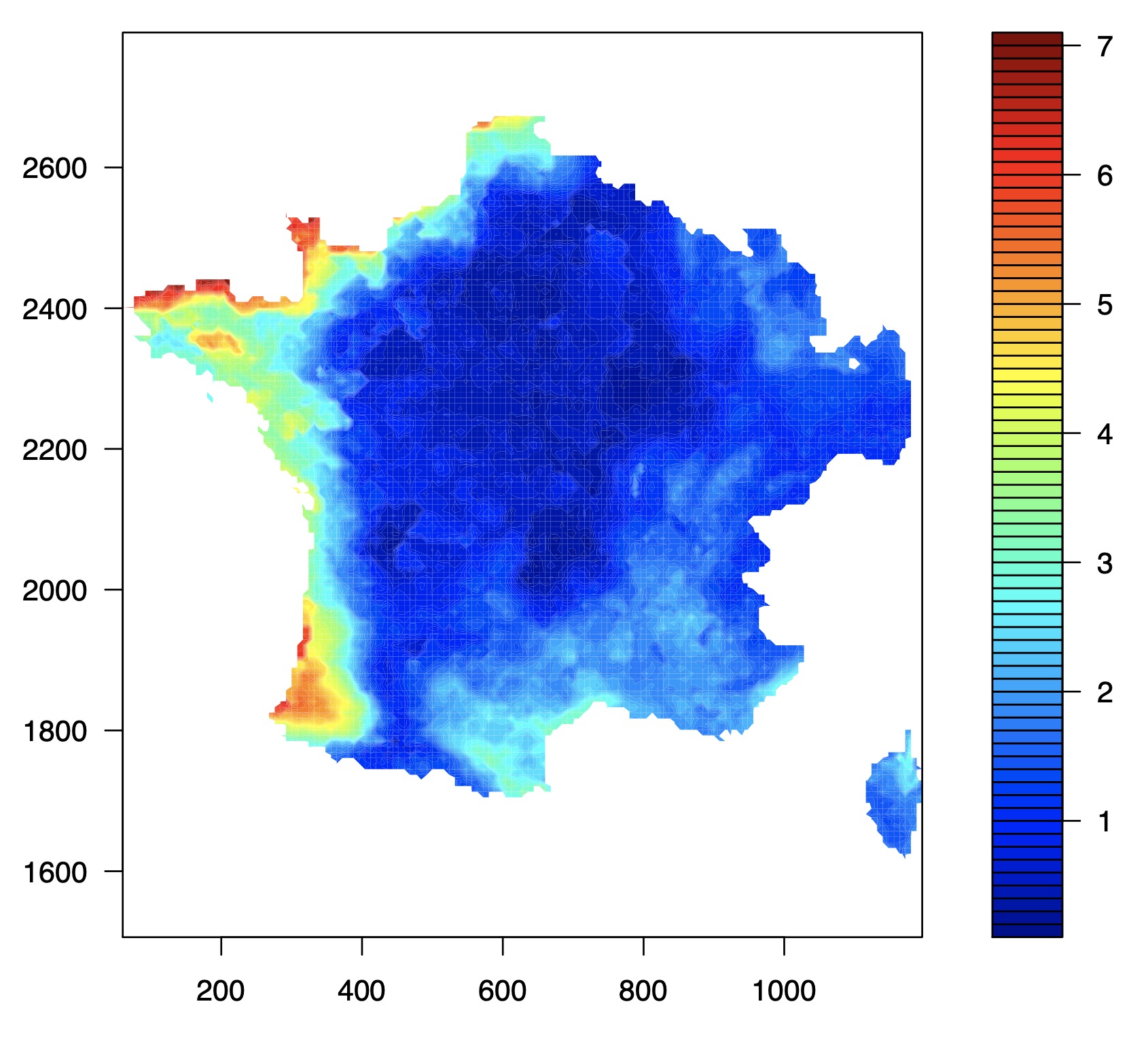}
    \caption{Mean and standard deviation fields computed over $100$ simulations of  extreme heatwaves.}
    \label{fig:mean_sd_fields}
\end{figure}

We chose to simulate new events with a return period of $10$ years. Considering that there are $122$ summer days (corresponding to the period of June 1st to Sept 30th) per year, we therefore set the lower bound $u_1 = -\hat{\theta}_r/1220$ for the simulated median summaries $r(t)$,  and we fix the upper bound to $u_2=0$. \modif{There may be non-negligible day-to-day dependence in extreme temperature events. We point out that this does not modify the daily return period associated to a return level with on average one exceedance of the level during the period, but the exceedances may arrive in a temporally clustered way, such that a relatively low proportion of periods contain several daily events while a relatively high proportion of periods contain no event.}
The estimate $\hat{\theta}_r$ of the  $r$-extremal coefficient was obtained following the method described in Section~\ref{sec:extcoef} where we used the empirical $95\%$-quantile of the median summaries $r(t)$ as threshold. In a first stage, the $6$ selected episodes are modified through uplifting  using Algorithm~\ref{algo:r-Pareto2}, which results in  a  new training dataset of uplifted fields with values in the range $[-1,0]$. In a second stage, a nonparametric resampling algorithm is run on this uplifted dataset, and is able to generate as many as desired new spatial versions of those. We here opt for the Direct Sampling algorithm \citep[DS,][]{Mariethoz.al.2010} on the transformed scale.  As discussed in the introduction, DS computes a distance between the simulated spatial patterns and observed ones. The definition of this distance is rather flexible and allows for covariates. Here, in order  to account for the nonstationarity of correlation structure across the French territory, we further introduce a $10$\% weight on the spatial coordinates, leaving a $90$\% weight for the spatial pattern. We parameterize DS such as to consider spatial patterns consisting of $20$ values. In a third stage, the resampled "uniform" scores are transformed back to the original scale using the locally estimated probability distributions. Figure \ref{fig:6new_episodes} shows $6$ new realizations of  extreme events. One can see that the general pattern is reproduced thanks to the local modelling of the marginal distributions, and that there appears an interesting variety of local patterns among these realizations. Our approach thus takes into account quite naturally the highly nonstationary marginal distributions across the French territory. Figure \ref{fig:mean_sd_fields} shows the mean and standard deviation fields computed over $100$ realizations of new extreme heatwaves generated through our approach. The standard deviation map shows clearly that the ocean shore has a much higher variability during heatwaves than inland zones or the Mediterranean shore, due to the fact that these regions are less likely to be hit by heatwaves striking the large zones of the inland territory. 

\subsection{Validation}
\begin{figure}
    \centering
    \includegraphics[width=4cm]{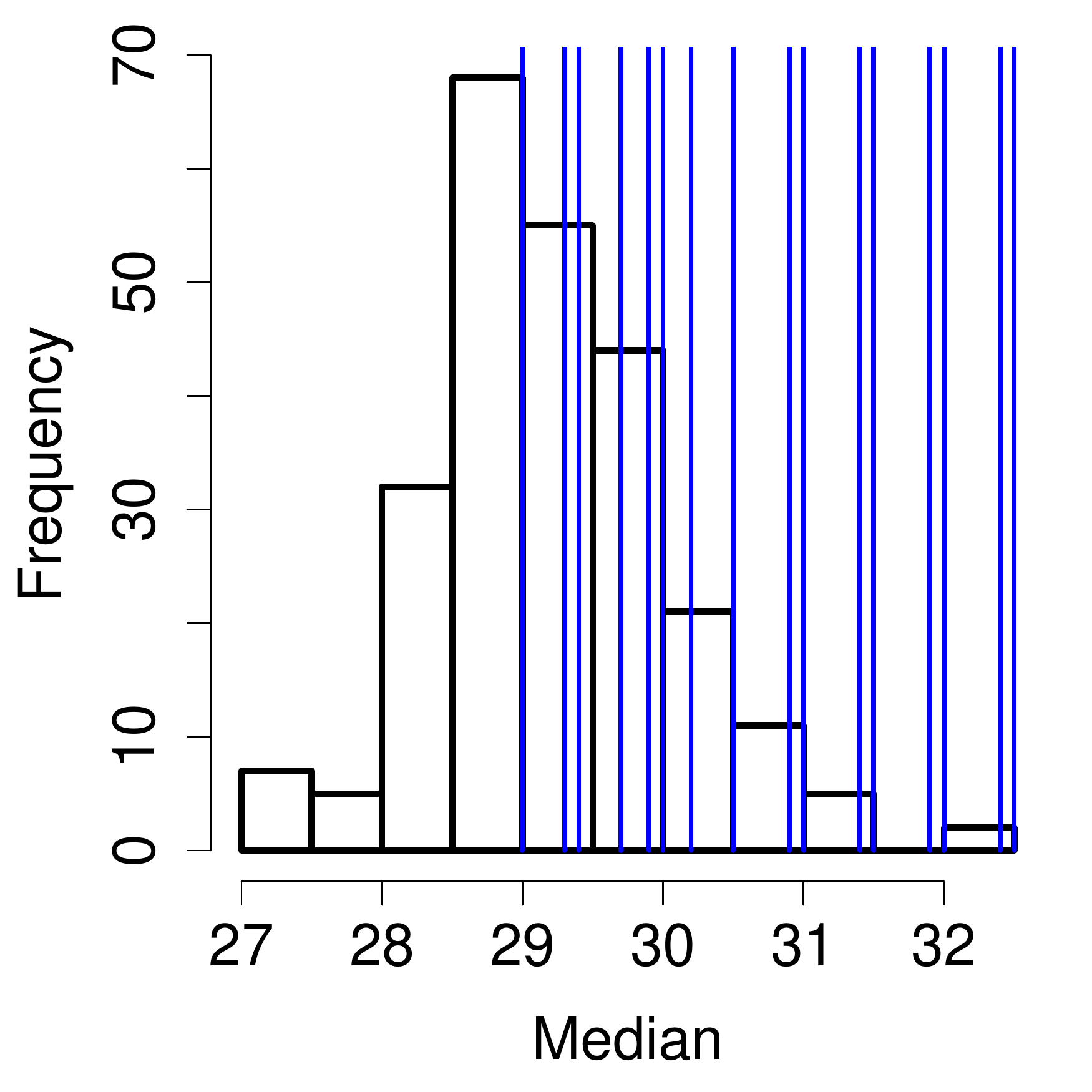}   \qquad \includegraphics[width=4cm]{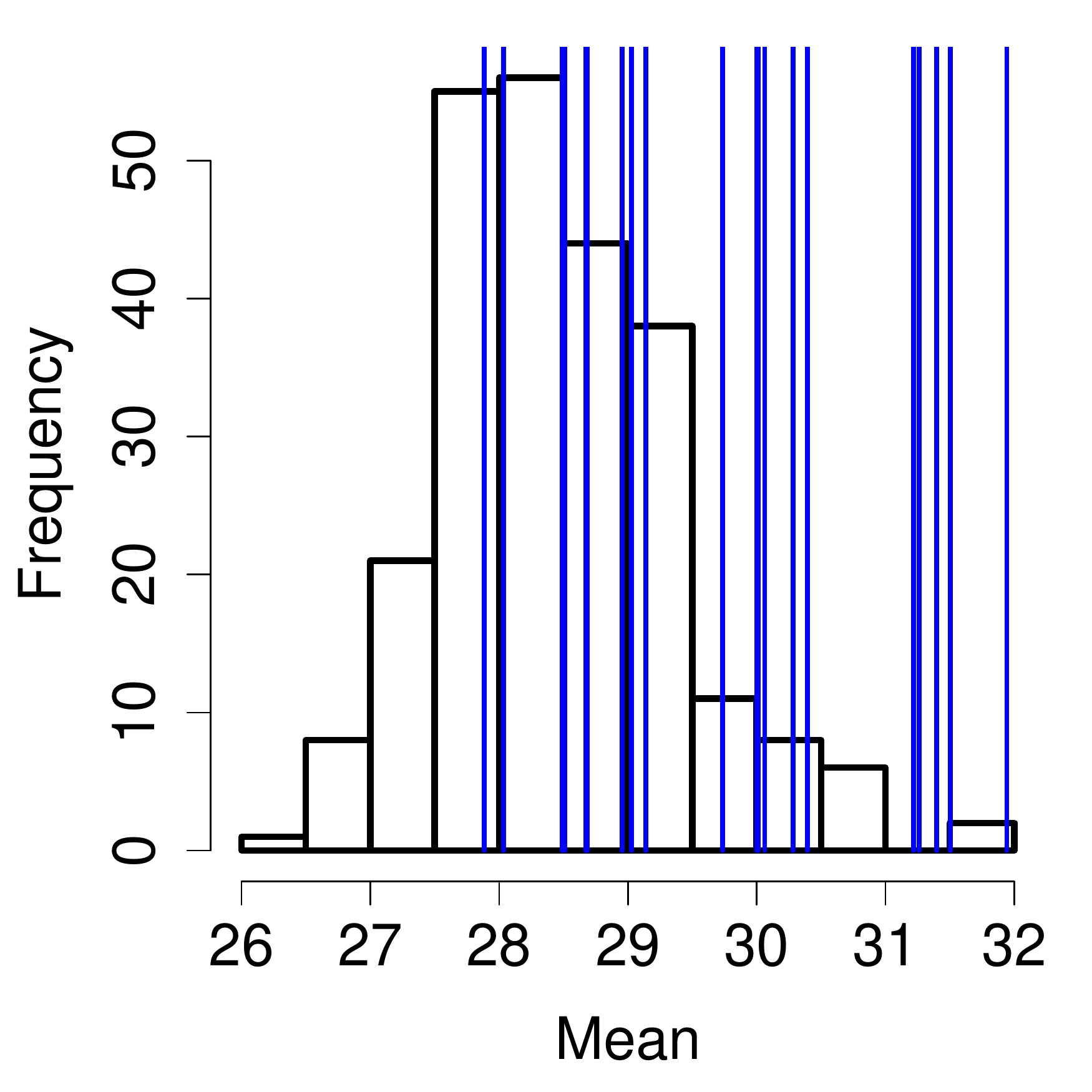} 
    \qquad
    \includegraphics[width=4cm]{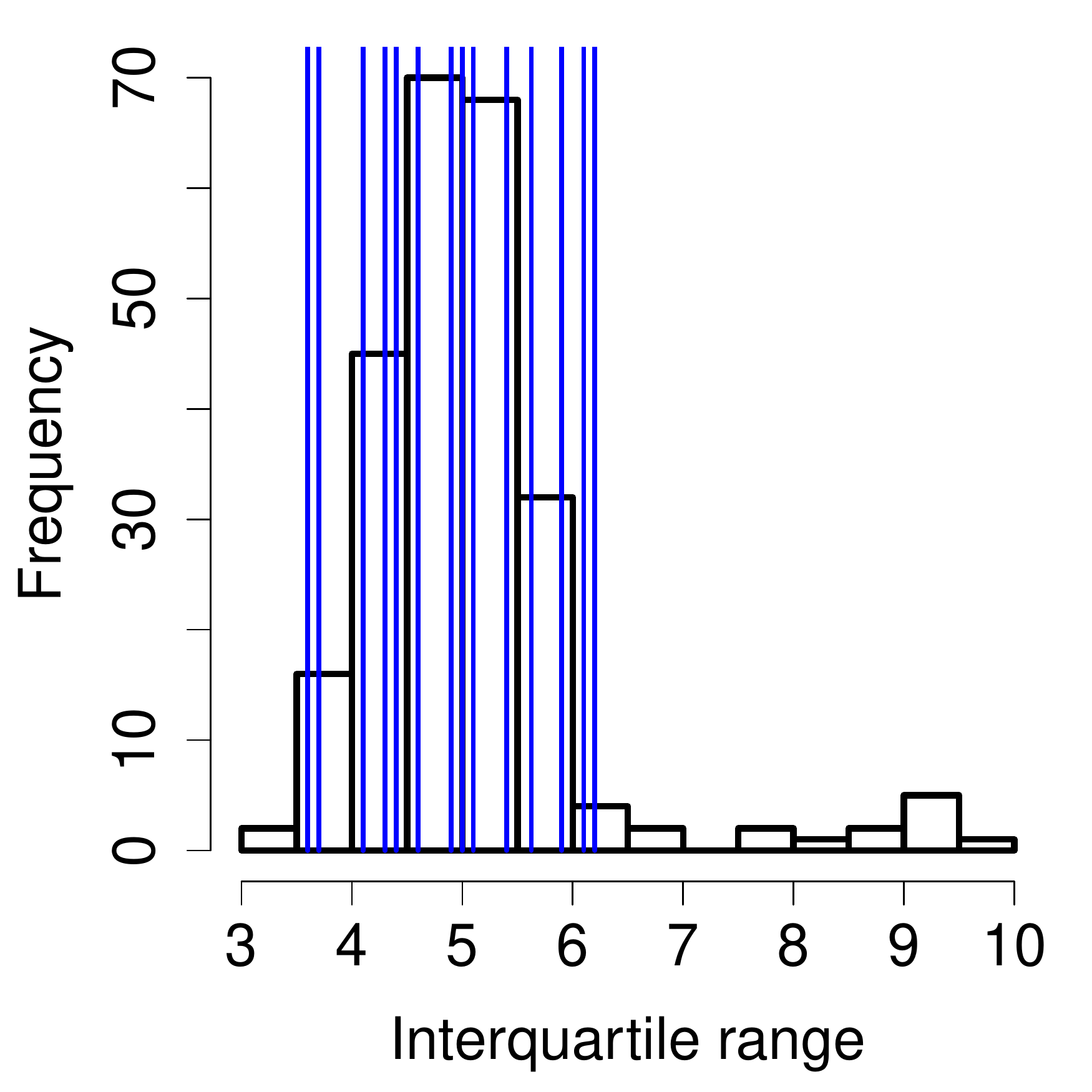}  \\
        \includegraphics[width=4cm]{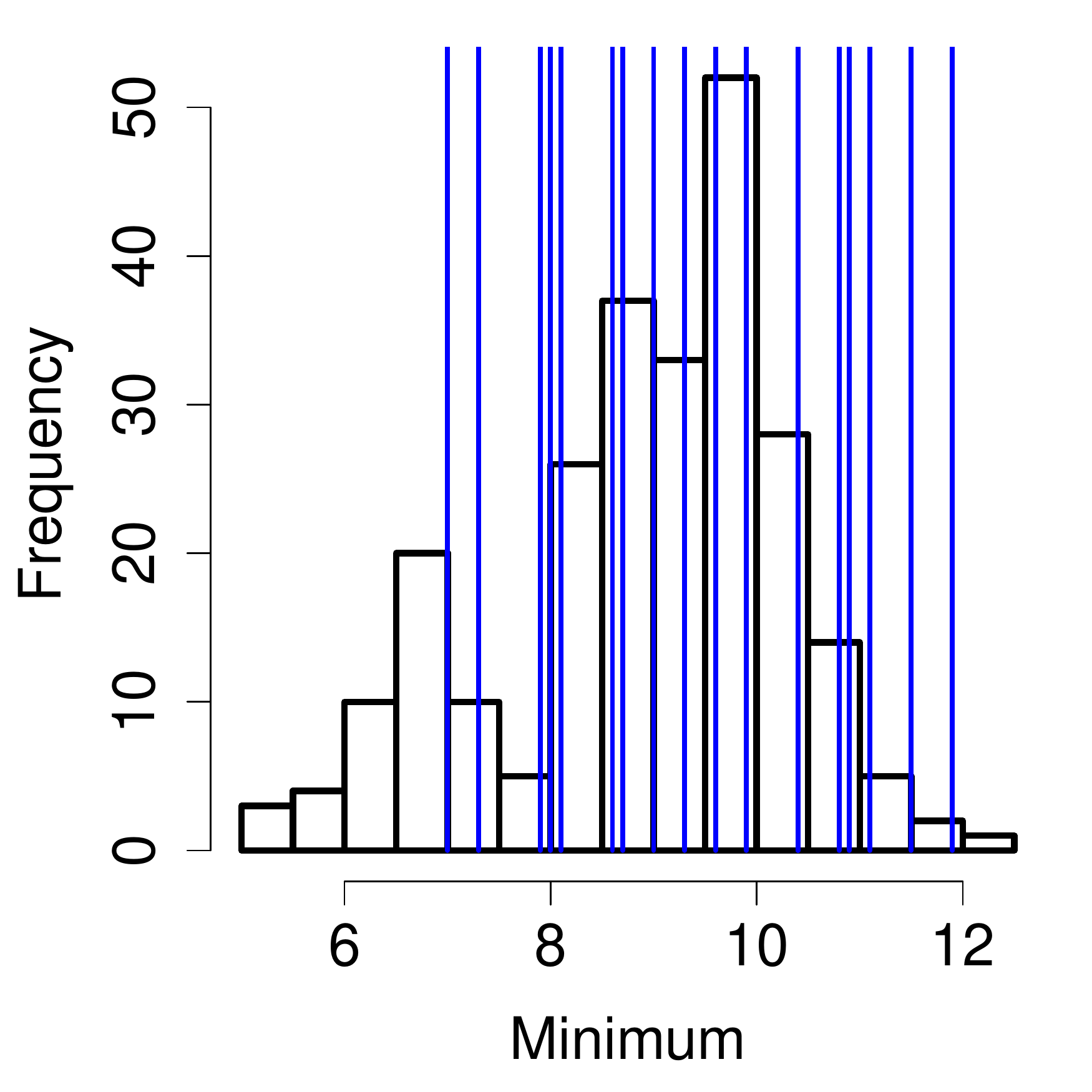} \qquad \includegraphics[width=4cm]{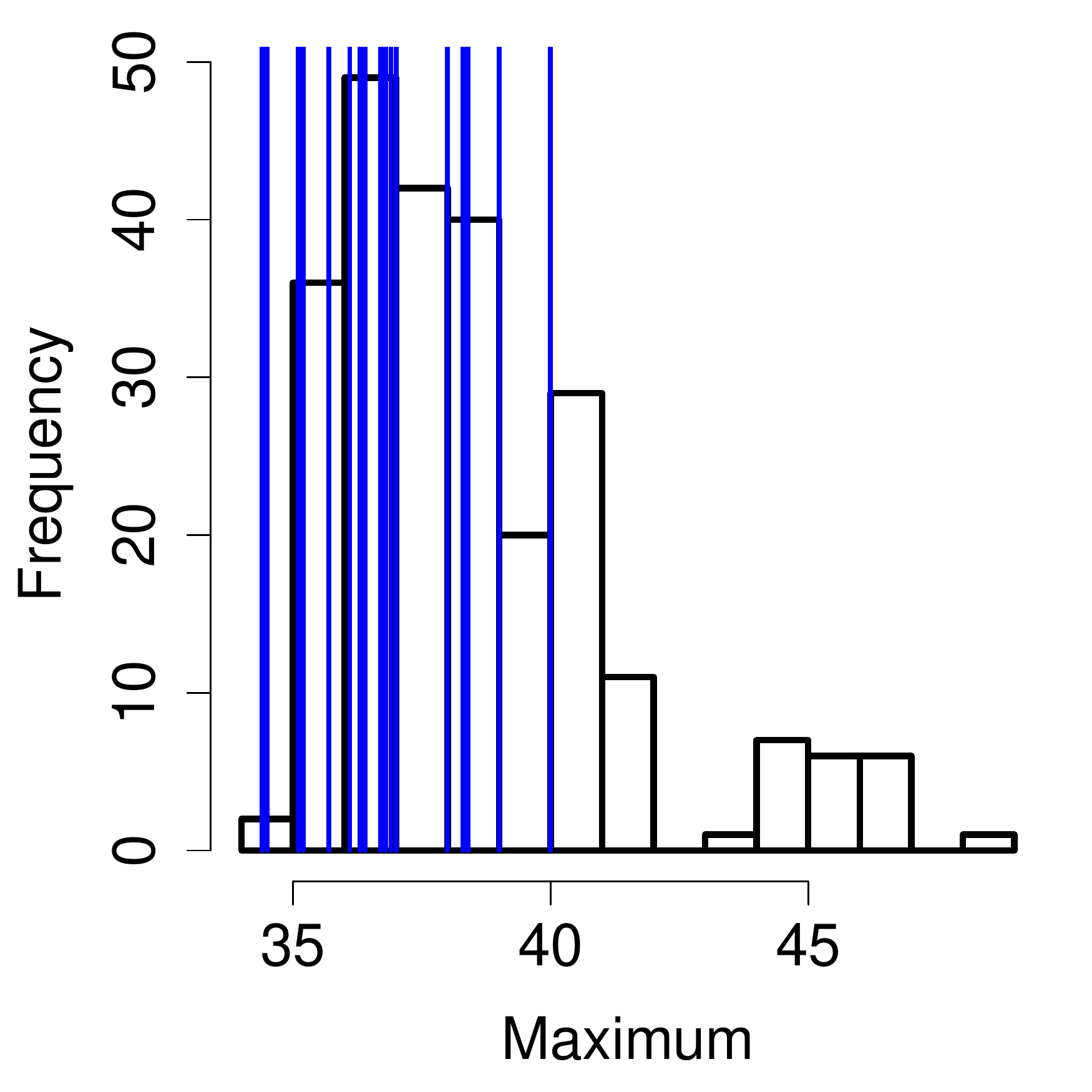}  \qquad
            \includegraphics[width=4cm]{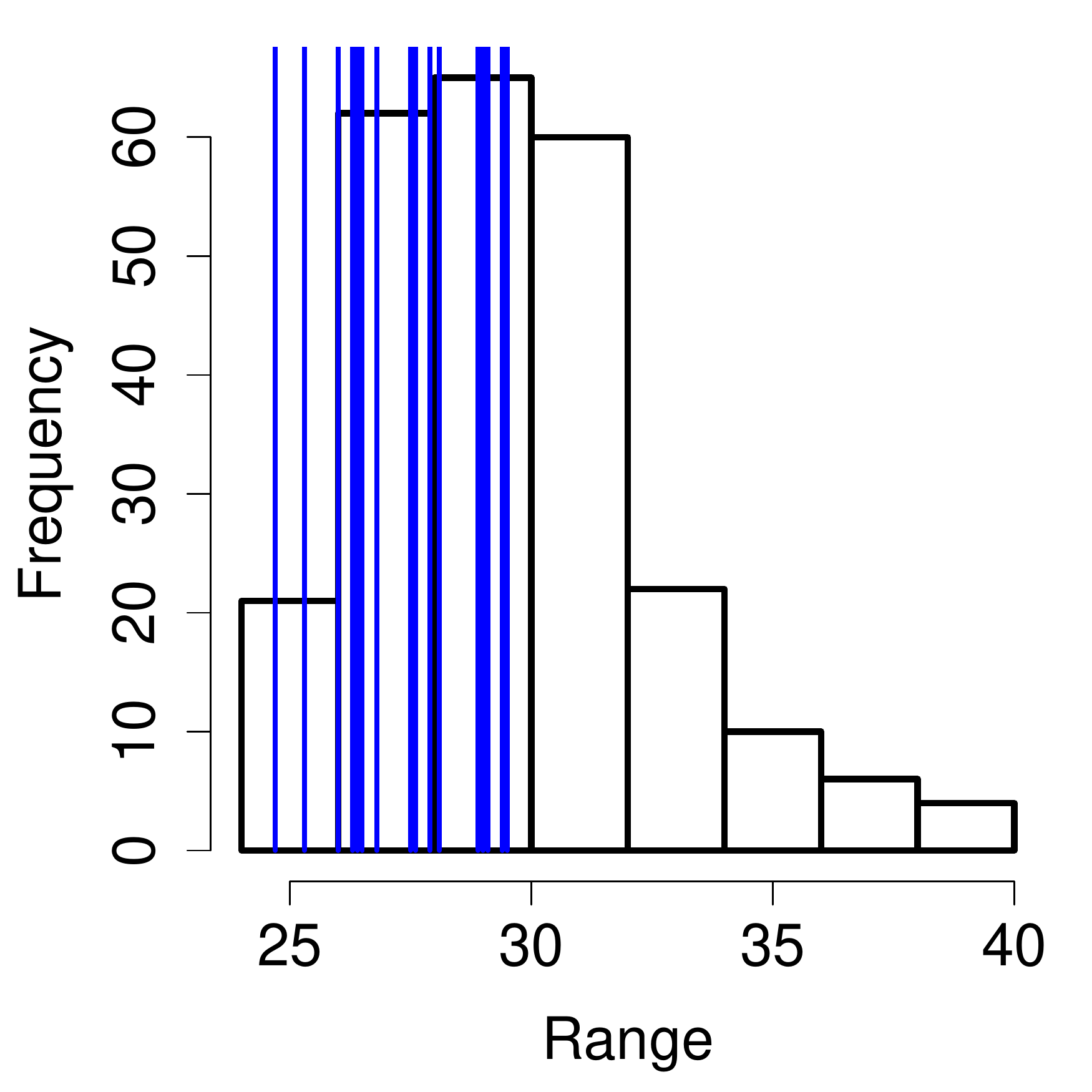}  
    \caption{Validation of summary statistics of simulated lifted extreme episodes. Top row (left to right): median, mean, interquartile range; bottom row(left to right): minimum, maximum, difference maximum $-$ minimum. Histograms show summary values for $250$ simulations using Direct Sampling on enriched data. Blue lines show   values of the $20$ most extreme observed events, held out for validation.}
    \label{fig:valid}
\end{figure}

Can simulated data, obtained by data enrichment and nonparametric resampling, realistically reproduce important data features? Validation of extremal data features and of extrapolation performance is notoriously intricate owing to \modif{a too small} data base. Here, we study this question by considering the $30$ most extreme days retained after temporal declustering, of which the $10$ least extreme events are used for generating $250$ simulations, and the $20$ most extreme events are kept as validation data for comparing their summary statistics against the simulations. In the data enrichment step, we have lifted the $10$ training days to return levels of the median $r(t)$ of standardized data larger than $-0.08$, corresponding to an approximate lower bound for the $20$ validation days, and further corresponding to a return period of approximately $17$ days within the four-month summer period.  Figure~\ref{fig:valid} shows a histogram of the distribution of summary statistics (mean, median, interquartile range, minimum, maximum, range $=$ maximum $-$ minimum) calculated for the $250$ simulations, and values of the $20$ most extreme observed events.       
In all cases, the $20$ validation values are covered by the histogram bars of the simulations. The simulations tend to show  stronger variability  (\emph{e.g.}, for minima, maxima and ranges), and slightly lower values for central tendencies (median, mean), but we have to keep in mind that the number of $20$ validation days is small in comparison to the number of simulations. When taking into account the  heatwave during the end of June 2019, some of the seeming extrapolation biases are mitigated. To date, the SAFRAN product is not yet available for this heatwave, but the maximum temperature of $45.9$ degrees observed over the M\'et\'eo France station network on June 27th of 2019 suggests that the maximum SAFRAN value for this event can be expected to lie in the far right tail of the maxima of resampling simulations reported in Figure~\ref{fig:valid}, where none of the 2010-2016 events occurred. 
Overall, we consider these validation plots as satisfactory, especially since we are in a setting of complete extrapolation where none of the training data fall within the range of the target return levels of the summary functional $r$. 

\section{Conclusion}\label{sec:conclusion}
We have extended nonparametric resampling techniques to allow for realistic extrapolation of extreme values beyond the observed range of data. An illustration of this approach was given through the generation of artificial, very extreme heat wave scenarios for mainland France. Especially with high-dimensional gridded data, it may be preferable to avoid using generative statistical models whose practical usefulness may be limited by strong assumptions on the structure of data, onerous model fitting procedures and the resulting compromises with respect to models' complexity and realism. When data are only weakly dependent in the extremes \modif{(\emph{i.e.}, with small spatial range of extremal dependence, in particular in the case of asymptotic independence),} our naive uplifting technique may work well, although the effective sample size in training data (\emph{i.e.}, the number of effectively independent variables) may increase strongly between training and resampling data and create some artefacts at extreme quantiles \newmodif{in particular if the spatial resolution is fine with respect to spatial range}.  
By working with replicated training data structures (\emph{e.g.}, a time series of spatial snapshots) and using assumptions from extreme value theory of stochastic processes, we have bypassed the shortcomings of naive resampling discussed in Section~\ref{sec:limitations}. In some cases, the data transformation steps (marginal transformations, scale-profile decomposition, post-processing of small values) may be based on rather strong assumptions or involve considerable uncertainty, for instance if we have only relatively few replicates, if the spatial range of dependence of extremes is very short, or if data seem to be asymptotically independent such that we have limit $0$ in the conditional exceedance probability \eqref{eq:ad}.  In such cases, the lifting procedure should be carefully validated.
\newmodif{In future work, we plan to study in more detail the use of alternative post-processing techniques, and to perform a comparison of the application of nonparametric resampling at different steps (before or after lifting) in the algorithm.}
Extensions to include temporal dependence in the simulated data are possible, for example by replacing single snapshots with a small series of consecutive snapshots, such that short-range temporal dependence is preserved. 

For clarity of presentation, we have concentrated on accurate data extrapolation in the upper tail, but our results also apply to lower tails modulo switching tails, for instance by considering $-X_{\cal I}$ instead of $X_{\cal I}$, and only minor adjustments would be necessary to adapt our algorithms to consider joint extrapolation in both upper and lower tails. 

The most challenging case remains the situation of training data without independent replication, where asymptotic extreme value theory does not apply. In principle, the lifting procedure using marginal standardization and the scale-profile decomposition suggested by Pareto processes remains applicable, but some rather strong assumptions must be made on the validity of asymptotic theory, and it may be difficult to determine return periods for resampled data with a  lifting step. 

A more general approach to remove some of the limitations of the methods in this paper could come from the conditional extremes framework \citep{Heffernan.Tawn.2004}, which provides theory and models based on how values evolve around and away from a fixed reference point, conditional to observing a very high value at this point. To apply this theory, one could initiate the simulation dataset with a new high value for a fixed reference point, either by fixing the value deterministically or by sampling it using an appropriate target distribution, and then one would fill the rest of the simulation  in accordance with theoretical constraints.  However, it remains to be explored how the semi-parametric asymptotic representations of conditional extremes could be estimated under moderate assumptions and then be transformed into appropriate and simple algorithmic steps during the resampling procedure. 

Our procedure heavily relies on a representation with two stochastically independent components: a univariate magnitude variable, from which it is easy to sample, and a profile process, for which classical resampling techniques can be applied. Similar decompositions arise with many other, nonasymptotic stochastic processes, and they offer a much larger pool of models for flexibly representing extremal dependence \citep[e.g.,][]{Huser.al.2017,Engelke.al.2018}. We envisage follow-up work to this paper to explore how our algorithms could be adapted to such representations. However, a major difficulty with such alternative constructions \modif{would arise if} the summary functional $r$ is parametrized through parameters of the dependence structure of the process, which then impedes direct application of nonparametric procedures. \modif{For an example,} consider random vectors following an elliptically contoured distribution, such as the multivariate Gaussian. Then there exists a scale-profile decomposition with the summary functional defined as the Mahalonobis norm of the inverse of the covariance matrix (or of the dispersion matrix if covariance is not defined), but the need to estimate the covariance structure compromises nonparametric resampling and requires further modifications.
\bibliographystyle{apalike}
\bibliography{biblio}

\appendix

\newmodif{
\section{Illustration of naive resampling}\label{sec:examples-naive}
}

\begin{figure}
    \centering
    \includegraphics[width=15cm]{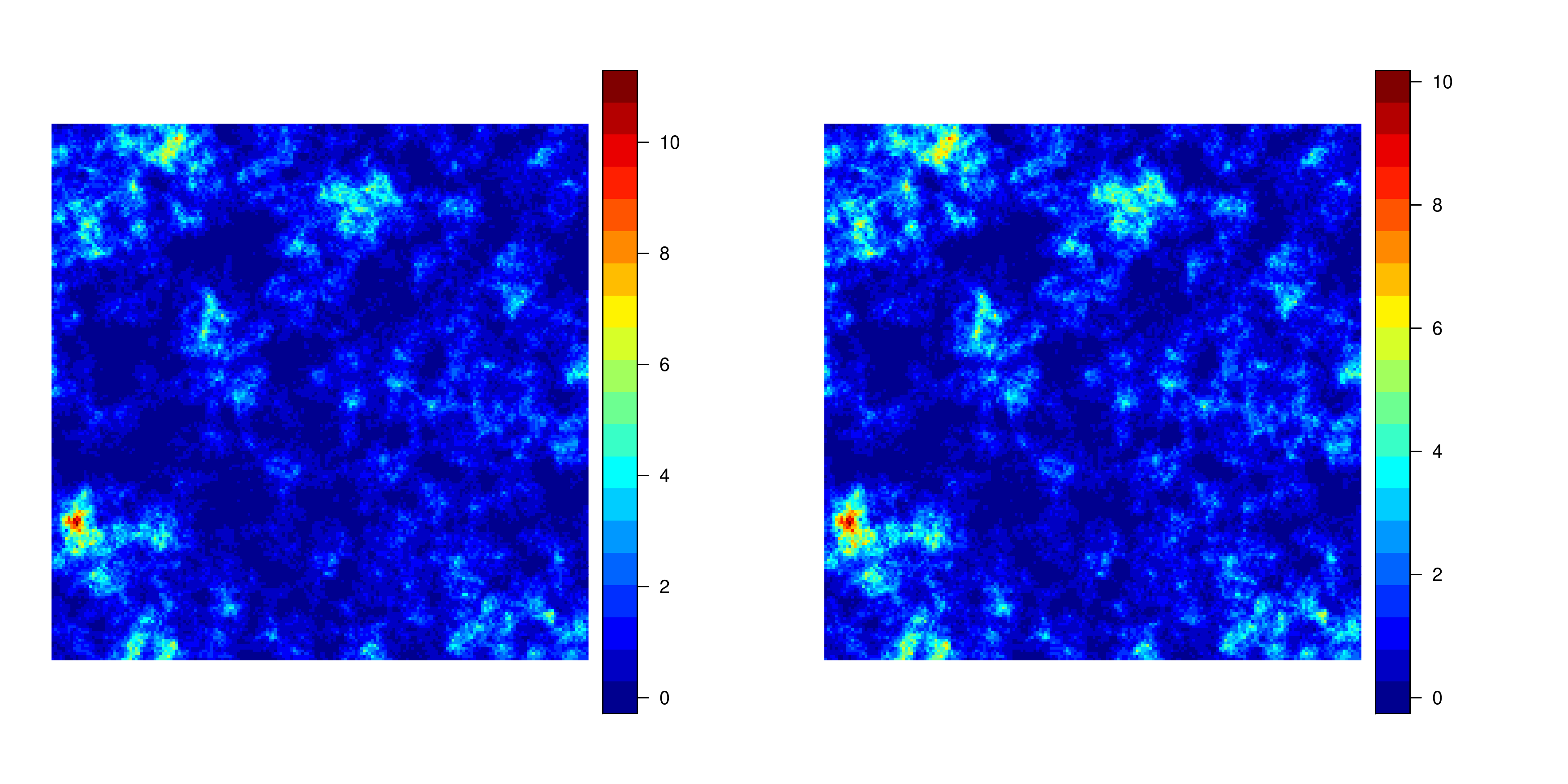}
    \caption{Left: Original transformed Gaussian random field with exponential margins. Right: naive resampling with identical spatial rank pattern.}
    \label{fig:naive-full}
\end{figure}

\newmodif{To illustrate the large spacings around the maximum, we first consider a setting with non-replicated data.} We simulated a transformed Gaussian random field with exponential marginals on a $120 \times 120$ grid covering the unit square. The covariance function is exponential with range equal to $1/8=0.125$. \modif{Here, the dependence range is relatively small with respect to the size of the domain, and we therefore are in the setting of a single realization with weak dependence at moderate to large distances. Due to the process being mixing, we expect the empirical distribution of values to be close to the theoretical marginal distribution.}
We apply the naive resampling approach following Algorithm~\ref{algo:naive}. In order to facilitate visual comparison, the spatial pattern of ranks, generated in Step~1 of the algorithm, remains identical to the one in the training data. The naive approach therefore consists in simply drawing independent standard exponential variables, one for each grid cell, and then re-attributing them to grid cells by matching ranks of training and simulation samples.
From a look at original and resampled data on the full grid, shown in  Figure~\ref{fig:naive-full}, no strong differences become obvious, except for slightly larger spatial extent of subregions with very high values. However, a zoom into the region around the global maximum tells a different story. The left panel of Figure \ref{sec:naive} represents original and resampled data for the $17 \times 17$ subgrid centered at the maximum of the original field. The ranks are preserved, but is also obvious that the simulated quantiles are much (unrealistically) higher: the maximum of the new image is above 10, while it was close to 5 in the original image. Furthermore, the surface around the maximum is much less continuous. \newmodif{For the range of moderately extreme values where we still have many observations in the original data, such that the i.i.d. resampling idea may produce realistic simulations.}
        
As already mentioned above, these conclusions remain valid for any marginal distribution, theoretical or empirical.

\begin{figure}
    \centering
    \includegraphics[width=12cm]{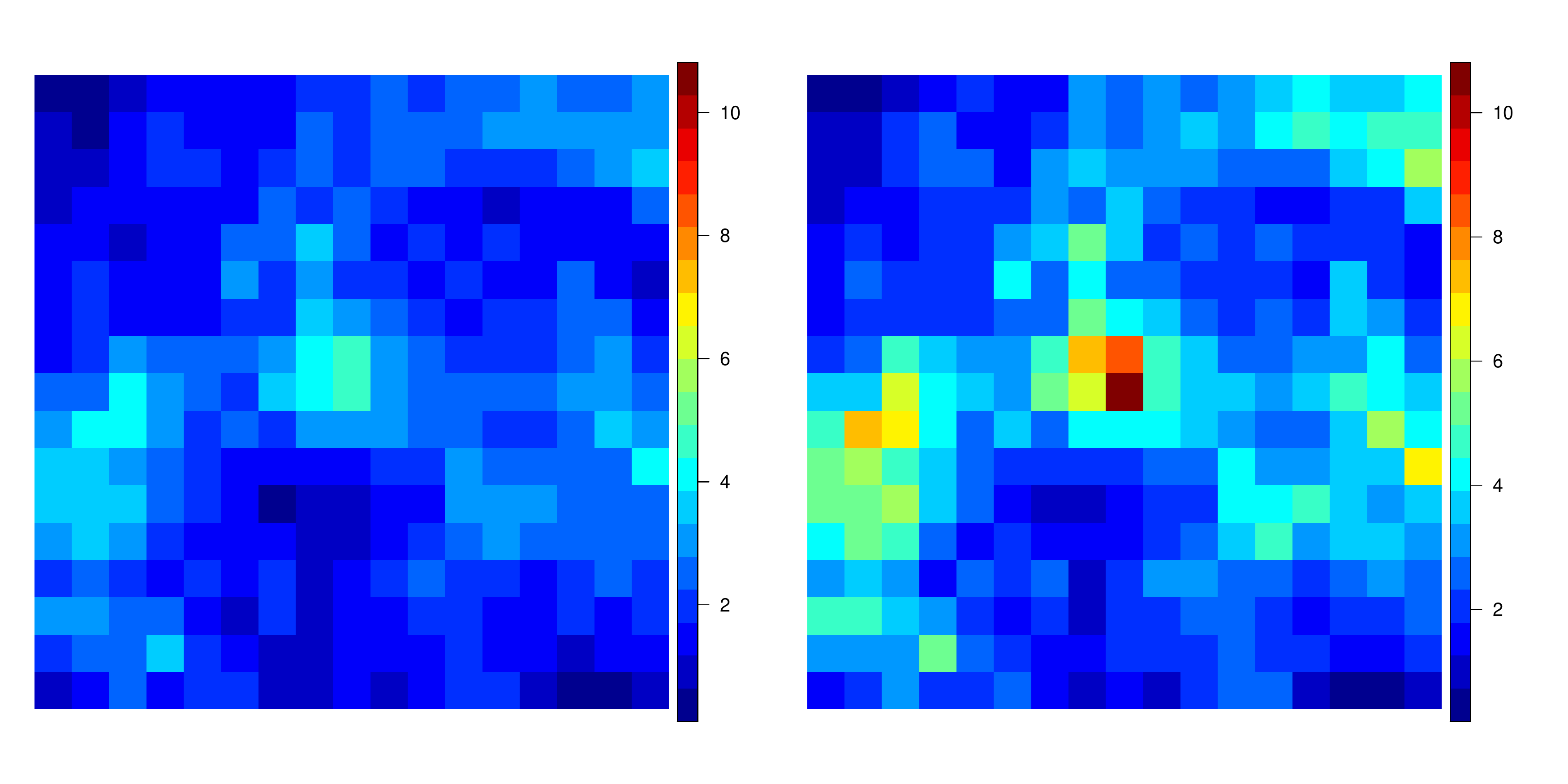}
    \caption{Left: values close to the global maximum in the original transformed Gaussian random field with exponential margins, with the global maximum  in the center of this subgrid. Right: naive resampling with identical spatial rank pattern.}
    \label{fig:naive}
\end{figure}

\begin{figure}
    \centering
    \includegraphics[width=5cm]{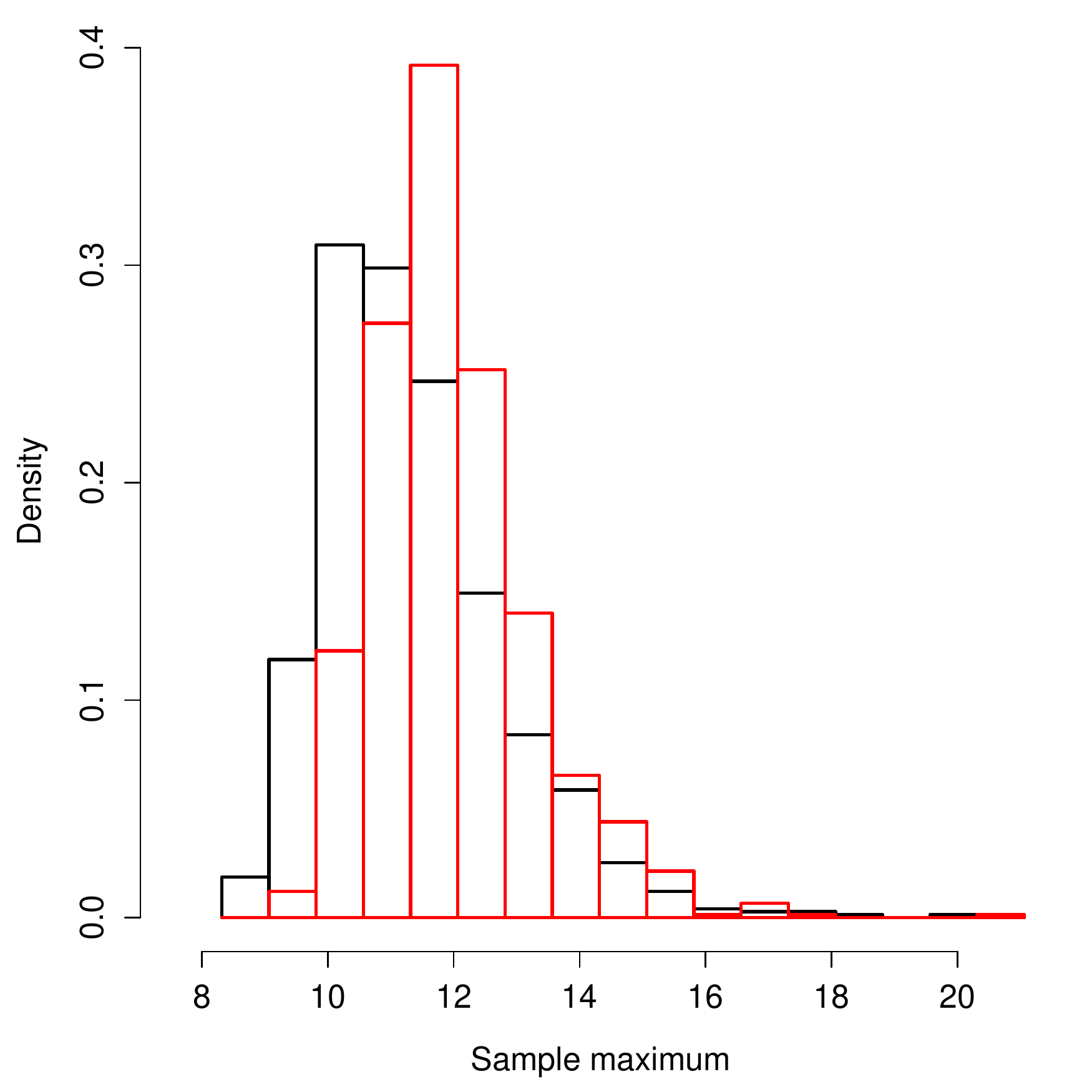} \qquad  \includegraphics[width=5cm]{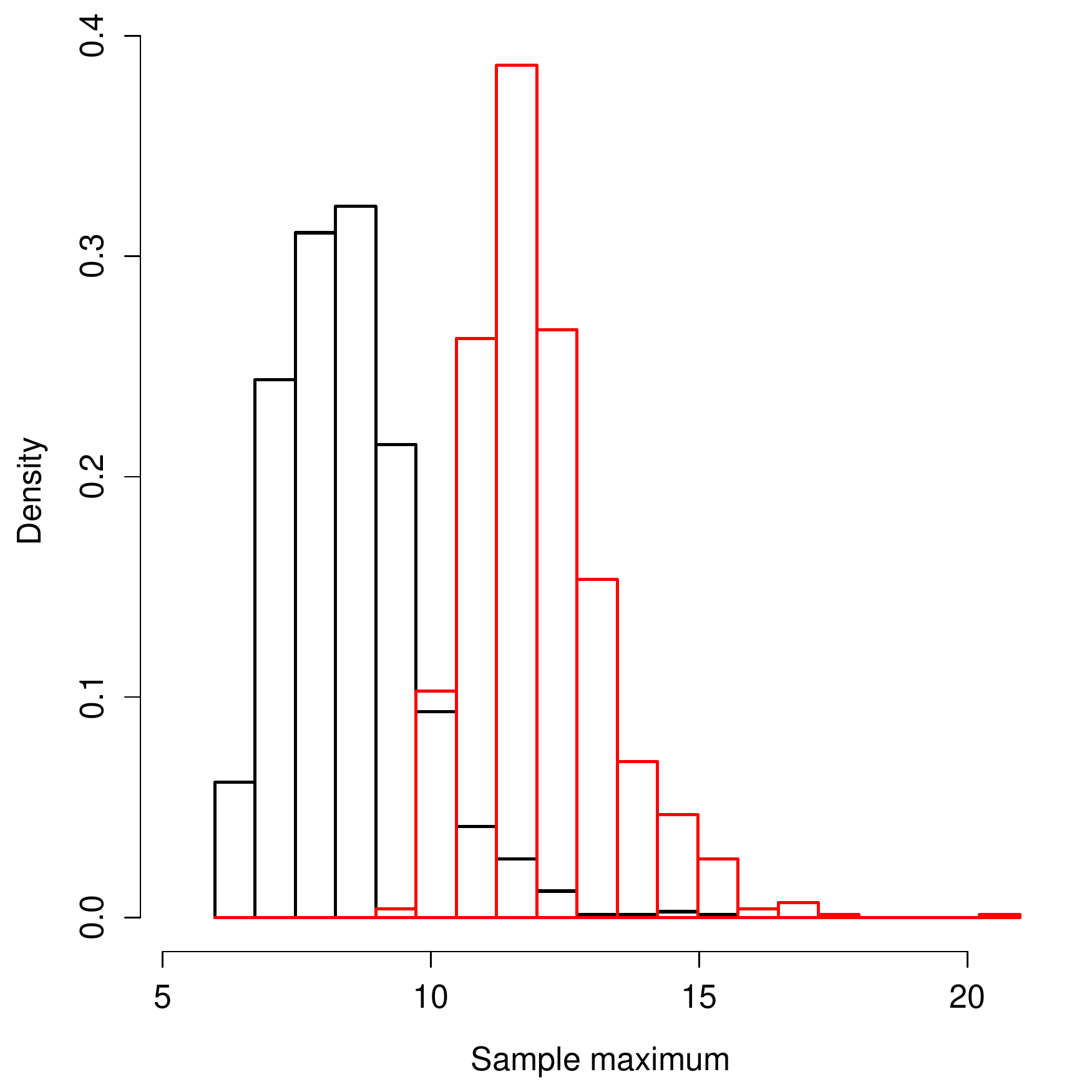} \\ \includegraphics[width=5cm]{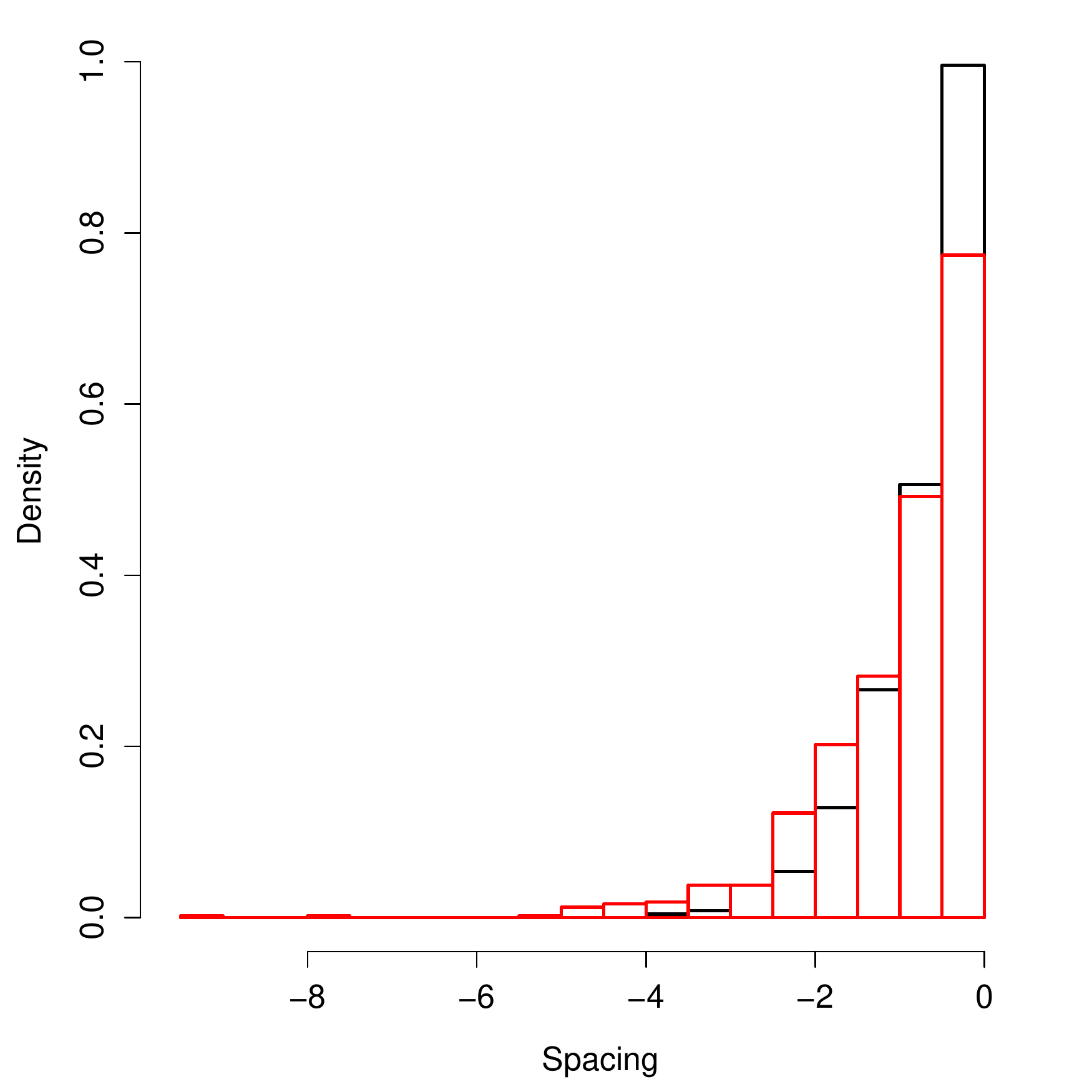} \qquad  \includegraphics[width=5cm]{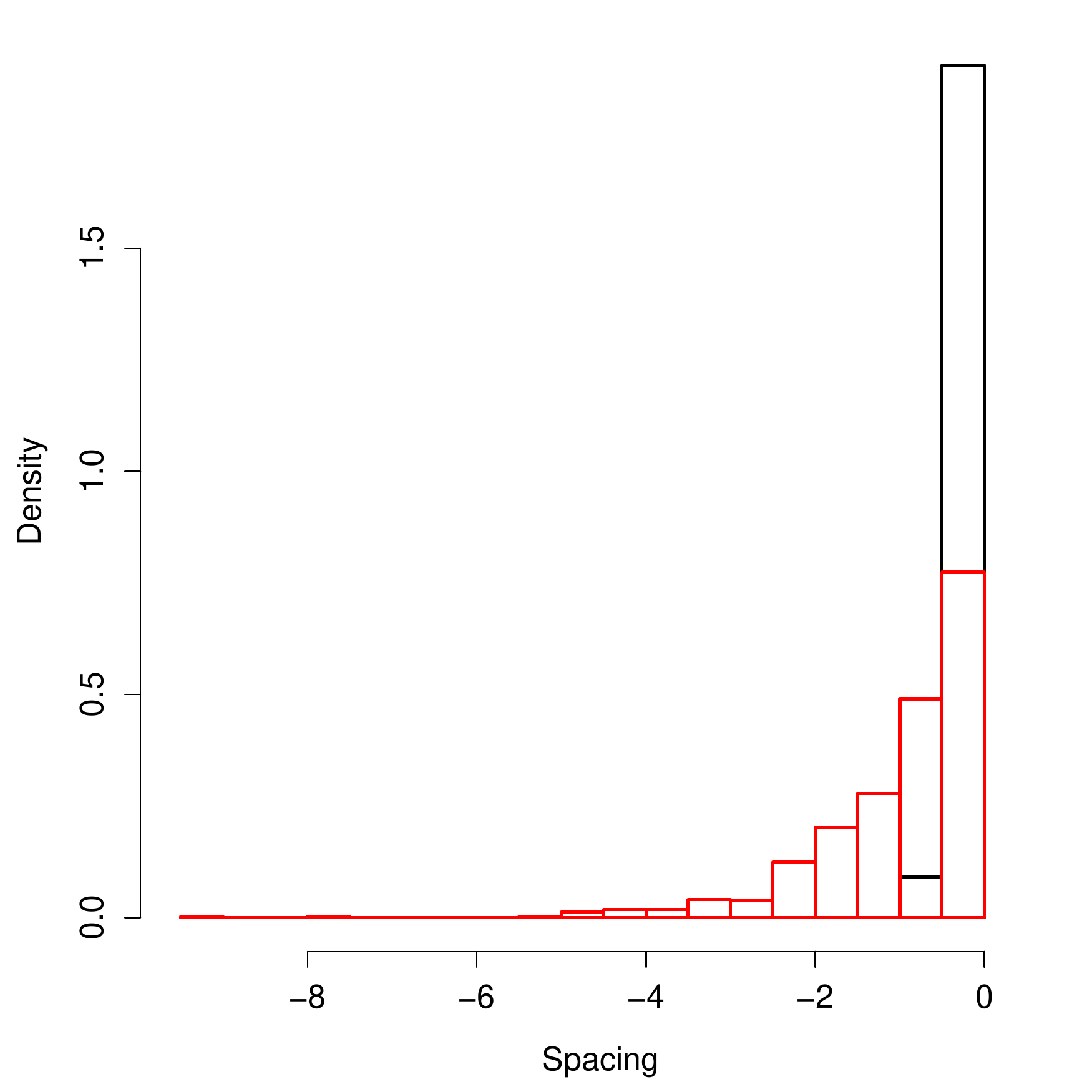} 
    \caption{Illustration of naive resampling in the replicated data setting. Histograms of maxima (first row) and spacings (second row) for $1000$ samples, with Gaussian dependence (left column) and $t$-dependence with degree of freedom 3 (right column). In black color: original sample with spatial dependence. In red color: I.i.d. samples. }
    \label{fig:hist-naive}
\end{figure}

\newmodif{
We now illustrate in more detail the issue of too large maxima and too large spacings around the maximum on replicated data, using two simulation settings corresponding to asymptotic independence and asymptotic dependence, respectively;  recall Equation~\eqref{eq:ad} for the definition of asymptotic dependence.
\begin{description}
\item Setting 1 (asymptotic independence): Each training sample consists of $100$ centered Gaussian fields $W$ with unit variance and exponential correlation with range $0.2$, simulated on a regular $30\times 30$ grid covering the unit square$[0,1]^2$. 
\item Setting 2 (asymptotic dependence):  Same as setting 1, but now we study $t$-random fields with degree of freedom $\nu=3$. Each spatial Gaussian field is rescaled with a random variable $S$ where $(\nu/S)^2\sim \chi^2_\nu$. 
\end{description}
For each setting, we generated $1000$ training samples and transformed margins to the standard exponential scale using the exact, theoretical marginal distributions. We  also generated $1000$  i.i.d. standard exponential samples with the corresponding number of values ($100\times 30 \times 30$).
For each of the $1000$ samples on exponential scale, we then extracted the global maximum, as well as the spacing between the global maximum and the next largest value. Figure~\ref{fig:hist-naive} shows histograms of these summaries (spacings have negative sign here). In the Gaussian case (setting 1), the histograms are overall relatively similar between Gaussian dependence and independent resampling, but stronger differences arise for the occurrence of large absolute values. In the $t$-case, we see striking differences between samples with dependent and independent data: the i.i.d. sample produces spacings and maxima that are far too high in absolute value. 
}

\end{document}